\documentclass[twocolumn]{aastex63}

\usepackage[T1]{fontenc}
\usepackage{amsmath}
\newcommand{\bl}[1]{\mbox{\boldmath$ #1 $}}

\shorttitle{Dynamical freeze-out in protoplanetary disks}
\shortauthors{Molyarova et al.}
\graphicspath{{./}{figures/}}

\begin{document}

\title{Gravitoviscous protoplanetary disks with a dust component. \\ V. The dynamic model for freeze-out and sublimation of volatiles}

\correspondingauthor{Tamara Molyarova}
\email{molyarova@inasan.ru}

\author[0000-0003-0448-6354]{Tamara Molyarova}
\affiliation{Institute of Astronomy, Russian Academy of Sciences, 48 Pyatnitskaya St., Moscow, 119017, Russia}

\author[0000-0002-6045-0359]{Eduard I. Vorobyov}
\affiliation{Institute of Astronomy, Russian Academy of Sciences, 48 Pyatnitskaya St., Moscow, 119017, Russia}
\affiliation{Department of Astrophysics, University of Vienna, Vienna, 1180, Austria}

\author[0000-0002-4324-3809]{Vitaly Akimkin}
\affiliation{Institute of Astronomy, Russian Academy of Sciences, 48 Pyatnitskaya St., Moscow, 119017, Russia}

\author[0000-0002-2479-3370]{Aleksandr Skliarevskii}
\affiliation{Institute of Astronomy, Russian Academy of Sciences, 48 Pyatnitskaya St., Moscow, 119017, Russia}

\author[0000-0002-5111-0395]{Dmitri Wiebe}
\affiliation{Institute of Astronomy, Russian Academy of Sciences, 48 Pyatnitskaya St., Moscow, 119017, Russia}

\author[0000-0001-9818-0588]{Manuel G\"{u}del}
\affiliation{Department of Astrophysics, University of Vienna, Vienna, 1180, Austria}

\begin{abstract}
The snowlines of various volatile species in protoplanetary disks are associated with abrupt changes in gas composition and dust physical properties. Volatiles may affect dust growth, as they cover grains with icy mantles that can change the fragmentation velocity of the grains. In turn, dust coagulation, fragmentation, and drift through the gas disk can contribute to the redistribution of volatiles between the ice and gas phases. Here we present the hydrodynamic model FEOSAD for protoplanetary disks with two dust populations and volatile dynamics. We compute the spatial distributions of major volatile molecules (H$_2$O, CO$_2$, CH$_4$, and CO) in the gas, on small and  grown dust, and analyze the composition of icy mantles over the initial 0.5~Myr of disk evolution. We show that most of ice arrives to the grown dust through coagulation with small grains. Spiral structures and dust rings forming in the disk, as well as photodissociation in the outer regions, lead to the formation of complex snowline shapes and  multiple snowlines for each volatile species. During the considered disk evolution, the snowlines shift closer to the star, with their final position being a factor 4--5 smaller than that at the disk formation epoch. 
We demonstrate that volatiles tend to collect in the vicinity of their snowlines, both in the ice and gas phases, leading to the formation of thick icy mantles potentially important for dust dynamics. The dust size is affected by a lower fragmentation velocity of bare grains in the model with a higher turbulent viscosity.
\end{abstract}


\section{Introduction} \label{sec:intro}

Volatile chemical species are substantial components of protostellar disks, residing both in the gas phase and in icy mantles of dust grains. The amount of ice in cold regions of protostellar disks is comparable to that of rock \citep{1981PThPS..70...35H,1985Icar...62....4S,2003ApJ...591.1220L,2014prpl.conf..363P}, so icy mantles can contribute significantly to the mass of solid material, affecting grain aerodynamic properties, collisional evolution, as well as dust emission.

The distribution of ices in protoplanetary disks is often qualitatively described via a concept of a snowline. A snowline can be broadly defined as a border between a disk region where a certain volatile mostly resides in the gas-phase, and a region where it mostly resides in the solid (ice) phase. Snowlines are often defined in terms of a so-called ``freeze-out temperature'', but the dependence of this temperature on local disk conditions makes this definition of limited use. More accurately positions of snowlines can be determined from the comparison of freeze-out and sublimation rates \citep{2011ApJ...735..131S, 2016ApJ...821...82O, 2015A&A...582A..41H}.

Snowlines of major disk volatiles reflect changes in physical conditions that can significantly alter the dust properties and cause sharp jumps in the dust distribution. The changes in dust properties due to the presence of snowlines can be caused by various mechanisms. Ices can contribute to dust evolution as they raise the dust fragmentation velocity and therefore soften the dust fragmentation barrier \citep{2009ApJ...702.1490W,2019ApJ...878..132O}, causing observable changes in dust emission \citep{2015ApJ...815L..15B}. Specifically, higher stickiness or fragility of dust grains in the vicinity of snowlines, especially that of water, is suggested as a possible explanation of observed ring-like structures in dust continuum emission in protoplanetary disks \citep{2015ApJ...815..109P,2016ApJ...821...82O,2017ApJ...845...68P}. Later in the disk evolution, snowlines may favor dust growth and planetesimal formation \citep{2012ApJ...752..106O,2015A&A...577A..65B}.

Snowlines are not fixed at their positions in the disk. They can move gradually or respond abruptly to various changes in the disk. In particular, their locations can be affected by the dust radial drift \citep{2010ApJ...719.1633H,2015ApJ...815..109P}. In turn, snowlines themselves affect the dust drift \citep{2017ApJ...845...68P}. In disks around FU~Orionis-type stars (FUors), snowline dynamics caused by an intense luminosity outburst can also be a trigger for a significant shake-up in dust evolution processes and is claimed to be detected in observations \citep{2016Natur.535..258C}. The burst can induce a phenomenon known as preferential recondensation of ice, followed by oligarchic growth of icy grains \citep{2017MNRAS.465.1910H}. Dynamics of volatiles, particularly in disks around FUors, was previously considered by \citet{2013A&A...557A..35V} for CO$_2$ and CO.

The tight connection between the dust growth and destruction and grain mantle properties implies that the evolution of volatiles should not be considered in a simplified snowline framework. One rather has to treat the volatile evolution alongside with the disk evolution, taking into account various connections between the disk parameters and its chemical inventory.

Combining astrochemical modeling with a dynamical disk model is a computationally challenging task. There are a number of studies exploring chemical composition of a self-gravitating disk within a hydrodynamic model \citep{2011MNRAS.417.2950I,2015MNRAS.453.1147E,2017MNRAS.472..189I}. The authors show how the gravitational instability may affect the gas chemical composition. \citet{2011MNRAS.417.2950I} show that for many species, adsorption and desorption are the most important processes determining their gas-phase abundances. A number of studies specifically address the behavior of volatiles in the vicinity of their snowlines, combining freeze-out and evaporation with dust dynamics and/or evolution \citep{1988Icar...75..146S,2004ApJ...614..490C,2017A&A...608A..92D,2017A&A...600A.140S,2018ApJ...864...78K}.

In this paper, we model the evolution of volatiles in a protoplanetary disk using an upgrade of the 2D hydrodynamic code FEOSAD with two evolving dust populations \citep{2018A&A...614A..98V}  hereinafter referred to as small and grown dust. The modified code includes time-dependent adsorption and desorption of four volatiles (H$_2$O, CO$_2$, CH$_4$, and CO), an updated dust growth model, and accounts for the effect of icy mantles on the dust fragmentation velocity. We calculate the distributions of the volatiles in the gas phase and on the surface of small and grown dust grains and discuss the effects of gas and dust dynamics and evolution on the distribution of ices in protostellar disks. The distributions of ices obtained in our work can be used to explore the possible effects of icy mantles, their composition, and temperature on dust evolution via the dust fragmentation velocity.

The paper is organized as follows.
In Section~\ref{sec:model} we describe the model, specifically addressing evolution of icy mantles in Section~\ref{sec:accdes}, rates of freeze-out and desorption in Section~\ref{sec:desorption}, and the impact of the ices on fragmentation in Section~\ref{sec:vfrag}. In Section~\ref{sec:results} we present and analyze the results of two simulation runs: global disk properties and evolution are described in Sections~\ref{sec:2d} and~\ref{sec:space-time}, the volatiles and the effects associated with the snowlines in Sections~\ref{sec:volatiles} and~\ref{sec:properties}, and dust rings formation in Section~\ref{sec:dustrings}. The conclusions are summarized in Section~\ref{sec:conclusions}.

\section{Model description}
\label{sec:model}

We use the FEOSAD code (Formation and Evolution Of Stars And Disks), which is a numerical hydrodynamics code written in the thin-disk limit and designed to model the co-evolution of gas and dust in a protoplanetary disk over thousands of orbital periods~\citep{2018A&A...614A..98V}. The simulations start from the gravitational collapse of a flattened and rotating prestellar core, proceed through the disk formation phase, and end within the T~Tauri stage when most of the core material has already accreted onto the forming star plus disk system. The simulations capture spatial scales from sub-au to thousands of au with a numerical resolution that is sufficient to resolve disk substructures on au-scales (e.g., spiral arms). Long evolutionary times and enormously different spatial scales make fully three-dimensional simulations prohibitively expensive and a one-dimensional viscous evolution equation for the surface density of gas, as introduced in \citet{1981ARA&A..19..137P}, is often employed to address this problem \citep[e.g.,][]{2016MNRAS.461.2257K}. The obvious advantage of the thin-disk models is that they, unlike the viscous evolution equation, use the full set of hydrodynamic equations, being at the same time computationally inexpensive in comparison to the full three-dimensional approach. The thin-disk models, therefore, present an indispensable tool for studying the long-term evolution of protoplanetary disks for a large number of model realizations with a much higher realism than is offered by the simple one-dimensional viscous disk models. 

FEOSAD considers the following physical processes: disk self-gravity, turbulent viscosity parameterized using the $\alpha$-approach of \citet{1973A&A....24..337S}, radiative cooling, stellar and background heating, and dust drift through the gas including backreaction of dust on gas.
The hydrodynamic equations of mass, momentum, and internal energy for the gas component read
\begin{equation}
\label{cont}
\frac{{\partial \Sigma_{\rm g} }}{{\partial t}}   + \nabla_p  \cdot 
\left( \Sigma_{\rm g} \bl{v}_p \right) =0,  
\end{equation}

\begin{eqnarray}
\label{mom}
\frac{\partial}{\partial t} \left( \Sigma_{\rm g} \bl{v}_p \right) +  [\nabla \cdot \left( \Sigma_{\rm
g} \bl{v}_p \otimes \bl{v}_p \right)]_p & =&   - \nabla_p {\cal P}  + \Sigma_{\rm g} \, \bl{g}_p + \nonumber
\\ 
+ (\nabla \cdot \mathbf{\Pi})_p  - \Sigma_{\rm d,gr} \bl{f}_p,
\end{eqnarray}
\begin{equation}
\frac{\partial e}{\partial t} +\nabla_p \cdot \left( e \bl{v}_p \right) = -{\cal P} 
(\nabla_p \cdot \bl{v}_{p}) -\Lambda +\Gamma + 
\left(\nabla \bl{v}\right)_{pp^\prime}:\Pi_{pp^\prime}, 
\label{energ}
\end{equation}
where subscripts $p$ and $p^\prime$ denote the planar components $(r,\phi)$ in polar coordinates, $\Sigma_{\rm g}$ is the gas mass surface density,  $e$ is the internal energy per surface area, ${\cal P}$ is the vertically integrated gas pressure calculated via the ideal  equation of state as ${\cal P}=(\gamma-1) e$ with $\gamma=7/5$, $f_p$ is the friction force between gas and dust, $\bl{v}_{p}=v_r \hat{\bl r}+ v_\phi \hat{\bl \phi}$  is the gas velocity in the disk plane, and is $\nabla_p=\hat{\bl r} \partial / \partial r + \hat{\bl \phi} r^{-1} \partial / \partial \phi $ the gradient along the planar coordinates of the disk. The gravitational acceleration in the disk plane,  $\bl{g}_{p}=g_r \hat{\bl r} +g_\phi \hat{\bl \phi}$, includes the gravity of the central protostar when formed and takes into account disk self-gravity of both gas and dust found by solving the Poisson integral \citep{1987gady.book.....B}. Turbulent viscosity is taken into account via the viscous stress tensor  $\mathbf{\Pi}$ and the magnitude of kinematic viscosity $\nu=\alpha c_{\rm s} H$ is parameterized using the $\alpha$-prescription of \citet{1973A&A....24..337S}, where $c_{\rm s}$ is the sound speed  and $H$ is the vertical scale height of the gas disk calculated using an assumption of local hydrostatic equilibrium. The expressions for radiative cooling $\Lambda$ and irradiation heating $\Gamma$ (the latter includes the stellar and background irradiation) can be found in \citet{2018A&A...614A..98V}. The temperature of background radiation is set equal to 15~K.

\begin{table*}
\caption{Model parameters. $M_{\rm core}$ is the initial core mass, $\beta$ is the ratio of rotational to gravitational energy, $T_{\rm init}$ is the initial gas temperature, $\Omega_0$ and $\Sigma_0$ are the angular velocity and gas surface density at the core center, respectively, $r_0$ is the radius of the central plateau in the initial core, $R_{\rm out}$ is the initial radius of the core outer boundary, and $\alpha$ is the viscous parameter. $M_{\star}$ and $M_{\rm disk}$ are the masses of the central star and the disk  at the end of simulations (500\,kyr). Note that about 10\% of the initial core mass was assumed to be evacuated with jets and outflows in our model. A small amount of mass remains in the envelope by the end of simulations. }
\label{tab:parameters}
 \begin{tabular}{lcccccccccc}
 \hline\noalign{\smallskip}
Model  & $M_{\rm core}$ & $\beta$ & $T_{\rm init}$ & $\Omega_0$ & $\Sigma_0$ & $r_0$ & $R_{\rm out}$ & $\alpha$ & $M_{\star}$ & $M_{\rm disk}$ \\
       & ($M_{\odot}$)  & (\%)   & (K) & (km s$^{-1}$ pc$^{-1}$) & (g cm$^{-2}$) & (au) & (au)       &   --     & ($M_{\odot}$) & ($M_{\odot}$) \\
    \hline\noalign{\smallskip}
1  & 0.66 & 0.23 & 15 & 2.0 & $1.73\times 10^{-1}$ & 1029 & 6082 & $10^{-2}$ & 0.46 & 0.15 \\
2  & 0.66 & 0.23 & 15 & 2.0 & $1.73\times 10^{-1}$ & 1029 & 6082 & $10^{-4}$ & 0.42 & 0.20 \\
  \noalign{\smallskip}\hline
\end{tabular}
\end{table*}

The gas and dust components co-evolve and interact with each other via the common gravity and friction force that takes the backreaction of dust on gas into account following the analytic method described and extensively tested in \citet{2018ARep...62..455S}. Initially all dust is in the form of small submicron grains, but is allowed to grow as the collapse proceeds and the disk forms and evolves. Small submicron dust is rigidly linked to the gas, while grown dust can be dynamically decoupled from gas. In this work, we upgrade the FEOSAD code to consider the evolution of icy mantles on the surface of dust grains and we also use an improved model for dust growth as described in Section~\ref{sec:newgrowth}. The evolution of icy mantles includes freeze-out and sublimation of volatiles (Sections~\ref{sec:accdes} and~\ref{sec:desorption}), exchange of ices between small and grown dust populations due to coagulation and fragmentation of dust grains (Section~\ref{sec:growth}), and the transport of volatile species with gas and dust populations. The presence of icy mantles also affects dust evolution, as described in Section~\ref{sec:vfrag}. We consider the most abundant species, namely water, carbon monoxide, carbon dioxide, and methane, which can be present in their gas form or as icy mantles on the two dust populations. No surface reactions are taken into account.

The transport of mass and angular momentum in our numerical model is controlled by the strength of gravitational and viscous torques. The former arise at the early evolution when the disk is gravitationally unstable to develop a spiral structure, and are taken into account self-consistently through the gravity force ${\bl g}_p$, which includes disk self-gravity. The viscous torques operate through the entire disk evolution (but begin to dominate over gravitational torques once the spiral structure diminishes) and are taken into account via the viscous stress tensor ${\bl\Pi}$. Here,  we adopt two values of the $\alpha$-parameter: $10^{-2}$ and $10^{-4}$, which can be expected for a disk with developed and suppressed magnetorotational instability, respectively \citep[e.g.,][]{2018ApJ...868...27Y}.

As initial conditions,  we consider flattened prestellar cores that are expected to form through the slow expulsion of magnetic field due to ambipolar diffusion, with the angular momentum remaining constant during axially symmetric core compression \citep{1997ApJ...485..240B}. The corresponding gas surface density $\Sigma_{\rm g}$ and angular velocity $\Omega$ of the prestellar core can be expressed as follows
\begin{equation}
\Sigma_{\rm g}=\frac{r_{0}\Sigma_{\rm g,0}}{\sqrt{r^{2}+r_{0}^{2}}},
\label{eq:sigma}
\end{equation}
\begin{equation}
\Omega=2\Omega_{0}\left(\frac{r}{r_{0}}\right)^{-2}\left[\sqrt{1+\left(\frac{r}{r_{0}}\right)^{2}}-1\right],
\label{eq:omega}
\end{equation}
where $\Sigma_{\rm g,0}$ and $\Omega_{0}$ are the gas surface density and angular velocity at the center of the core and $r_{0}$ is the radius of the central plateau. The initial gas temperature in collapsing cores is $T_{\mathrm{init}}=15\,\mathrm{K}$. The initial dust-to-gas mass ratio is 1:100. We consider two models, their initial parameters being presented in Table~\ref{tab:parameters}. These models differ mainly in the values of the $\alpha$-parameter, $\alpha=10^{-2}$ for  Model~1 and $\alpha=10^{-4}$ for Model~2.

The simulations were performed on the polar grid ($r,\phi$) with $256\times 256$ grid cells using the operator-split procedure similar in methodology to the ZEUS code \citep{1992ApJS...80..753S}. The radial grids are logarithmically spaced, while the azimuthal grids have equal spacing. The central disk region of 0.8~au in radius is carved out and replaced with the sink cell to avoid too small time steps imposed by the Courant condition. The minimum size of the grid cell that is adjacent to the inner boundary is 0.029~au, while the sizes of grid cells at 10 and 100~au are 0.35 and 3.5~au, respectively. We impose the carefully designed inflow-outflow boundary condition at the sink--disk interface, which helps us to minimize the effects of the boundary on the gas and dust flow. More details on the boundary conditions and solution procedure can be found in \citet{2018A&A...614A..98V}.

\subsection{Updated dust growth model}
\label{sec:newgrowth}

In this section, we describe the dust growth scheme, which is an updated version of the scheme first presented in \citet{2018A&A...614A..98V}. In the FEOSAD code, small and grown  dust components are represented by their surface densities $\Sigma_{\rm d,sm}$ and $\Sigma_{\rm d,gr}$, respectively. Each dust population has the size distribution $N(a)$ described by a simple power-law function $N(a)= C a^{-p}$ with a fixed exponent $p=3.5$ and a normalization constant $C$. For small dust, the minimum size is $a_{\rm min}=5\times 10^{-7}$\,cm and the maximum size is $a_*=10^{-4}$\,cm. For grown dust, $a_*$ is the minimum size and $a_{\rm max}$ is the maximum size, which can vary due to dust coagulation and fragmentation.

The dynamics of small and grown dust grains is described by the following continuity and momentum equations (note that small dust is assumed to be strictly dynamically coupled to gas)
\begin{equation}
\label{contDsmall}
\frac{{\partial \Sigma_{\rm d,sm} }}{{\partial t}}  + \nabla_p  \cdot 
\left( \Sigma_{\rm d,sm} \bl{v}_{ p} \right) = - S(a_{\rm max}),  
\end{equation}
\begin{equation}
\label{contDlarge}
\frac{{\partial \Sigma_{\rm d,gr} }}{{\partial t}}  + \nabla_p  \cdot 
\left( \Sigma_{\rm d,gr} \bl{u}_{ p} \right) = S(a_{\rm max}),  
\end{equation}
\begin{eqnarray}
\label{momDlarge}
\frac{\partial}{\partial t} \left( \Sigma_{\rm d,gr} \bl{u}_{ p} \right) +  \left[\nabla \cdot \left( \Sigma_{\rm
d,gr} \bl{u}_{ p} \otimes \bl{u}_{p} \right)\right]_{ p}  &=&   \Sigma_{\rm d,gr} \, \bl{g}_{p} + \nonumber \\
 + \Sigma_{\rm d,gr} \bl{f}_{ p} + S(a_{\rm max}) \bl{v}_{p},
\end{eqnarray}
where $\bl{u}_{ p}$ are the planar components ($r,\phi$) of the grown dust velocity.

\begin{figure*}
\begin{centering}
\includegraphics[width=0.66\columnwidth]{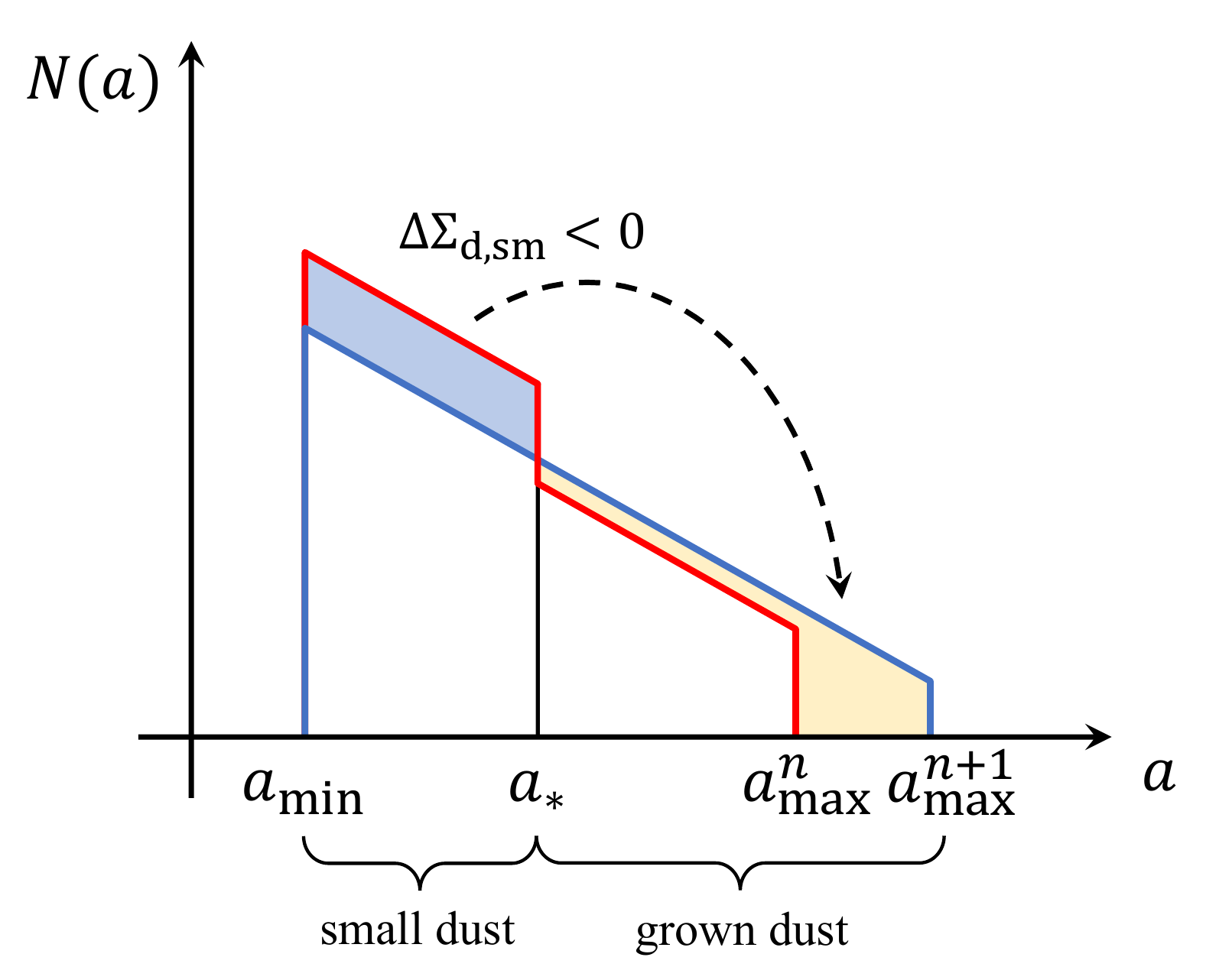}
\includegraphics[width=0.66\columnwidth]{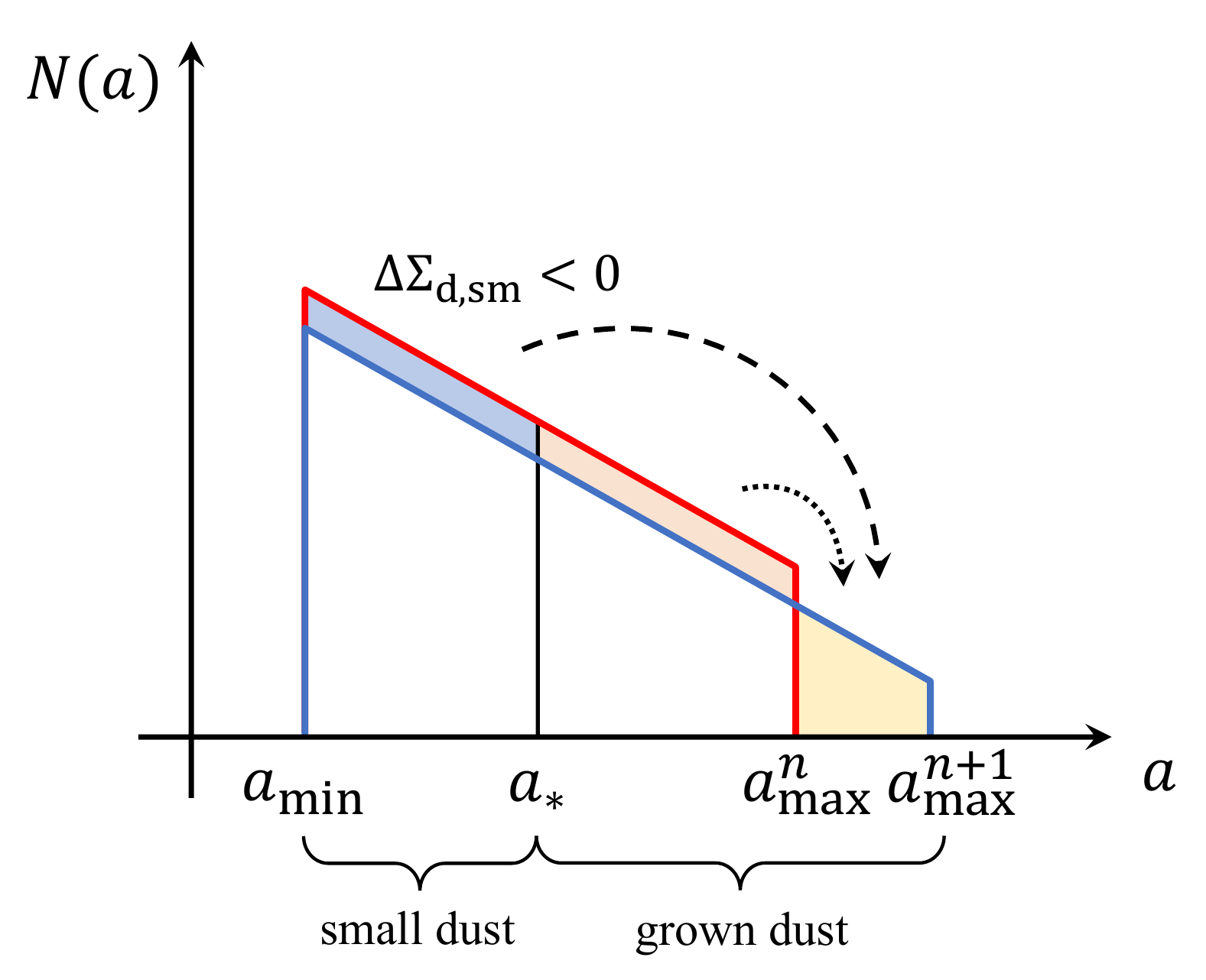}
\includegraphics[width=0.66\columnwidth]{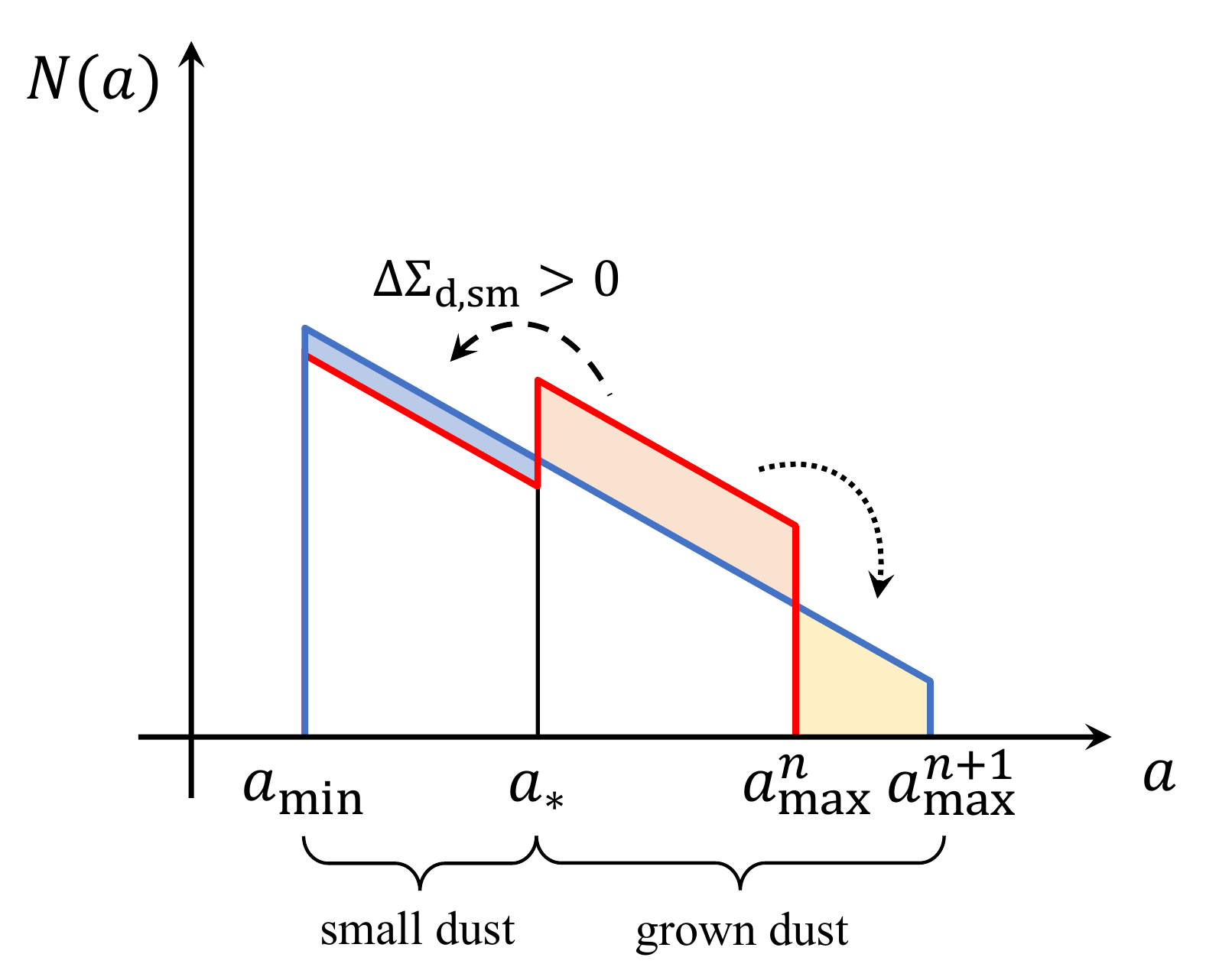}
\par\end{centering}
\caption{\label{fig:0} Illustration of the adopted scheme for dust evolution. Examples of dust size distribution before (red lines) and after (blue lines) dust evolution step. The change in small dust surface density $\Delta \Sigma_{\rm d,sm}$ is shaded with blue, added and removed grown dust is marked by yellow and orange colors, respectively. Different values of the gap at the discontinuity between small and grown dust populations are shown: in favor of small dust (left), no discontinuity (center), in favor of grown dust (right). Only the case of $a_{\rm max}^{n+1}>a_{\rm max}^{n}$ is presented.}
\end{figure*}

The quantity $S(a_{\rm max})$ is the rate of small-to-grown dust conversion per disk surface area (in g~s$^{-1}$~cm$^{-2}$). The general idea behind our dust conversion scheme is illustrated in Figure~\ref{fig:0}, showing the dust size distributions at $n$ and $n+1$ time steps with the red and blue lines, respectively. The blue area schematically represents the amount of small dust (per surface area) converted into grown dust during one hydrodynamic time step. We note that the area highlighted in orange is transferred above $a^n_{\rm max}$, but it does not change the mass of grown dust. The change in the surface density of small dust due to conversion into grown dust $\Delta \Sigma_{\rm d,sm}=\Sigma_{\rm d,sm}^{n+1}-\Sigma_{\rm d,sm}^{n}$ can be expressed as
\begin{equation}
    \Delta\Sigma_{\mathrm{d,sm}} =  \Sigma_{\mathrm{d,tot}}^n 
    \frac
    { I_1
    \left( 
    C_{\rm sm}^{n+1} C_{\rm gr}^n \, I_2 - 
    C_{\rm sm}^n C_{\rm gr}^{n+1} I_3
    \right)
    }
    {
    \left( 
    C_{\rm sm}^{n+1} I_1+  C_{\rm gr}^{n+1} I_3
    \right)  
    \left( 
    C_{\rm sm}^{n} I_1 + C_{\rm gr}^{n} I_2
    \right)
    \label{growth1}
    },
\end{equation}
where $\Sigma_{\rm d,tot}^n=\Sigma_{\rm d,sm}^n+\Sigma_{\rm d,gr}^n$ is the total dust surface density, $C_{\rm sm}$ and $C_{\rm gr}$ are the normalization constants for small and grown dust size distributions at the current ($n$) and next ($n+1$) time steps, and the integrals $I_1$, $I_2$, and $I_3$ are defined as
\begin{eqnarray}
   I_1= \int_{a_{\rm min}}^{a_*} a^{3-\mathrm{p}}da,\nonumber
   \\
   I_2=\int_{a_*}^{a_{\mathrm{max}}^{n}} a^{3-\mathrm{p}}da,
   \\
   I_3=\int_{a_*}^{a_{\mathrm{max}}^{n+1}} a^{3-\mathrm{p}}da.\nonumber
\end{eqnarray}

By introducing different normalization constants for small and grown dust we implicitly assume that the dust size distribution can be discontinuous at $a_\ast$. In the original paper \citep[see eq. 12 in][]{2018A&A...614A..98V}, we  set $C_{\rm sm}^{n} = C_{\rm gr}^{n}$ and $C_{\rm sm}^{n+1} = C_{\rm gr}^{n+1}$, effectively suggesting that the dust grows in such a manner that the distribution is continuous across $a_\ast$. However, the dust surface density in a given computational cell can change not only due to growth (the right-hand side term in Eqs.~\ref{contDsmall} and \ref{contDlarge}), but also due to dust flow through the cell (the second term on the left-hand side). Because of different dynamics of small and grown dust, a discontinuity may develop at $a_\ast$ after we account for advection.

To account for this effect, we modify our approach in the following manner. We suggest that the dust size distribution can develop a discontinuity at $a_\ast$ due to differential dust dynamics. However, dust growth due to the $S$ term smooths out the discontinuity each time it could occur after the advection step. Physically this assumption corresponds to the dominant role of dust evolution over the dust flow in setting the fixed shape of the dust size distribution. 
This can be achieved by setting $C_{\rm sm}^{n+1}=C_{\rm gr}^{n+1}$ in Equation~(\ref{growth1}), while keeping $C_{\rm sm}^{n}$ and $C_{\rm gr}^{n}$. The resulting expression is then
\begin{equation}
    \Delta\Sigma_{\mathrm{d,sm}} =  \Sigma_{\mathrm{tot}}^n 
    \frac
    {
    I_1
    \left( 
    C_{\rm gr}^n I_2 -  C_{\rm sm}^n I_3
    \right)
    }
    { I_4
    \left(  C_{\rm sm}^{n} I_1 + C_{\rm gr}^{n} I_2
    \right)
    },
    \label{growth2}
\end{equation}
where the integral $I_4$ is defined as
\begin{equation}
    I_4= \int_{a_{\mathrm{min}}}^{a_{\rm max}^{n+1}} a^{3-\mathrm{p}}da,
\end{equation}
and the normalization constants can then be found as
\begin{equation}
\label{c1}
    C_{\rm sm}^{n} = 
    \frac
    {
    3 \Sigma_{\rm d,sm}^{n} \Delta s
    }
    {
    4 \pi \rho _{\rm s} I_1
    }, \,\,\,\,
        C_{\rm gr}^{n} = 
    \frac
    {
    3 \Sigma_{\rm d,gr}^{n} \Delta s
    }
    {
    4 \pi \rho _{\rm s} I_2
    },
\end{equation}
where $\rho_{\rm s}=3$\,g\,cm$^{-3}$ is the material density of dust grains and $\Delta s$ is the total dust surface area in a given computational cell. Substituting the normalization constants in Equation~(\ref{growth2}), we finally obtain
\begin{equation}
\label{final}
    \Delta\Sigma_{\mathrm{d,sm}} = 
    \frac
    {
    \Sigma_{\rm d,gr}^n I_1  - 
    \Sigma_{\rm d,sm}^n I_3
    }
    {
    I_4
    }.
\end{equation}
The rate of small-to-grown dust conversion during one time step $\Delta t$ is then written as 
\begin{equation}
S(a_{\rm max})=-\Delta \Sigma_{\rm d,sm}/\Delta t.
\end{equation}

To finalize the calculation of $S(a_{\rm max})$, the maximum radius of grown dust $a_{\rm max}$ must be computed at each time step and in each computational cell. The evolution of $a_{\rm max}$ is described as
\begin{equation}
{\frac{\partial a_{\rm max}}{\partial t}} + (u_{\rm p} \cdot \nabla_p ) a_{\rm max} = \cal{D},
\label{dustA}
\end{equation}
where the growth rate $\cal{D}$ accounts for the change in $a_{\rm max}$ due to coagulation and the second term on the left-hand side accounts for the change of $a_{\rm max}$ due to dust flow through the cell.  We write the source term $\cal{D}$ as
\begin{equation}
\cal{D}=\frac{\rho_{\rm d} {\it v}_{\rm rel}}{\rho_{\rm s}},
\label{GrowthRateD}
\end{equation}
where $\rho_{\rm d}$ is the total dust volume density, $\rho_{\rm s}$ is the material density of dust grains, and $v_{\rm rel}$ is the dust-to-dust collision velocity. We consider Brownian and turbulence-induced particle velocities, but not the drift velocity, which is of less importance for the grain sizes relevant for our study \citep[see, e.g., Figure 3 in][]{2014prpl.conf..339T}. The adopted approach is similar to the monodisperse model of \citet{1997A&A...319.1007S} and is described in more detail in \citet{2018A&A...614A..98V}. The value of $a_{\rm max}$ is capped by the fragmentation barrier \citep{2012A&A...539A.148B} defined as
\begin{equation}
 a_{\rm frag}=\frac{2\Sigma_{\rm g}v_{\rm frag}^2}{3\pi\rho_{\rm s}\alpha c_{\rm s}^2},
 \label{eq:afrag}
\end{equation}
where $v_{\rm frag}$ is the fragmentation velocity (see Eq.~\ref{eq:vfrag}). Whenever $a_{\rm max}$ exceeds $a_{\rm frag}$, the growth rate $\cal{D}$ is set to zero. We note that the assumption that any discontinuity in the slope of the dust size distribution (which may develop as a result of dust drift) smooths out owing to dust growth implies also that grown-to-small dust conversion can occur even if $a_{\rm max}<a_{\rm frag}$.

As dust grows, the span in the dust sizes and in the corresponding Stokes numbers covered by the grown component may become substantial. However, we calculate the stopping time using $a_{\rm max}$, meaning that our model tracks the dynamics of the upper end of the dust size distribution where most of the dust mass reservoir is located if $p=3.5$. We introduced an effective Stokes number for grown dust population in the bidisperse approach in \citet{2020MNRAS.499.5578A}. A more rigorous approach to studying dust dynamics requires the use of multi-bins with narrower ranges of dust sizes and this model is currently under development.

\subsection{Modeling the evolution of volatiles}
\label{sec:accdes}

The amount of major volatiles in the gas phase and on the icy mantles of the two dust populations is defined by their surface densities: $\Sigma_{s}^{\rm gas}$ for the gas phase, $\Sigma_{s}^{\rm sm}$ for the ice on the surface of small dust, and $\Sigma_{s}^{\rm gr}$ for the ice on the surface of grown dust ($s$ stands for the index of a species). We assume that initially all the volatiles are on the surface of small grains (which are the only grains present at the onset of prestellar core collapse). 

The surface densities of the considered species in the gas and ice phases can change via three main physical processes that are considered at each time step and for every grid cell. First, the surface densities $\Sigma_{s}^{\rm gas}$, $\Sigma_{s}^{\rm sm}$, and $\Sigma_{s}^{\rm gr}$ are updated taking into account freeze-out and sublimation of volatile species for both dust populations. Second, dust evolution (coagulation and fragmentation) is considered, which redistributes the solids (including ices) between small and grown dust fractions.
Finally, the surface densities are updated to take into account the transport of volatile species with gas, small dust grains, and grown dust grains. The advection of volatiles is calculated using the same third-order-accurate piece-wise parabolic method as for the gas and dust components \citep{2018A&A...614A..98V}. The processes concerning the first and second steps are described in detail below.

Expanding the method from \cite{2013A&A...557A..35V}, we describe the phase transitions of volatiles using the following equations

\begin{eqnarray}
\frac{{\rm d}\Sigma_{s}^{\rm gas}}{{\rm d} t}&=&-\lambda_s\Sigma_{s}^{\rm gas}+\eta_s^{\rm sm}+\eta_s^{\rm gr},\label{eq:sig1}\\
\frac{{\rm d}\Sigma_{s}^{\rm sm}}{{\rm d} t}&=&\lambda_s^{\rm sm}\Sigma_{s}^{\rm gas}-\eta^s_{\rm sm},\label{eq:sig2}\\
\frac{{\rm d}\Sigma_{s}^{\rm gr}}{{\rm d} t}&=&\lambda_s^{\rm gr}\Sigma_{s}^{\rm gas}-\eta_s^{\rm gr},\label{eq:sig3}
\end{eqnarray}
where $\lambda_s^{\rm sm}$ and $\lambda_s^{\rm gr}$ (s$^{-1}$), are the adsorption rates onto small and grown dust grains and  $\lambda_s=\lambda_s^{\rm sm}+\lambda_s^{\rm gr}$. We also define $\eta_s=\eta_s^{\rm sm}+\eta_s^{\rm gr}$ (g\,cm$^{-2}$\,s$^{-1}$), where $\eta_s^{\rm sm}$ and $\eta_s^{\rm gr}$ are the desorption rates from small and grown dust populations, respectively. Each of desorption rates is a sum of thermal and photodesorption rates $\eta_s^{\rm sm}=\eta_s^{\rm th,sm}+\eta_s^{\rm ph,sm}$ and $\eta_s^{\rm gr}=\eta_s^{\rm th,gr}+\eta_s^{\rm ph,gr}$. We note that the desorption rates do not depend on $\Sigma_{s}^{\rm sm}$ and $\Sigma_{s}^{\rm gr}$, thus implying the zeroth-order desorption model. This approach implies that only the upper layer of the icy mantle can sublimate, which is reasonable for multilayer mantles. Laboratory studies show that in simulated interstellar conditions the desorption of various molecules is rather described by zeroth-order kinetics \citep{2001MNRAS.327.1165F,2005ApJ...621L..33O,2006A&A...449.1297B}.

For the initial composition of volatiles we use the ice abundances measured with \textit{Spitzer} in protostellar cores \citep{2011ApJ...740..109O}. The initial surface densities of volatile species relative to the initial surface density of gas ($\Sigma_{\rm g}^{\rm init}$) in the models are shown in Table~\ref{tab:abundances}. Although there are estimates of ice mass being comparable to the rock mass \citep{2014prpl.conf..363P}, we adopt a total ice mass from \citet{2011ApJ...740..109O}, which is an order of magnitude lower than that of rock, so that we can suggest that icy mantles are thin and do not affect the dust size. As the total surface density of volatiles at the starting moment is only about 8.6\% of the dust surface density, this approximation is likely valid along the disk evolution. We will check if at some point mass of ice notably exceeds that of rock and discuss possible effects in Section~\ref{sec:properties}.

\begin{table}
\caption{Binding energies, molecular weights, and initial abundances for the considered volatiles. Binding energies for H$_2$O, CO$_2$, and CO are based on the experimental data from \citet{2017SSRv..212....1C} for desorption from crystalline water ice, the binding energy for methane is taken from \citet{1996ApJ...467..684A}. Initial surface densities of ice on small dust relative to total gas surface density $\Sigma_{s}^{\rm sm}/\Sigma_{\rm g}^{\rm init}$ are based on ice abundances relative to water $f_{\rm ice}$ determined for low-mass protostars \citep{2011ApJ...740..109O}.}
\label{tab:abundances}
 \begin{tabular}{lcccc}
 \hline\noalign{\smallskip}
Species  & $E_{\rm b}/k_{\rm B}$ & $m_{\rm sp}$ & $f_{\rm ice}$ & $\Sigma_{s}^{\rm sm}/\Sigma_{\rm g}^{\rm init}$ \\
         & (K)                         & (amu)       & (\%)          &  \\
    \hline\noalign{\smallskip}
H$_2$O & 5770 & 18 & 100 & 3.90$\times10^{-4}$  \\
CO$_2$ & 2360 & 44 & 29  & 2.77$\times10^{-4}$  \\
CO     & 850  & 28 & 29  & 1.76$\times10^{-4}$  \\
CH$_4$ & 1100 & 16 & 5   & 1.74$\times10^{-5}$  \\
  \noalign{\smallskip}\hline
\end{tabular}
\end{table}

The system of Equations~\ref{eq:sig1}, \ref{eq:sig2}, and \ref{eq:sig3} has an analytic solution. We use the explicit expressions for the solution, restricting ice abundances to positive values. The analytic solutions, its asymptotic behavior, and the results of tests using a single-point model are presented in Appendix~\ref{sec:solutions}.

\subsection{Adsorption and desorption rates}
\label{sec:desorption}

Volatiles can be adsorbed from the gas phase to become icy mantles on the surface of dust grains; they can also evaporate from these mantles by thermal or photo desorption. Here we describe the parameterization of these processes adopted in our model. For simplicity, we drop species index $s$ in this subsection.

\subsubsection{Thermal desorption}
\label{sec:thermaldesorption}

The rate coefficient $k_{{\rm td}}$ (s$^{-1}$) for thermal desorption of species from the dust surface can be written as \citep{1987ASSL..134..397T}
\begin{equation}
k_{{\rm td}}=\sqrt{\frac{2 N_{\rm ss} E_{\rm b}}{\pi^2 m_{\rm sp}}} \exp{\left(-\frac{E_{\rm b}}{k_{\rm B}T}\right)},
\label{eq:therm_des_rate_coef}
\end{equation}
where $N_{\rm ss}$ (cm$^{-2}$) is the surface density of binding sites, $E_{\rm b}$ (erg) is the binding energy of the species to the surface, $m_{\rm sp}$ (g) is the species mass, $k_{\rm B}$ is the Boltzmann constant, $T$ is the midplane temperature. In the case of the zeroth-order desorption, the mass desorption rate per disk unit area $\eta_{\rm td}$ (g cm$^{-2}$\,s$^{-1}$) can be approximated as:
\begin{equation}
\eta_{\rm td} =  \chi n_{\rm act}\widetilde{\sigma}_{\rm tot} N_{\rm ss} m_{\rm sp} k_{{\rm td}},
\label{eq:therm_des_volrate}
\end{equation}
where $\chi$ is the fractional abundance of the species in the icy mantle, $n_{\rm act}$ is the number of actively evaporating monolayers, and $\widetilde{\sigma}_{\rm tot}$ (cm$^2$\,cm$^{-2}$) is the total surface area of dust grains (small or grown) per disk unit area. Typically, $2-4$ upper mono-layers can evaporate freely. And, as all considered species have comparable abundances, we assume $\chi n_{\rm act}=1$ as a typical constant value across the disk. For the surface density of binding sites we take $N_{\rm ss}=10^{15}$\,cm$^{-2}$, which is a characteristic value for amorphous water ice \citep{2017SSRv..212....1C}. The adopted values of binding energies (see Table~\ref{tab:abundances}) are based on the experimental data for sublimation from an icy surface \citep{2017SSRv..212....1C}. As most of gas and dust mass lies in high-density midplane regions, we assume equal gas and dust temperatures. The thermal balance includes heating by the stellar and background radiation, dust radiative cooling as well as gas adiabatic and viscous terms \citep{2018A&A...614A..98V}. As we track volatiles on small and grown dust, the corresponding values of $\widetilde{\sigma}_{\rm tot}$ are taken for these two populations. For the power-law size distribution ($p=3.5$) and dust size not dependent on the vertical height $a\ne a(z)$ the total small and grown dust surface areas are:
\begin{eqnarray}
\label{eq:surface}
\widetilde{\sigma}_{\rm tot}^{\rm sm}&=&\frac{3 \Sigma_{\rm d, sm}}{\rho_{\rm s}\sqrt{a_{\rm min} a_*}},\\
\label{eq:surface1}
\widetilde{\sigma}_{\rm tot}^{\rm gr}&=&\frac{3 \Sigma_{\rm d, gr}}{\rho_{\rm s}\sqrt{a_* a_{\rm max}}}.
\end{eqnarray}
While the smallest grains in the disk, $a_{\rm min}=0.005\mu$m, and the boundary between small and grown dust ensembles, $a_*=1\mu$m, do not vary along the vertical height, the maximum grain size $a_{\rm max}$ does vary as dust tend to sediment towards the midplane. Our dust evolution model estimates $a_{\rm max}$ in the disk midplane, where most dust is located. Thus, the value of $\widetilde{\sigma}_{\rm tot}^{\rm gr}$ can be underestimated by a factor of a few, which is acceptable in the context of our relatively simple dust evolution model.

\subsubsection{Photodesorption}
\label{sec:photodes}

In translucent regions ices can be additionally transferred to the gas phase through photodesorption by stellar or interstellar ultraviolet radiation. In the cold regions of the outer disk and envelope, this effect can be notable.
The photodesorption rate coefficient $k_{\rm pd}$ (s$^{-1}$) can be estimated as
\begin{equation}
    k_{\rm pd}=Y F_{\rm UV} \sigma_{\rm m}, 
\label{eq:k_pd}
\end{equation}
where $Y$ (mol~photon$^{-1}$) is the photodesorption yield, $F_{\rm UV}$ (photons~cm$^{-2}$\,s$^{-1}$) is the UV photon flux, and $\sigma_{\rm m}$ (cm$^{2}$) is the molecule UV cross-section.

The corresponding mass rate of photodesorption per disk unit  area $\eta_{\rm pd}$ (g cm$^{-2}$\,s$^{-1}$) is
\begin{equation}
    \eta_{\rm pd} = \chi n_{\rm act}\widetilde{\sigma}_{\rm tot}N_{\rm ss} m_{\rm sp}k_{\rm pd}.
\end{equation}
assuming again $\chi n_{\rm act}=1$ and approximating $\sigma_{\rm m}\approx N_{\rm ss}^{-1}$ one gets
\begin{equation}
    \eta_{\rm pd} =\widetilde{\sigma}_{\rm tot} m_{\rm sp}Y F_{\rm UV}.
\end{equation}
We assume that the photodesorption yield is equal to one of water $Y = 3.5 \times 10^{-3} + 0.13  \exp\left({-336/T_{\rm mp}}\right)$ \citep[][note the misprint in the Y$_0$ value, caption to their Fig.3, see also \cite{1999A&A...342..542W}]{1995Natur.373..405W}, which is probably the upper boundary for experimentally measured values \citep{2020RuCRv..89..430M}.

It is convenient to measure the radiation flux $F_{\rm UV}$ in the units of standard UV field, $F_{\rm UV}=F_0^{\rm UV}G$, where dimensionless parameter $G$ characterizes the strength of the field. The standard interstellar radiation intensity  $I_0(E)$ (photon cm$^{-2}$\,s$^{-1}$\,sr$^{-1}$\,eV$^{-1}$) is approximated in the following form \citep[Eq.11]{1978ApJS...36..595D}
\begin{eqnarray}
I_0(E)= 1.658\times10^{6}\left(\frac{E}{\rm eV}\right) - 2.152\times10^{5}\left(\frac{E}{\rm eV}\right)^2 +\nonumber
\\
+ 6.919\times10^{3}\left(\frac{E}{\rm eV}\right)^3.
\label{eq:stdUV_Draine}
\end{eqnarray}
The corresponding integrated UV intensity $I^{\rm UV}_0$ (photon cm$^{-2}$\,s$^{-1}$\,sr$^{-1}$) in the energy range $6.2-13.6$\,eV ($912-2000$\,\AA) is equal to
\begin{equation}
    I^{\rm UV}_0=\int\limits_{6.2\,{\rm eV}}^{13.6\,{\rm eV}} I_0(E)\,{\rm d}E=1.474\times10^7 {\rm photon~cm^{-2}\,s^{-1}\,sr^{-1}}.
\end{equation}
The flux through the unit area from one hemisphere in case of isotropic  radiation is equal to
\begin{equation}
    F^{\rm UV}_0=\pi I^{\rm UV}_0=4.63\times10^7 {\rm photon~cm^{-2}\,s^{-1}.}
\end{equation}
In this work, we only consider the interstellar irradiation penetrating the disk and envelope in the vertical direction and neglect the possible input from the central star as photodesorption is only important at the outer disk boundary, which is shadowed from the star by the disk itself.

We assume a slightly elevated unattenuated interstellar UV field $G_{\rm env}=5.5G_0$ as disks are born in star forming regions. For the disk midplane, which is illuminated from above and below, this field scales with the UV optical depth $\tau_{\rm UV}$ as:
\begin{equation}
G_{\rm UV}=0.5 G_{\rm env} e^{-\tau_{\rm UV}}.
\label{eq:gfactoruv}
\end{equation}
The optical depth in UV towards the disk midplane can be calculated from the surface densities of small and grown dust as

\begin{equation}
\tau_{\rm UV}= 0.5 \left(\varkappa_{\rm sm} \Sigma_{\rm d,sm} + \varkappa_{\rm gr} \Sigma_{\rm d,gr}\right),
\label{eq:tauUV}
\end{equation}
where $\varkappa_{\rm sm}=10^4$\,cm$^2$\,g$^{-1}$, $\varkappa_{\rm gr}=2\times 10^2$\,cm$^2$\,g$^{-1}$ are typical values for small and grown dust absorption coefficients in the UV~\citep[Fig.~1]{2019MNRAS.486.3907P}. The factor 0.5 in Equation~\ref{eq:gfactoruv} reflects our assumption on dominant radiation transfer in vertical direction (i.e. optical depth in radial directions $\gg$ optical depth in vertical direction).

\subsubsection{Adsorption}

The adsorption rate $\lambda$\,(s$^{-1}$) of a species on a monodisperse non-charged dust grain ensemble with number density $n$ is \citep{1990MNRAS.244..432B}:
\begin{equation}
    \lambda=\pi a^2 n \sqrt{\frac{8k_{\rm B}T}{\pi m_{\rm sp}}}.
    \label{eq:lambda0}
\end{equation}
For a multidisperse dust ensemble, the term $\pi a^2 n$ in Equation~\ref{eq:lambda0} becomes the total cross-section of dust grains in the unit volume, which is four times smaller than $\widetilde{\sigma}_{\rm tot}$ in Equations~\ref{eq:surface} and~\ref{eq:surface1}:
\begin{equation}
\lambda= \frac{\sigma_{\rm tot}}{4} \sqrt{\frac{8k_{\rm B}T}{\pi m_{\rm sp}}}.
\label{eq:lambda}
\end{equation}

\subsection{Equilibrium snowline positions}
\label{sec:equilibrium}

To demonstrate the effect of advection of volatiles on their snowlines, we need to compare the snowline positions obtained in our model with their equilibrium positions. Solutions for Eqs.~\ref{eq:sig1}--\ref{eq:sig3} at $\Delta t \rightarrow \infty$ (see Appendix~\ref{sec:solutions}) for given dust and gas distributions at some disk evolution moment would represent the equilibrium distribution of volatiles, as if timescales of adsorption and desorption were negligibly short compared to local dynamic time scales. We define \textit{an equilibrium snowline} of a species as a location where its equilibrium gas-phase abundance is equal to the sum of the equilibrium abundances in the ices.

In the presented model, the equilibrium between the gas and ice phases is not necessarily sustained, as we solve time-dependent equations for adsorption and desorption (see Section~\ref{sec:accdes}). Thus, the obtained distribution of matter is not necessarily consistent with the equilibrium snowline positions. We will use equilibrium snowlines to relate them to the distributions of volatiles and to assess the importance of volatile dynamics.

\subsection{Exchange of ices due to dust growth}
\label{sec:growth}

Dust evolution processes lead to the small dust conversion to grown dust and vice versa. Icy mantles should be transported from one dust population to the other as well. We redistribute ices proportionally to dust redistribution: if certain mass fraction of small dust turns into grown dust, then the same fraction of every ice species on small dust turns into ice on grown dust.

Surface densities of species on dust at the time step $n+1$,  $\Sigma_{s, n+1}^{\rm sm}$ and $\Sigma_{s, n+1}^{\rm gr}$, can be found from their surface densities at time step $n$:
\begin{equation}
\Sigma_{s,n+1}^{\rm sm}=\Sigma_{s,n}^{\rm sm}+\Delta \Sigma_{\rm d,sm} A_{s}, 
\end{equation}
\begin{equation}
\Sigma_{s,n+1}^{\rm gr}=\Sigma_{s,n}^{\rm gr}-\Delta \Sigma_{\rm d,sm}  A_{s}, 
\end{equation}
where $\Delta \Sigma_{\rm d,sm} $ is defined by Eq.~\ref{growth1} and $A_s$ is the mass fraction of a given ice on the relevant dust component, which is to be transported:
\begin{equation}
A_s=
\begin{cases}
\Sigma_{s,n}^{\rm sm}/\Sigma^n_{\rm d,sm}, & \text{ if } \Delta \Sigma_{\rm d,sm} < 0; \\
\Sigma_{s,n}^{\rm gr}/\Sigma^n_{\rm d,gr}, & \text{ if } \Delta \Sigma_{\rm d,sm} \geq 0.
\end{cases}
\end{equation}

\subsection{Fragmentation velocity of mantled grains}
\label{sec:vfrag}

To assess the effect of icy mantles on dust evolution, we use the fragmentation velocity $v_{\rm frag}$ as a parameter depending on the presence of icy mantle on the surface of grown dust grains. At each time step, after the species surface densities are calculated, we update the values of fragmentation velocity $v_{\rm frag}$. To do that, we check if the total amount of ices at a given location is sufficient to cover all the grown dust grains present there with at least one monolayer of icy mantle. If so, the dust is treated as ``sticky'' with high fragmentation velocity, otherwise it is ``bare'' and fragile, with lower fragmentation velocity. Although there are laboratory studies that show that water ice might not improve stickiness of the grains \citep{2019ApJ...873...58M}, other works show that the presence of icy mantles helps to overcome the fragmentation barrier \citep{2009ApJ...702.1490W,2015ApJ...798...34G,2019ApJ...878..132O}. In this work, we choose to assume that the icy mantles in general, regardless of their composition, increase $v_{\rm frag}$.

Total surface density of the four considered ices on grown dust is
\begin{equation}
\Sigma_{\rm ice}=\sum_{s=1}^{4} \Sigma_{s}^{\rm gr}.
\end{equation}

We assign a specific value to $v_{\rm frag}$, comparing the ratio between $\Sigma_{\rm ice}$ and $\Sigma_{\rm d,gr}$ to the threshold value $K$. To define the value of the threshold $K$, we use $\Sigma_{\rm ice}^{\rm min}$, which is a minimum surface density of ices on grown dust that is sufficient to cover their surfaces with one monolayer. The average size of grown dust can be expressed as $\overline{a_{\rm gr}}=\sqrt{a_*a_{\rm max}}$. For the monolayer thickness estimated as the size of the water molecule $a_{\rm ml}=3\times10^{-8}$\,cm, and volume densities of ice and dust $\rho_{\rm ice}=1$\,g\,cm$^{-3}$ and $\rho_{\rm s}=3$\,g\,cm$^{-3}$, the threshold value is

\begin{equation}
K = \frac{\Sigma_{\rm ice}^{\rm min}}{\Sigma_{\rm d,gr}} = \frac{4\pi \overline{a_{\rm gr}}^2 a_{\rm ml}\rho_{\rm ice} }{4/3 \pi \overline{a_{\rm gr}}^3 \rho_{\rm s}}=\frac{3 a_{\rm ml}\rho_{\rm ice}}{\overline{a_{\rm gr}} \rho_{\rm s}} = \frac{3 \times 10^{-8} \rm cm}{\sqrt{a_*a_{\rm max}}}
\label{eq:threshold}
\end{equation}

For example, for the grown dust with $a_*=10^{-4}$\,cm and $a_{\rm max}=10^{-2}$\,cm we find $K = 3\times10^{-5}$. So, for a reasonable size of grown dust grains, a very small amount of ice is needed to cover them with one molecular layer. Then we use $K$ to define local fragmentation velocity 

\begin{equation}
v_{\rm frag}=
\begin{cases}
15\text{\,m\,s$^{-1}$,} & \text{ if } \Sigma_{\rm ice}/\Sigma_{\rm d,gr} > K ; \\
1.5\text{\,m\,s$^{-1}$,} & \text{ if } \Sigma_{\rm ice}/\Sigma_{\rm d,gr} \leq K .
\end{cases}
\label{eq:vfrag}
\end{equation}
The values of fragmentation velocities for bare and icy dust grains are chosen based on experimental results and are typical for icy and bare grains \citep{2009ApJ...702.1490W,2015ApJ...798...34G}.

\section{Gas, dust, and volatiles in young protoplanetary disks}
\label{sec:results}

\subsection{Global disk structure}
\label{sec:2d}
The evolution of dust and volatiles starts simultaneously with the collapse of the prestellar core. In the predisk phase the dust growth is slow due to low densities and rare collisions between grains. After $t=55$\,kyr, the disk is formed, providing a favorable environment for efficient dust growth and initiating decoupled dynamics of grown dust grains.

\begin{figure*}
\centering
\includegraphics[width=1.89\columnwidth]{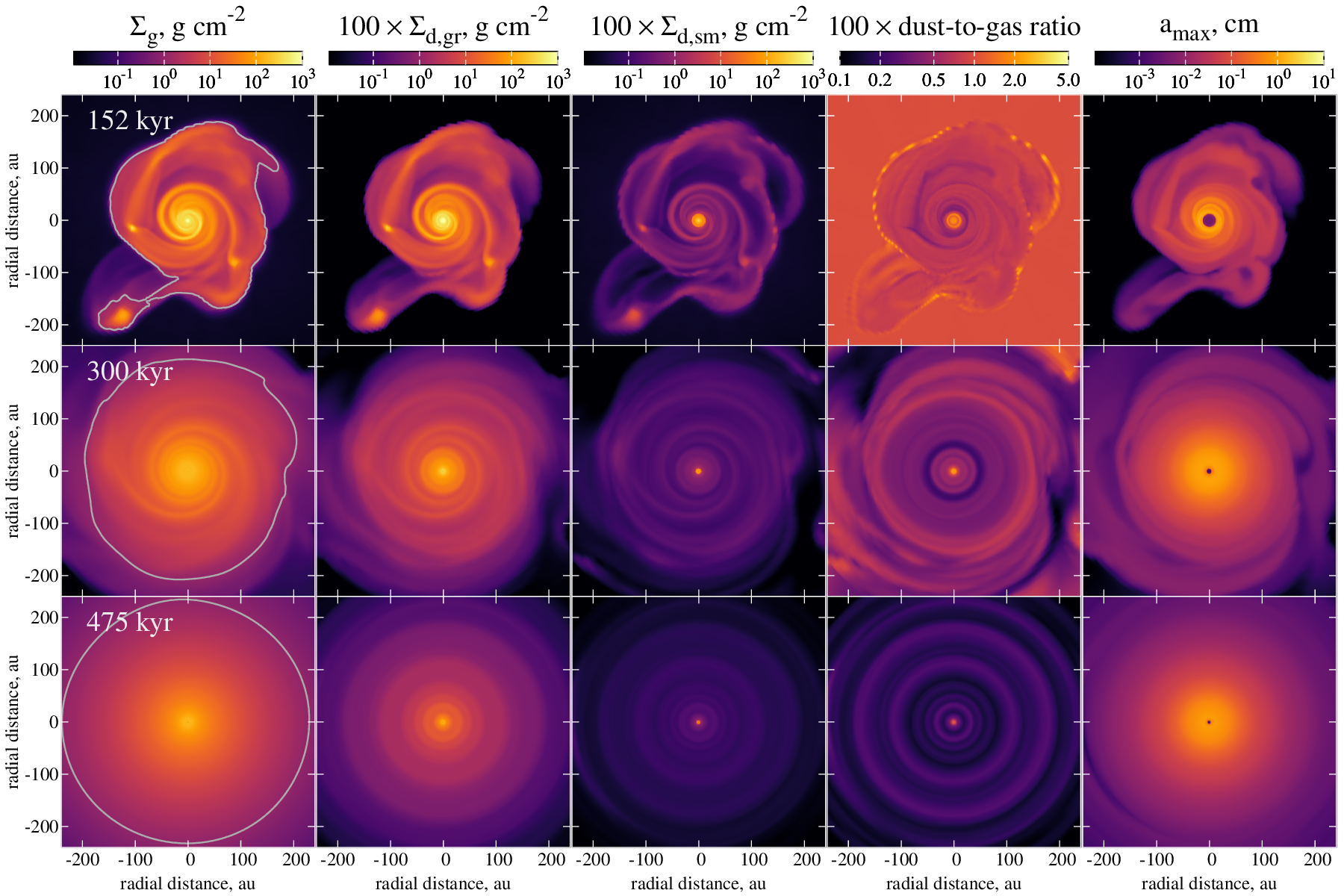}
\includegraphics[width=1.89\columnwidth]{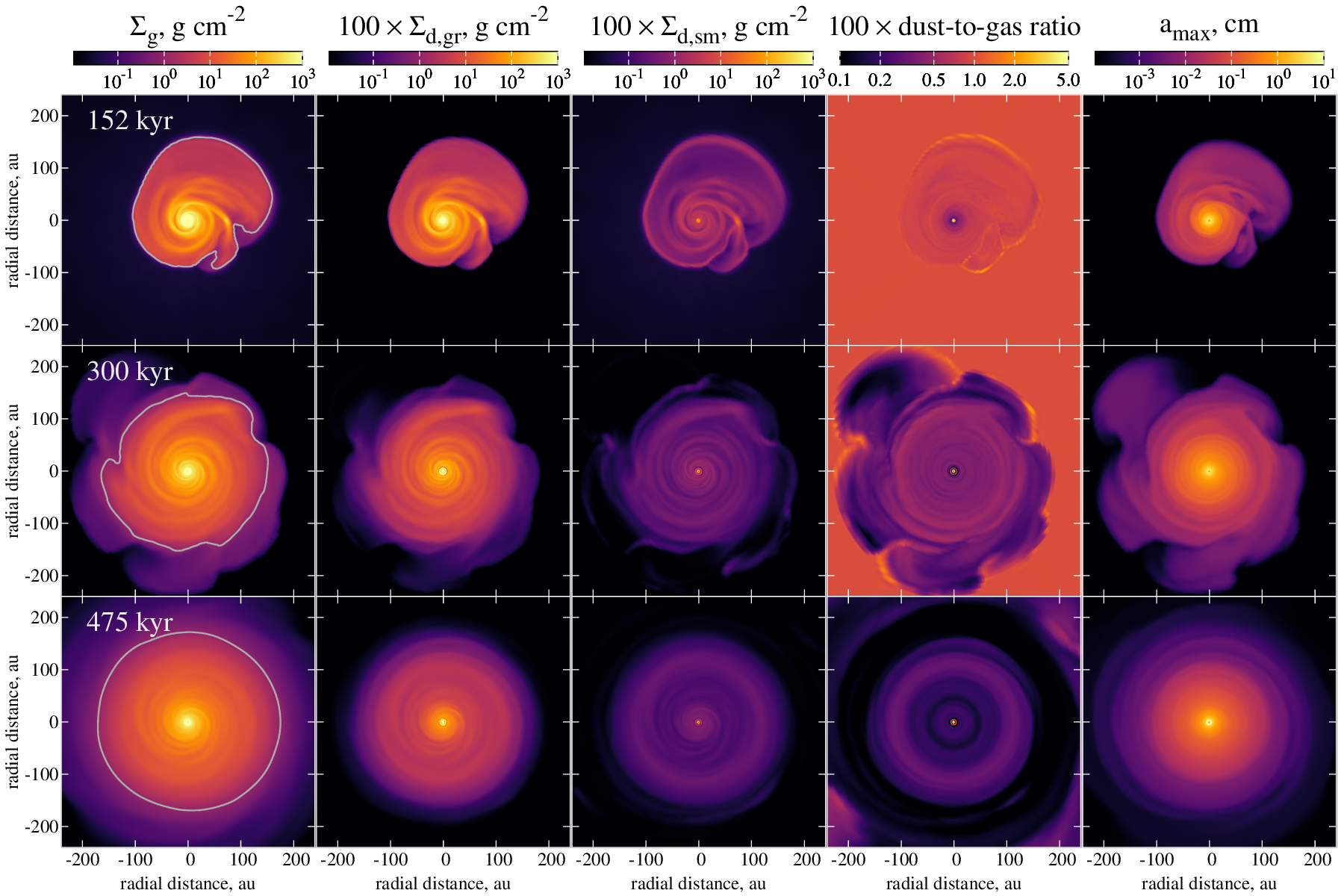}
 \caption{Surface density of gas, grown dust, and small dust, dust-to-gas mass ratio and maximum dust size in Model~1 ($\alpha=10^{-2}$, upper panels) and Model~2 ($\alpha=10^{-4}$, lower panels)} at selected time moments. The light gray line shows the $\Sigma_{\rm g}=1$\,g\,cm$^{-2}$ isocontour as an approximate outer radius of the gas disk.
  \label{fig:2d-2}
\end{figure*}

Figure~\ref{fig:2d-2} shows the distribution of gas and dust and main dust properties in Model~1 ($\alpha=10^{-2}$) and Model~2 ($\alpha=10^{-4}$) at three selected time moments. When calculating the dust-to-gas mass ratio we do not consider ices as part of dust. Hence, this value represents in fact the rock-to-gas mass ratio (ice masses are not shown). The three shown time moments represent the main stages in disk evolution. At the early time, 152\,kyr, the disk is relatively compact and gravitationally unstable. It is highly asymmetric and exhibits a distinct spiral pattern. Later, at an age of 300\,kyr, the disk becomes more stable, although weaker spiral arms  are still present. At 475\,kyr the disk becomes more uniform and nearly axisymmetric in Model~1 and shows only traces of spiral arms in Model~2. The gas disk gradually grows in size due to mass-loading from the infalling envelope (at the early stages) and spreads out due to viscous spreading (at the later stages). While throughout the disk evolution some part of gas is lost in the accretion onto the star and with jets, the dust mass loss is even more profound due to its additional radial drift. This results in dust-to-gas mass ratio $<10^{-2}$ in most of the disk volume at later evolutionary stages.

Dust grains grow much faster in the disk than in the envelope. Most of the dust in the disk is found in the form of grown grains, and small dust is depleted. The exception is the very inner 5--20\,au of the disk where the fragmentation barrier is low due to the absence of icy mantles. In this region, the dust density can be quite high due to the formation of dust rings (Model~2, see Sect.~\ref{sec:dustrings}). Generally, dust size increases towards the center of the disk. However, in Model~1 there is a sharp drop in dust size in the inner iceless region. In Model~2, dust size reaches its maximum inside the inner dust rings. Grown dust tends to drift towards the star, decreasing the dust-to-gas ratio throughout most of the disk extent and elevating it in the inner parts (see Sect.~\ref{sec:space-time} for the details of dust evolution). Large-scale (tens and hundreds au) ring-like structures form at the latest stages, appearing especially bright in the dust-to-gas ratio maps. A more detailed description of the dust rings formation mechanisms can be found in Sect.~\ref{sec:dustrings}.

A notable difference between the two models is the rate at which the disk spreads out. A more viscous disk in Model~1 ($\alpha=10^{-2}$) tends to spread faster and is larger in size. It also tends to evolve faster and becomes completely axially symmetric at 500\,kyr. In Model~2 ($\alpha=10^{-4}$) the disk still possesses spiral substructures in both gas and dust at the end of the simulation. The thermal structure of the disks is also affected by the $\alpha$-value, as viscous heating is one of the main heating mechanisms in the inner 10 au of the disk. As a result, the disk in Model~1 is somewhat hotter.

Another important process affected by the $\alpha$-value is the formation of gravity-bound and pressure-supported clumps, which emerge as a result of gravitational fragmentation in massive disks. Both considered models produce massive gravitationally unstable disks with masses around 0.2\,$M_\odot$, however, in our simulations, long-lived clumps form only at the early stages of disk evolution in Model~1. The 152\,kyr snapshot for Model~2 in Figure~\ref{fig:2d-2} also shows signs of clump formation but the resulting object does not survive for more than 1000\,yr. Apart from the gravitational instability described by the Toomre criterion \citep{1964ApJ...139.1217T}, the formation of long-lived clumps requires the cooling rate to be greater than the characteristic growth rate of the gravitational instability \citep{2003ApJ...597..131J}. This condition is usually fulfilled only at large distances from the star, $\gtrsim$100\,au \citep{2005ApJ...621L..69R}. In the evolutionary stage when the disk is most massive, its size in Model~2 is more compact due to lower turbulent viscosity than in Model~1. As a result, the regions with appropriate conditions for clump formation are hardly present in Model~2.

\subsection{Long-term evolution of the disk}
\label{sec:space-time}

The ability to compute the long-term evolution of a protoplanetary disk is one of the major advantages of the FEOSAD code. The numerical simulations start with a cloud core collapse and the formation of a gravitationally unstable disk and continue to a 0.5\,Myr old axially-symmetric disk, which represents a Class~II young stellar object. In this Section, we consider the azimuthally averaged distribution of gas, dust, and the volatiles throughout the disk early history and reveal the long-term effects of icy mantles on the dust evolution.
\begin{figure*}
\centering
\includegraphics[width=2\columnwidth]{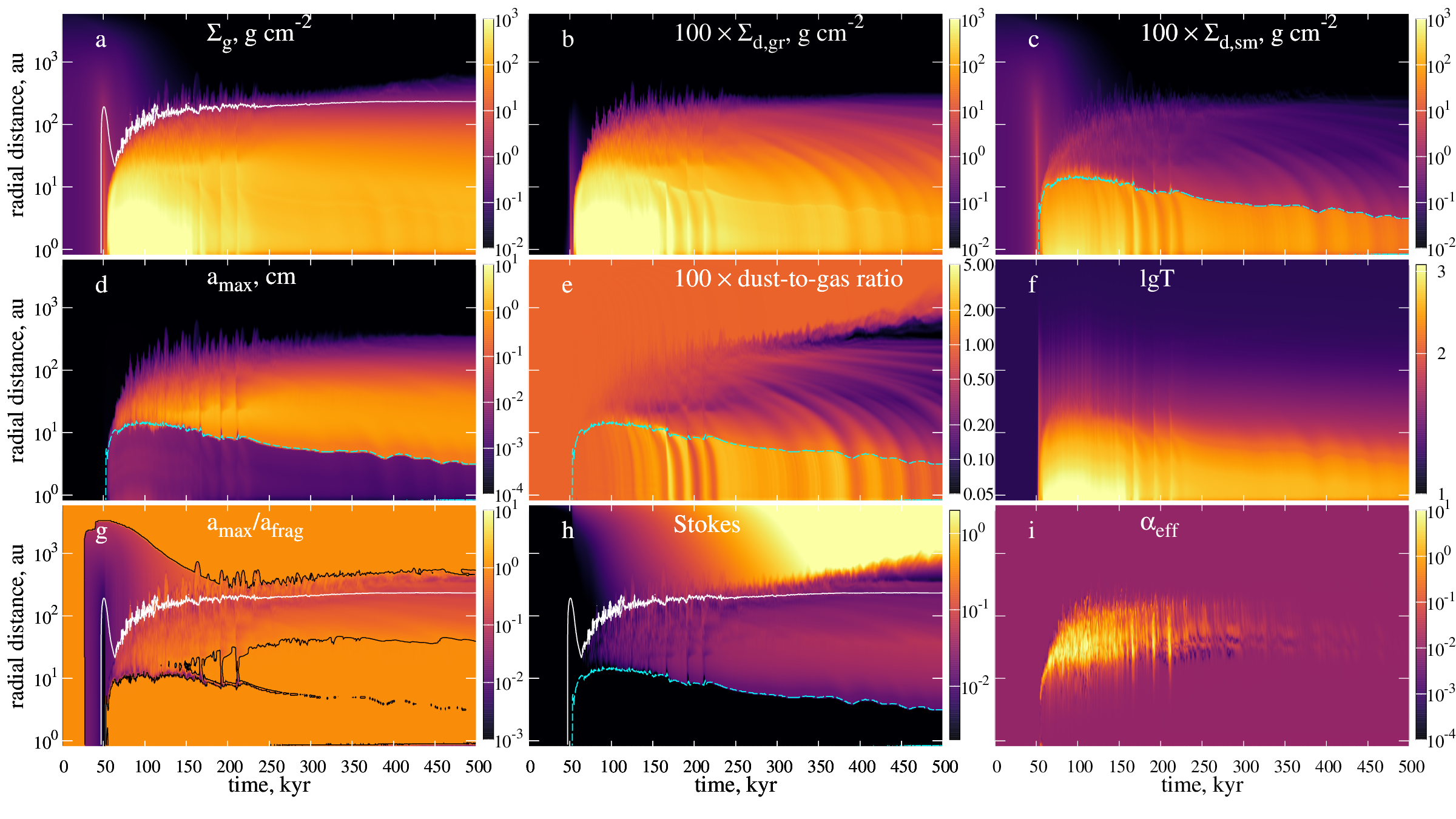}
 \caption{Time evolution of azimuthally averaged radial distributions of gas and dust parameters in Model~1 ($\alpha=10^{-2}$). Solid white line indicates the approximate boundary of the disk at $\Sigma_{\rm g}=1$\,g\,cm$^{-2}$, dashed cyan line shows the position of a ``thermal'' water snowline (see Section~\ref{sec:volatilesgen}). Black contour in panel (g)  corresponds to $a_{\rm max}/a_{\rm frag}=0.9$ and outlines approximately the regions where dust size is restricted mainly by fragmentation. The panels show surface density of gas (a), grown dust (b), small dust (c), as well as maximum dust size (d), dust-to-gas ratio (e), temperature logarithm (f), dust maximum size relative to the fragmentation barrier (g), Stokes number (h), effective $\alpha$-parameter caused by gravitational instability (i) (see Section~\ref{sec:dustrings} for details.)}
  \label{fig:st-2}
\end{figure*}

\begin{figure*}
\centering
\includegraphics[width=2\columnwidth]{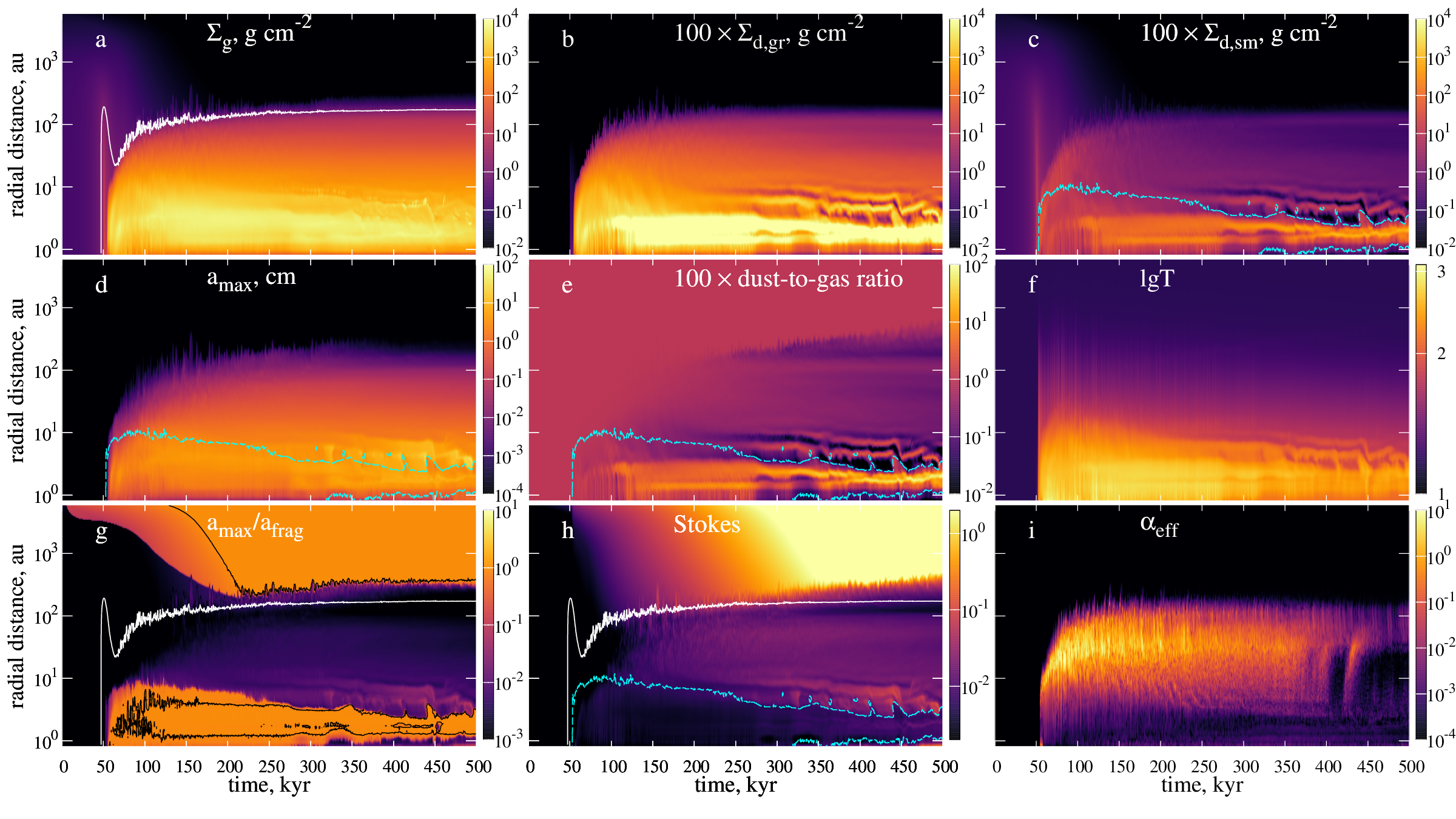}
 \caption{Same as Figure~\ref{fig:st-2} but for Model~2 with $\alpha=10^{-4}$.}
  \label{fig:st-4}
\end{figure*}
Figures~\ref{fig:st-2} and~\ref{fig:st-4} present the position-time diagrams showing the azimuthally averaged quantities that describe the evolution of gas and dust for Models~1 and~2 ($\alpha=10^{-2}$ and $\alpha=10^{-4}$, respectively). The radial distributions are shown for all time moments up to 500\,kyr. Both long- and short-term trends are observed in the models. The disk forms at around 50~kyr after the onset of gravitational collapse, and during the subsequent $\approx450$\,kyr the gas disk grows in size and spreads out. Concurrently, the disk loses the material through accretion to the star and gradually cools down. During the initial 200\,kyr the disk experiences accretion bursts responsible for short-term disk heating events with a duration of several hundred years. Such bursts are most prominent in Model~1. Diluted regions beyond $\sim$200\,au represent the infalling envelope and supply the disk with material during the initial stages of evolution. Dust does not grow above a minimum value of $10^{-4}$\,cm in the envelope because of rare collisions and low fragmentation barrier. The Stokes numbers increase in time in this region, mostly due to decreasing gas densities in the heavily depleted envelope.

In the disk where the gas density is high and collisions between grains are frequent, dust grows efficiently up to centimeters in size (more than decimeter in Model~2), and  drifts towards the star \citep{1977MNRAS.180...57W}. This leads to a decrease in the dust-to-gas ratio to $10^{-3}$ and lower in the outer disk after 300\,kyr, when the envelope is depleted and the supply of small dust is terminated. However, in the inner $\approx$10\,au of the disk, both small and grown dust accumulate and dust-to-gas ratio is elevated. The mechanisms of dust accumulation in the two models are different. In Model~1, one of the main factors is a decrease in the Stokes numbers caused by a drop in $a_{\rm max}$. The maximum size of dust grains drops because the fragmentation velocity is altered by the desorption of ices in this inner disk region. In Model~2, dust rings develop in the inner disk due to the formation of a dead zone described in more detail in Sect.~\ref{sec:dustrings}.

In Model~1, the effect of water snowline on dust properties is evident. Indicated by the cyan dashed line in Figure~\ref{fig:st-2} is the classic snowline associated with thermal desorption and temperature gradient in the disk (see Sect.~\ref{sec:volatilesgen} for more details on the snowlines). Right inside this snowline, the dust-to-gas ratio is elevated due to the drift of larger dust from outside the snowline. Besides, the content of small dust there is comparable to that of grown dust, and the maximum size of grains is around $10-100$\,$\mu$m, contrasting with centimeter grains outside the snowline.

This is the effect of fragmentation velocity, which drops to $v_{\rm frag}=1.5$\,m\,s$^{-1}$ in this region, as dust grains are relieved of their icy mantles in the warm inner disk regions. A factor of 100 contrast in the dust size across the snowline is directly dictated by a factor of 10 change in $v_{\rm frag}$ (see Eqs.~\eqref{eq:afrag} and~\eqref{eq:vfrag}). Efficient fragmentation of bare grains lowers the dust size, thus slowing down its radial drift. It also produces more small dust, which is not affected by radial drift. Therefore, both factors, decrease of $a_{\rm max}$ and increase of the small dust fraction, favor dust accumulation in this region.

Another notable, albeit not very prominent, feature is a thin ring in gas and grown dust, coincident with the position of the water snowline. A mechanism of this ring formation is described in Section~\ref{sec:dustrings} along with a more thorough illustration of dust and gas distribution around this ring.

The effect of ice mantles on dust in Model~1 is prominent due to the principal role of collisional fragmentation in restricting the dust growth. As panel (g) in Figure~\ref{fig:st-2} suggests, in the model with $\alpha=10^{-2}$, the maximum dust size in most of the disk is very close to the dust size limited by the fragmentation barrier $a_{\rm frag}$. The solid black lines outline the (orange shaded) disk areas dominated mostly by fragmentation.

It should be noted that at the snowline itself the dust size does not reach the fragmentation barrier. This effect arises from a sharp change in $a_{\rm frag}$ combined with azimuthal variations in the gas and dust radial velocities. Although dust on average drifts radially inward, it can nevertheless experience cyclic motions around a given radial position during the orbital period. At some azimuths dust grains with lower $a_{\rm max}$ move from inside the snowline to the region with higher $a_{\rm frag}$.  Outside the snowline these grains could have coagulated and grown up to $a_{\rm frag}$, if they stayed there long enough. However, they stay there only for a fraction of their Keplerian period (then moving back inside the snowline), which is comparable to the coagulation timescale in this region ($\sim$10\,yr), so that the fragmentation limit is mostly not reached. This is a purely multidimensional effect and cannot be observed in one-dimensional disk simulations. More details about azimuthal variations of gas and dust velocities are presented in Appendix~\ref{sec:radvels}.

In Model~2, however, the disk is less turbulent and $a_{\rm frag}$ is two orders of magnitude higher, as suggested by Eq.~\eqref{eq:afrag}. As a result, collisional fragmentation limits the dust size only inside the water snowline, where $v_{\rm frag}$ is lowered, while in the bulk of the disk the radial drift imposes stronger constraints (see panel (g) in Figure~\ref{fig:st-4}). Similarly, \citet{2017A&A...600A.140S} do not see the effect of CO snowline on dust growth in their modeling due to the dominance of radial drift in this region. There is no sharp change in the dust size at the water snowline, as the border between the regions with drift- and fragmentation-governed dust size is several au interior to it, particularly at earlier times. Although the transition from drift- to fragmentation-governed regions also results in the change of dust properties, in this case it is less sharp and occurs not at the snowline (compare black contours in panel (g) with inner dust rings in panels (b), (c), and (e) in Figure~\ref{fig:st-4}). Besides, the snowline affects the Stokes numbers (panel (h) in Figure~\ref{fig:st-4}) and outlines the inner rings at later stages.

Several features of the gas-dust distribution are worth noting. For instance, episodic variations in the gas and dust surface densities are characteristic of Model~1. They are best seen in the dust-to-gas ratio distribution (panel (e) in Figure~\ref{fig:st-2}). At earlier times ($\le 200$~kyr) they are associated with transient episodes of gravitational instability, which is induced in the inner disk regions by gas-dust clumps that migrate inward. The clump arriving in the inner disk is disintegrated via the action of tidal torques, which increases the density and perturbs the inner disk regions, resulting in the formation of a transient spiral pattern and causing additional inward transport of matter associated with it. The clump also brings a dust-rich material from the outer disk. The rise in the protostellar accretion rate caused by the falling clumps creates a luminosity outburst, which heats the disk. Such events are seen as spikes around 150--200\,kyr in the gas temperature distribution (see panel (f) in Figure~\ref{fig:st-2}) and were first reported by \citet{2005ApJ...633L.137V}.  The variations in the dust surface density at later stages in Model~1 ($\ge 350$~kyr) are caused by a ``trapping'' of drifting dust grains, arising from radial variations in the gas surface density due to the presence of tight spirals in the outer disk. We defer a detailed study of the dust trapping to a follow-up study. In Model~2, the prominent feature is a system of dust rings in the inner disk, discussed in more detail in Sect.~\ref{sec:dustrings}.

\subsection{Dynamics of volatiles}
\label{sec:volatiles}

\subsubsection{Evolution of volatiles in the gas and in the ices}
\label{sec:volatilesgen}

\begin{figure*}
\centering
\includegraphics[width=2\columnwidth]{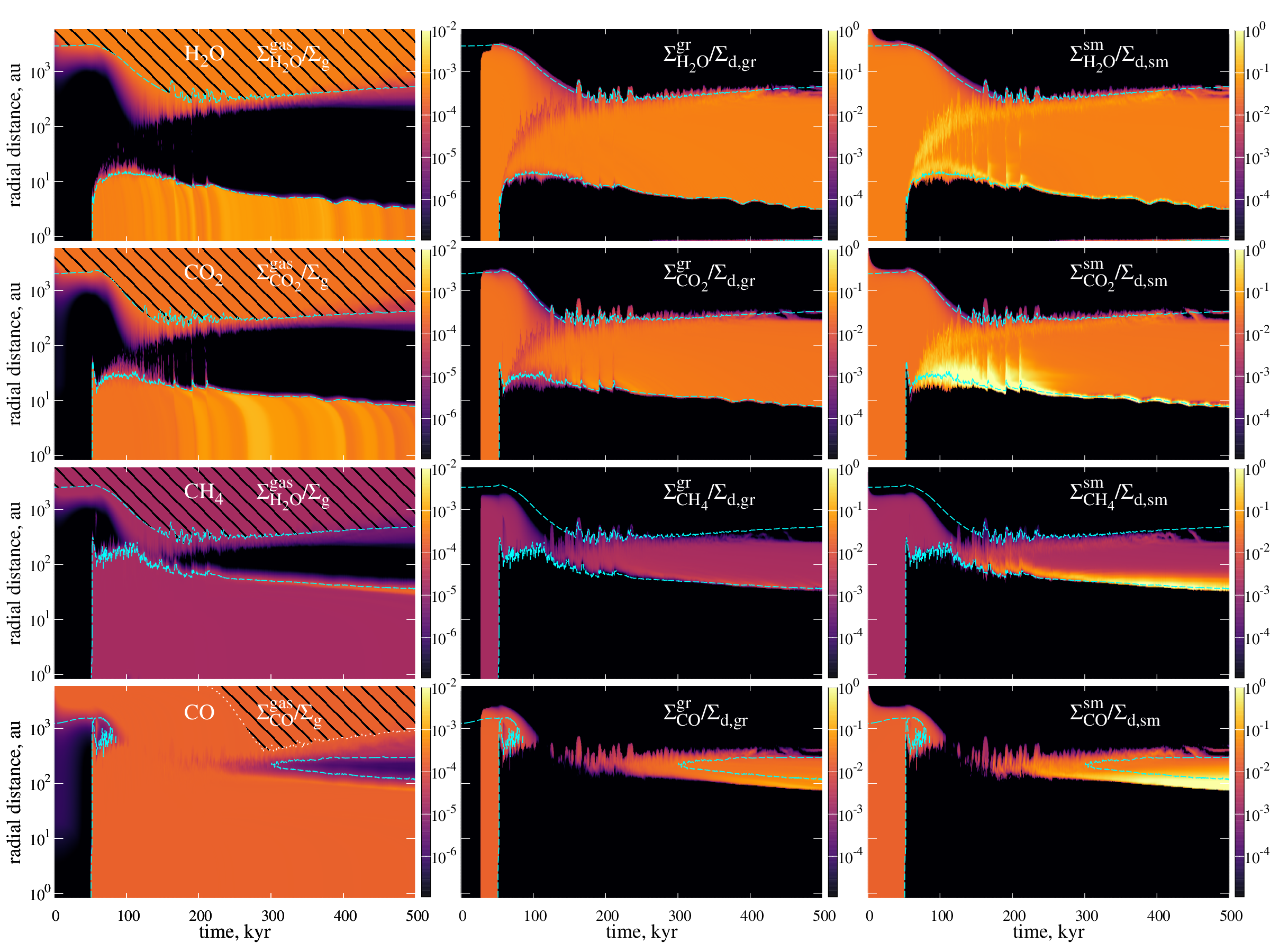}
 \caption{Time evolution of azimuthally averaged radial distributions of four volatiles in the gas and in the ice, in Model~1 ($\alpha=10^{-2}$). For each volatile, dashed cyan line shows the equilibrium snowline position. White dotted line marks the distance where the CO self-shielding coefficient is equal to 0.1.}
  \label{fig:st-2ch}
\end{figure*}

\begin{figure*}
\centering
\includegraphics[width=2\columnwidth]{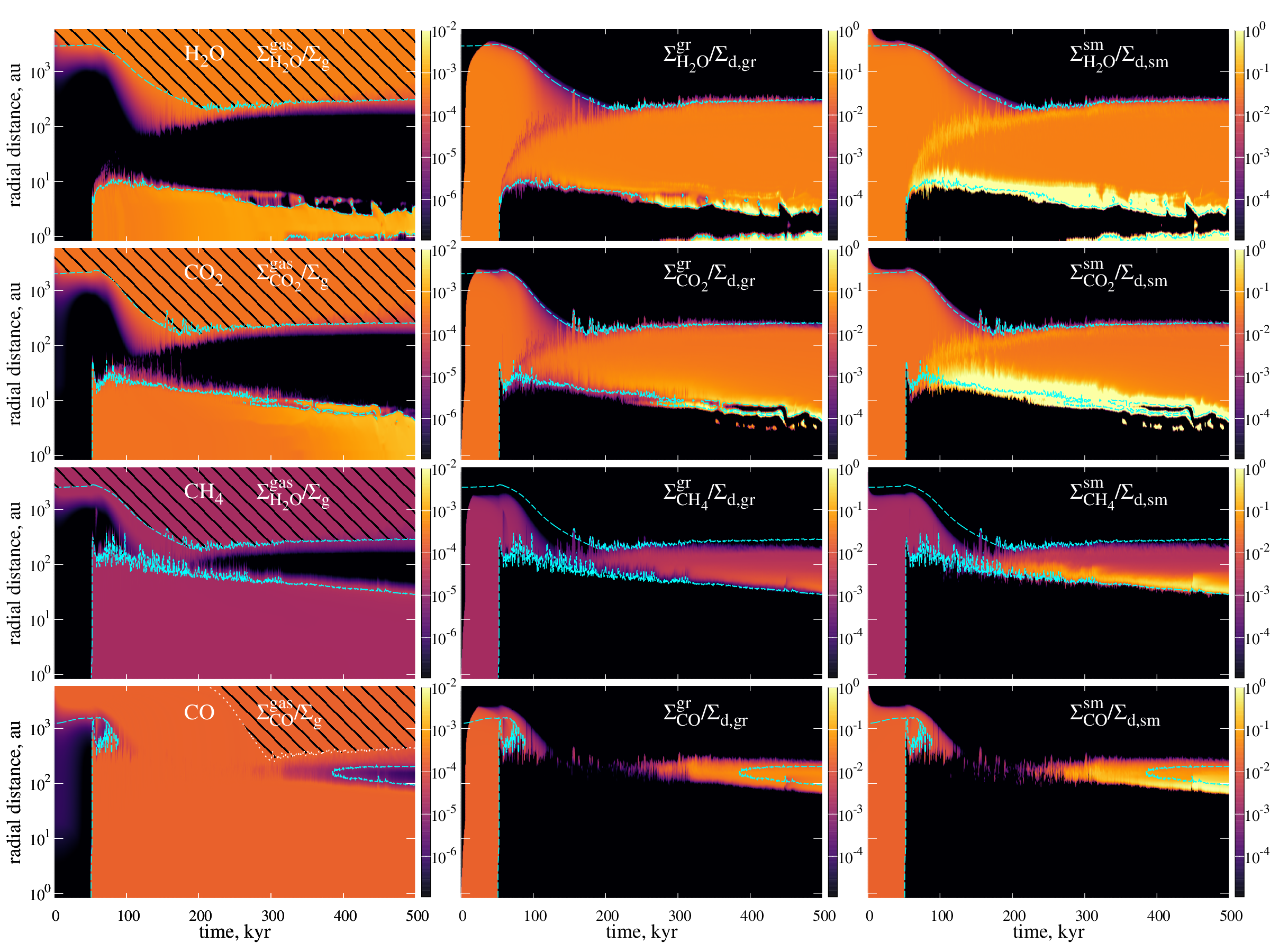}
 \caption{Same as Figure~\ref{fig:st-2ch}, but for Model~2 ($\alpha=10^{-4}$).}
  \label{fig:st-4ch}
\end{figure*}

In this Section, we consider the distribution of volatiles in different phases, the positions of their snowlines, and the effects arising at these snowlines due to collective dynamics of gas and dust. All four species in the gas and on the surface of grown and small dust grains are presented in Figures~\ref{fig:st-2ch} and~\ref{fig:st-4ch}. The surface densities of the species are shown with respect to the surface density of the corresponding carrier: the species in the gas phase are normalized to the gas surface density, while the species in the ice phase are normalized to the surface density of dust population on which they reside. Dashed cyan lines show the positions of equilibrium snowlines, which are found using the azimuthally averaged quantities for $t\rightarrow \infty$ as described in Section~\ref{sec:equilibrium}.

The equilibrium snowlines in Figures~\ref{fig:st-2ch} and~\ref{fig:st-4ch} reflect sharp changes in the distributions of gas- and ice-phase species. For each species we can indicate two snowlines: the one caused by thermal desorption and the one caused by photodesorption in the outer regions at hundreds of astronomical units from the star. These snowlines are further referred to as thermal and photo snowlines, respectively. The water snowline previously shown in Figures~\ref{fig:st-2} and~\ref{fig:st-4} is the thermal one. While thermal snowlines move closer to the star as the disk cools down due to both accretion luminosity and viscous heating waning in a disk with decreasing density \citep[see, e.g., ][]{2009SoSyR..43..508M}, photo snowlines stay at approximately the same distance of several hundred au. Their positions shift during the first 300\,kyr, when the dust surface densities determining UV illumination vary. After the supply of material from the envelope is drained and dust distributions stabilize, the photo snowlines become steady. 

Although the volatiles in the model are not necessarily at equilibrium, their resulting distribution is close to that suggested by the equilibrium snowlines. The agreement is quite good for H$_2$O, CO$_2$, and CH$_4$. Only for CO, the mismatch between the equilibrium snowlines and the actual distribution of the species is considerable. CO is the most volatile of the considered molecules, so its snowlines are virtually absent at the early disk evolution, and they appear only in the later evolution in the outermost disk regions. In the tenuous conditions typical for the outer disk and envelope, all  timescales are longer, thus, CO is far from equilibrium.

Although the equilibrium snowlines generally divide the ice- and gas-dominated regions of the disk quite sharply, at certain time moments the same phase can be present on both sides of the equilibrium snowlines. This is the effect of volatile dynamics, notable for H$_2$O and CO$_2$ in Figures~\ref{fig:st-2ch} and~\ref{fig:st-4ch}. At the early stages, up to 200\,kyr, the gas-phase species spread notably through the snowlines into the ice-dominated region. For the thermal snowline, this effect is explained by the presence of warm spiral arms that release water and carbon dioxide vapors in their wakes as they pass through the disk. The positions of equilibrium snowlines are derived using the azimuthally averaged quantities and the effect of spiral arms is washed out. The shape of snowlines in 2D and the effect of spiral arms on ice distribution are described in more detail in Section~\ref{sec:2dvolatiles}.

In the outer regions interstellar UV radiation responsible for the photodesorption of molecules from dust surface can also be a source of photodissociation of the gas-phase species, destroying the molecules. This effect is not considered in our model. The gas-phase abundance of H$_2$O, CO$_2$ and CH$_4$ beyond the photo-snowline must be much smaller than shown in the left columns of Figures~\ref{fig:st-2ch} and~\ref{fig:st-4ch}. The only exception is CO, for which the self-shielding may ensure large gas-phase abundances in disk outer regions beyond the photo-snowline. To quantify this effect, we show with the white dotted line the locations where the shielding factor $\theta$ from \citet[][Table 5]{1988ApJ...334..771V} is equal to 0.1. A CO molecule inside this line is likely to survive the photodissociation. 
Photodissociation could affect ice-phase species as well, but the corresponding reaction rates should probably be lower than those in the gas phase \citep{2020RuCRv..89..430M}. The details of this process are still unclear, so we chose to ignore it in our model.

The short-term spikes in the position of thermal snowlines (most visible in CO$_2$) are induced by protostellar accretion bursts, which occur in gravitationally unstable disks at the early stages of their evolution \citep[see, e.g.,][]{2015ApJ...805..115V}. In our models, the mass accretion rate is a few $\times 10^{-5}~M_\odot$~yr$^{-1}$ during these bursts, corresponding to an increase in the accretion luminosity by a factor of several and  the peak accretion luminosities reaching 10--20~$L_\odot$. While in the quiescent disk ices are mostly found on grown dust, after the luminosity bursts the volatiles resettle primarily on small grains, as they dominate in the total surface area, and these ices return to grown dust via coagulation only after tens of kyr (see also Section~\ref{sec:2dvolatiles}). A more detailed study of the effect of accretion bursts on the evolution of dust and volatiles will be presented in a follow-up study.

Model~1 exhibits variations in the water and carbon dioxide gas abundances in the inner disk. They correlate with the variations in the gas-to-dust ratio (panel (e) in Figure~\ref{fig:st-2}). At the same time, the distributions of ices are quite uniform and do not show signs of such variations. These variations are explained by the dynamics of gravitationally unstable disks. Pieces of dust-rich material, such as clumps or rings, migrate towards the star. The amount of dust in them is elevated but the amount of ice relative to dust does not change, so that the ice distribution is unaffected. When the dust-rich material crosses the snowline, the ice evaporates, enriching the gas with the corresponding species. For CO and CH$_4$ no effect is seen, as the clumps probably originate interior to the methane thermal snowline. In Model~2, clump formation is suppressed, as was explained in Section~\ref{sec:2d}, so that no variations in the gas-phase volatiles take place.

Another interesting feature of Model~2 is multiple thermal snowlines of water and carbon dioxide. Variations in the surface density and temperature arising due to the presence of dust rings inside 10\,au after 300\,kyr create multiple locations where the equilibrium abundances of gas and ice are equal. Dust is severely depleted between the rings, while the volatiles are still present there thanks to radial transport with the gas. This results in thick icy mantles with the ice surface density reaching and even exceeding the surface density of dust, in this case both for small and grown populations. The formation of the water snowline inside 1~au should be taken with caution. This snowline is formed because of the temperature drop near the sink-disk interface. This drop can be caused by a notable decrease in the disk optical depth thanks to dust depletion in this region. It may, however, be a boundary effect and further investigation is needed to clarify the nature of the innermost water snowline.   

In the vicinity of snowlines, the abundances of volatiles show local peaks, which are characterized by a variety of widths and amplitudes and occur both in the gas and ice phases. There are different mechanisms responsible for these effects, depending on the snowline and the model, including drift of mantled grains, azimuthally asymmetric radial velocities, and 2-D shape of snowlines (see Section~\ref{sec:2dvolatiles}). The effect of ice accumulation right beyond the snowline was first described by~\citet{1988Icar...75..146S} as a ``cold finger'' effect  \citep[see also][]{2004ApJ...614..490C}, it can also lead to accumulation of refractory material \citep{2003Icar..166..385C} in the gas-phase.

Accumulation of the gas-phase CO just interior to the snowline is caused by drifting grown dust, which brings solid CO to the inner disk regions, following the evaporation of icy mantles when crossing the corresponding snowline. This effect was also investigated by \citet{2017A&A...600A.140S} and \citet{2018ApJ...864...78K}, who showed that stronger turbulence leads to weaker enhancement of volatiles because of more efficient smearing of the excess peak by diffusion that scales with $\alpha$. In our model, the effect of turbulent $\alpha$ is the opposite, namely the gas-phase abundance peaks at the snowlines of CO and CH$_4$ are more prominent in Model~1 than in Model~2 (cf. Figures~\ref{fig:st-2ch} and~\ref{fig:st-4ch}).

In the presented simulations diffusion is not included in the model, so that the enhancement in the gas-phase CO and CH$_4$ is determined by the inward drift of the ice-mantled grains. Around the corresponding snowlines, the radial velocity of dust grains is not directly dependent on $\alpha$, because the grain size in these regions is determined by radial drift, and not by fragmentation. In our model, grown dust drifts faster in the model with $\alpha=10^{-2}$, and more volatiles are accumulated.
This is because in Model~1 dust size and consequently Stokes number in this region are higher than in Model~2.

Despite ignoring turbulent diffusion, our model still displays a diffusion-like redistribution of volatiles, which results in the accumulation of volatiles in the ice phase. This effect originates from azimuthal variations of the radial velocities of gas and dust. Even when the mass is transported inwards and azimuthally averaged radial velocities for both dust and gas are negative, the matter can still move locally outward within part of its quasi-Keplerian orbit. These oscillations lead to the effective diffusion of volatiles across the snowline and the amplitude of velocity variations is typically many times as high as the azimuthally averaged radial velocity (see Appendix~\ref{sec:radvels}). The effect seems to be stronger for lower $\alpha$, which could also contribute to the difference in the gas-phase enhancement.
A separate study of accumulation of volatiles at the snowlines is needed to provide a comprehensive description of this effect.

\subsubsection{Volatiles in gravitationally unstable disks}
\label{sec:2dvolatiles}

\begin{figure}
\centering
\includegraphics[width=0.93\columnwidth]{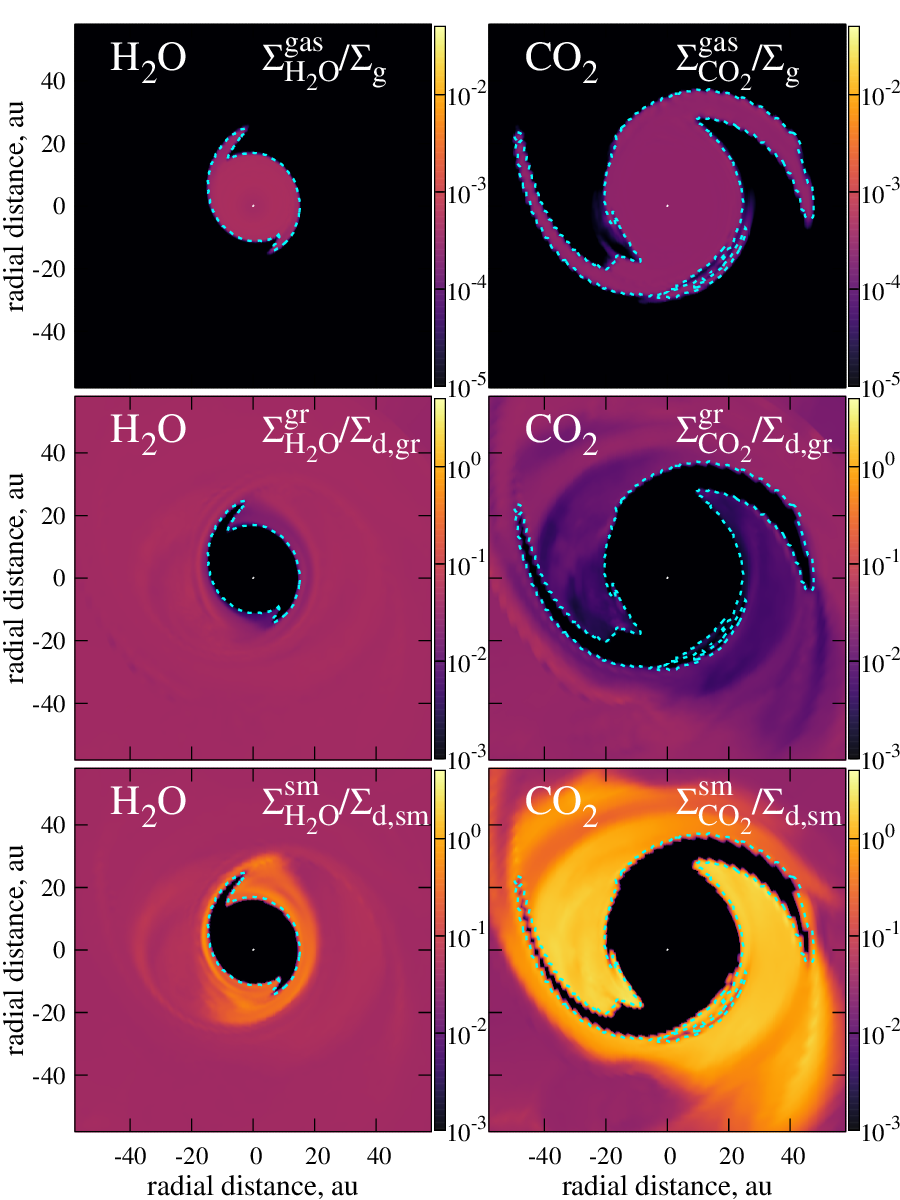}
\includegraphics[width=0.93\columnwidth]{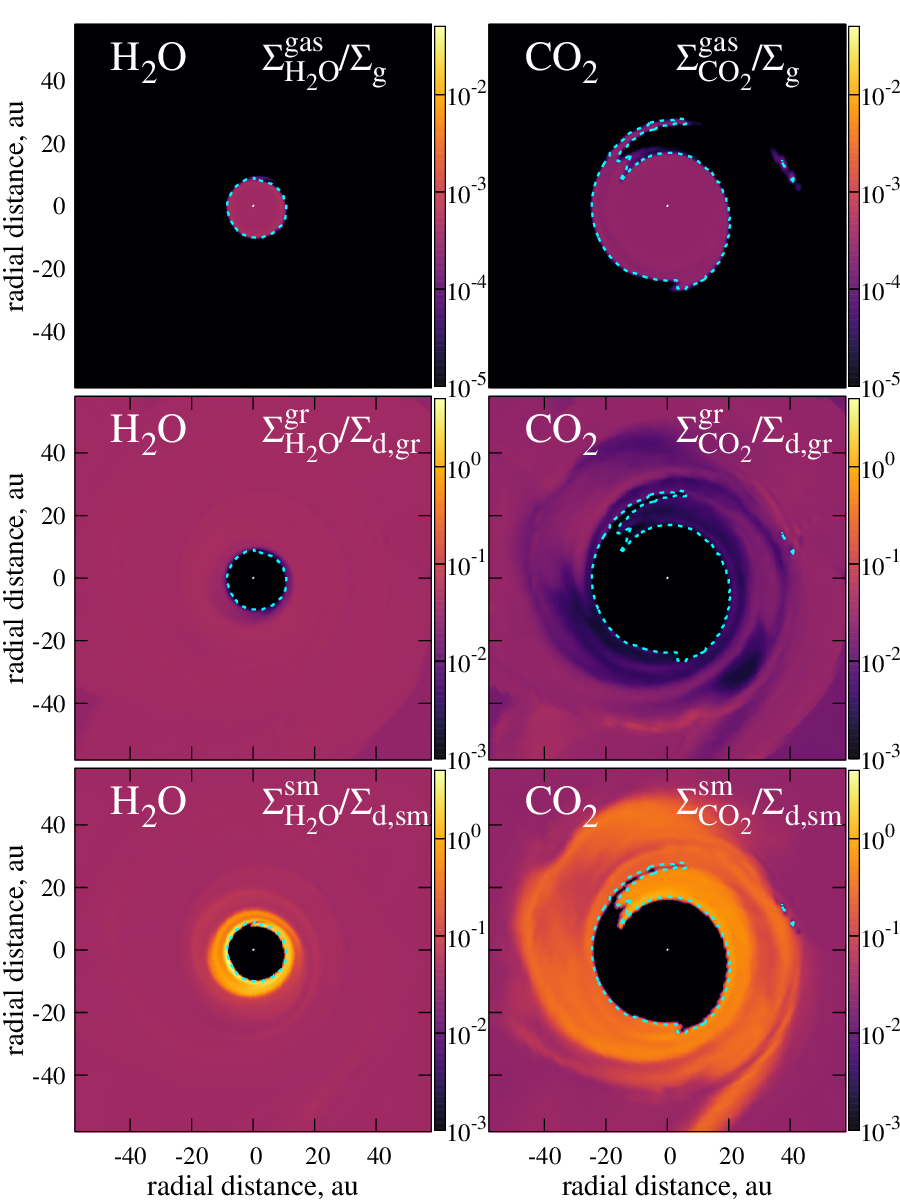}
 \caption{H$_2$O and CO$_2$ in the gas and in the ices at 100\,kyr, models with $\alpha=10^{-2}$ (upper six panels) and $\alpha=10^{-4}$ (lower six panels). Dashed cyan contours indicate the positions of equilibrium snowlines.}
  \label{fig:spiralices}
\end{figure}

Gravitational instability in young massive disks leads to the formation of a distinct spiral structure. At the early evolutionary stages, around 100--200\,kyr, the  spiral pattern is prominent both in the gas and in dust distributions. It notably affects the disk thermal structure, which is crucial for phase transitions of volatile molecules, and substantially alters the shape of the snowlines. Weaker spirals do not disappear completely even at the late stages and keep affecting gas and dust velocities.

Figure~\ref{fig:spiralices} shows the distributions of the two least volatile molecules under consideration, H$_2$O and CO$_2$, in the gas and on the surfaces of small and grown dust grains. The inner disk area encompassing the thermal snowlines is shown for a time moment of 100\,kyr. Clearly, the equilibrium snowline is not a perfect circle centered on the star. The snowline has a complicated shape that is determined and affected by the spiral pattern. We note that this effect diminishes once the disk gradually stabilizes and and becomes more axisymmetric at later evolutionary stages. The complex shape of the snowlines was also noted by \citet{2017MNRAS.472..189I} within a 3D hydrodynamic modeling with chemical evolution. They obtained the deviations of snowline shape from axial symmetry caused by clumps, spirals and shocks in a gravitationally unstable disk.

There is a minor mismatch between the spatial distribution of volatiles and the position of equilibrium snowlines, which are slightly exterior to the sharp changes in the gas and ice abundances of considered species. Thus, there is a contribution of time-dependent processes of adsorption and desorption in the positions of snowlines, albeit their effect is not strong.

In the bulk of the disk, ices are delivered to the surface of grown dust mainly through dust coagulation and growth, as small grains turn into grown ones, carrying along the icy species. Figure~\ref{fig:spiralices} demonstrates that in the regions affected by the spirals, the amount of ices on small dust increases, while the corresponding amount on grown dust decreases. As the surface density of small dust in these regions is two orders of magnitude lower due to dust growth, it means that most of the considered species are now on the surface of small dust instead of grown dust, while the total amount of ice stays approximately the same.

This peculiar pattern is created by the impact of the spiral density waves that warm up the gas as they run through the disk and sublimate the ices. As the spiral wave retreats, the gas cools down, and the sublimated species freeze back onto dust. The freeze-out occurs predominantly onto the small dust population, which has a greater total surface area despite much lower surface density. Indeed, the ratio between the total surface areas of small and grown dust grains can be expressed from Eqs.~\ref{eq:surface} and~\ref{eq:surface1} as $(\Sigma_{\rm d,sm}/\Sigma_{\rm d,gr} )\sqrt{a_{\rm max}/a_{\rm min}}$, which is $\approx10$ for the conditions between the thermal water snowline and the outer disk boundary. We note that the regions affected by spiral arms are characterized by $\Sigma_s^{\rm sm}/\Sigma_{\rm d,sm}\sim1$. Possible complications that can be caused by massive icy mantles are discussed in Section~\ref{sec:properties}.

The balance between the adsorption of volatiles to small and grown dust can be altered in favor of grown particles by the Kelvin curvature effect \citep{2013A&A...552A.137R}.
The saturated vapor pressure is higher for nanometer-size grains with their curved surface, leading to lower adsorption rate to small grains. The surface curvature effect is most important when adsorption and desorption are near balance.   \citet{2019A&A...629A..65R} argued that around the water snowline 0.1\,mm grains will lose their mantles 100 times as slow as micron-size grains. This effect is also responsible for sintering of ice in dust aggregates \citep{2011ApJ...735..131S,2016ApJ...821...82O}. In our model, small particles range from 5~to 1000\,nm, so that the freeze-out rate could be affected by the Kelvin curvature effect. We do not consider it in our model, but we note that it might lead to diminution of the ice mantles of small grains.

The effect of spiral arms is more pronounced for CO$_2$ than for H$_2$O. This is because the spiral arms are strongest in the intermediate disk regions (tens of au) where the Toomre Q-parameter is smallest. In the innermost disk regions, the Keplerian shear and the temperatures are too high to promote a strong gravitational instability. In the outer disk regions, the gas surface density quickly drops, also reducing the strength of gravitational instability. As a result, spiral arms efficiently heat the intermediate disk regions to the level that causes CO$_2$ desorption but their effect is reduced in the inner disk and water is less affected. In Model~2 ($\alpha=10^{-4}$), the disk is systematically colder because of smaller viscous heating, and snowlines move closer to the star. As a consequence, the effect of heating by spiral arms is less pronounced. 

\subsection{Dust rings}
\label{sec:dustrings}

In both models, we obtain multiple dust rings in the simulation results. In Model~1, there is a thin and faint ring in dust and gas, situated precisely at the thermal water snowline (see panels (a) and (b) in Fig.~\ref{fig:st-2}). There are also multiple dust rings descending to the star after 400\,kyr in this model (see panels (b), (c), and (e) in Fig.~\ref{fig:st-2}). In Model~2, a system of prominent dust rings is created inside 10\,au, which persist throughout the disk evolution (see panels (b), (c), and (e) in Fig.~\ref{fig:st-4}). All these rings are created by various mechanisms that we briefly discuss in this Section. 

\begin{figure}
\centering
\includegraphics[width=\columnwidth]{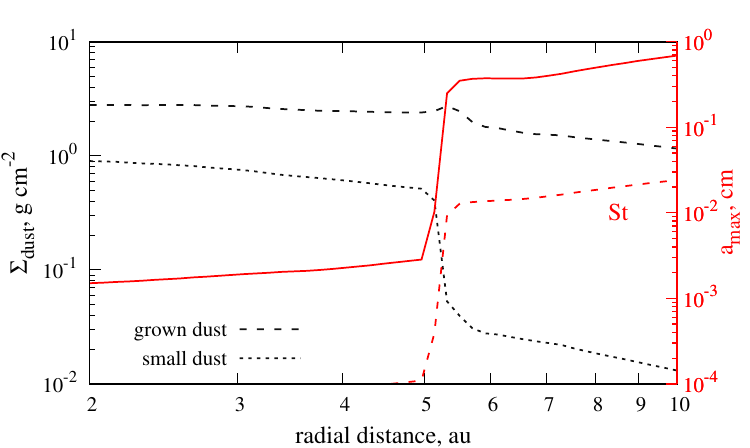}
 \caption{Radial profiles of main dust and gas parameters in the vicinity of the thin dust ring. Model~1 at $t=300$\,kyr. Stokes number scales with the left vertical axis.}
  \label{fig:ring1}
\end{figure}
The thin ring in grown dust and gas, coinciding with the position of the thermal snowline of water is connected to the volatiles and their effect on the fragmentation velocity. An example of the radial cross-section of this ring is shown in Figure~\ref{fig:ring1}. At this snowline, the fragmentation velocity changes and the dust maximum size decreases, resulting in a sharp drop in the Stokes number. With decreasing $St$ the inward drift of grown dust grains slows down and creates a factor of $\sim1.5$ increase in the grown dust surface density. Further accumulation is restricted by collisional fragmentation and transformation of grown dust into small dust. The formation of dust rings at the snowlines, which may assist in planetesimal formation, is described in many works \citep{2004ApJ...614..490C,2008A&A...487L...1B,2017A&A...608A..92D}. In our modeling, the ring due to this effect is only produced in the high-viscosity model and the increase in dust density is modest.

\begin{figure}
\centering
\includegraphics[width=\columnwidth]{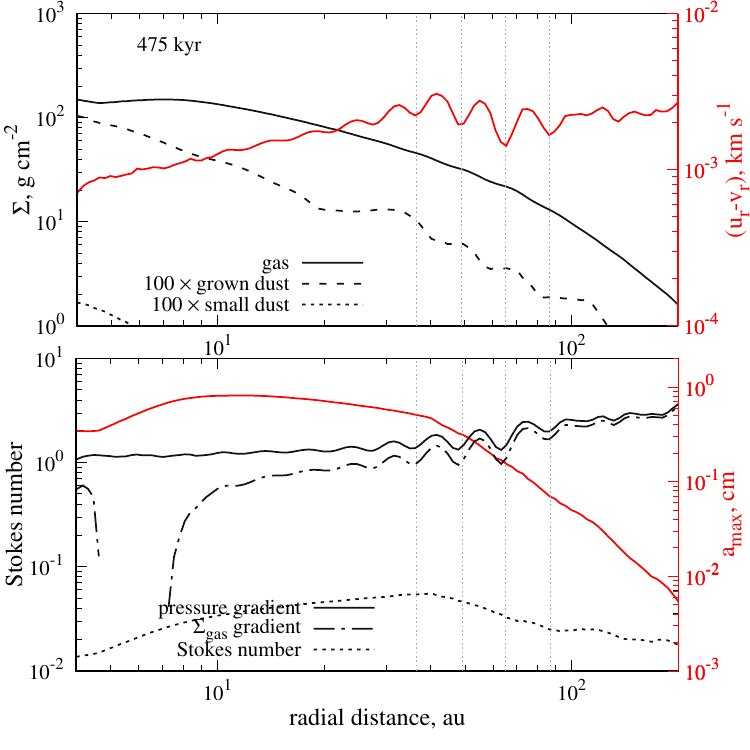}
 \caption{Radial profiles of disk parameters in Model~1 at $t=475$~kyr at a selected azimuthal cut. The top panel shows the gas surface density (solid black line), grown dust surface density (dashed black line), small dust surface density (dotted black line), and radial velocity of grown dust relative to that of gas (solid red line). The bottom panel presents the gas pressure gradient (solid black line), the gas surface density gradient (dash-dotted black line), Stokes number (dotted black line), and maximum size of grown dust (solid red line). The gradients correspond to the left y axis.}
  \label{fig:ring2}
\end{figure}

Large-scale rings that gradually migrate to the star at the late stages of disk evolution in Model~1 are likely formed thanks to the effect known as a ``traffic jam''. Figure~\ref{fig:ring2} illustrates the effect by showing the azimuthally averaged quantities at $t=475$~kyr. The step-like radial distribution of the grown dust surface density represents the dust rings (which are best seen in the dust-to-gas ratio in panel (e) of Fig.~\ref{fig:st-2}). We note that the gas surface density does not show such an expressed step-like behavior, although the gradient of $\Sigma_{\rm g}$  does show radial variations.
As a result, the gas pressure gradient that controls the dust drift velocity also shows local maxima and minima, which  correlate with the radial velocity of grown dust relative to that of gas $(u_r-v_r)$. These variations in the relative dust velocity create traffic jams, which result in the formation of the step-like distribution of $\Sigma_{\rm d,gr}$. The maximum size of dust grains and the Stokes numbers are consistent with the expectations of efficient radial drift of dust grains \citep[e.g.,][]{2016SSRv..205...41B}.
The radial variations in the gas surface density are caused by weak spiral density waves. It is not yet clear why similar ring structures do not form in Model~2. This low-$\alpha$ model has a stronger spiral pattern, which may smear out radial variations. We plan to explore this mechanism in more detail in follow-up studies.

Prominent rings in the inner disk regions in Model~2 are caused by the formation of a dead zone. The matter that is transported from the disk outer regions mainly by the action of gravitational torques in this low-$\alpha$ model hits a bottle neck in the innermost disk regions where the strength of the spiral pattern diminishes because of a rising disk temperature. We demonstrate this effect  by calculating the $\alpha$-parameter caused by gravitational instability. Following \citet{2016ARA&A..54..271K}, we express this quantity as
\begin{equation}
\alpha_{\rm GI} = \left| \frac{d \ln \Omega _{\rm K}}{d \ln r} \right| ^{-1}
\frac{1}{4 \pi G} 
\frac{\partial \Phi}{\partial r} 
\frac{\partial \Phi}{\partial \phi}
\end{equation}
where $G$ is the gravitational constant, $\Phi$ is the gravitational potential, and $\Omega_{\rm K}$ is the Keplerian angular velocity.  

Panels (i) in Figures~\ref{fig:st-2} and Figures~\ref{fig:st-4} present the space-time diagrams of the grown dust surface density and total effective $\alpha$-parameter (defined as the sum $\alpha_{\rm eff} = \alpha + \alpha_{\rm GI}$) for both considered models.
Model~2 is characterized by $\alpha_{\rm eff}$ that is a strong function of radial distance with a deep minimum in the inner 5--10 au. This is the region where prominent rings form in the grown dust density distribution. In Model~1 the effective $\alpha$-value is mostly set by the spatially and temporally constant viscous $\alpha=10^{-2}$. As a result, this model lacks a deep minimum in $\alpha_{\rm eff}$ in the inner disk regions and the dead zone does not form. Other mechanisms lead to the formation of weaker and more dynamic dust rings in this case, as discussed above.

Ring-like structures in dust continuum emission have been observed with high resolution in many protoplanetary disks \citep{2015ApJ...808L...3A,2018ApJ...869...17L,2018ApJ...869L..42H,2020arXiv201200189C}, although the apparent width and relative brightness of the rings would also be affected by the dust opacity variations \citep{2020MNRAS.499.5578A}. Nevertheless, among the observed mm continuum images of protoplanetary disks we can identify some structures analogous to weak-contrast dust rings obtained in our models (see Figure~\ref{fig:2d-2}), based on geometry of the rings and assuming that the apparent contrast in brightness is proportional to the surface density of grown dust. There are many analogues of large-scale evenly spaced multiple rings seen in Model~1, for example, HL~Tau and HD~143006, inner rings in RU~Lup and AS~209, outer rings in DoAr~25, DL~Tau, and GO~Tau. A single wide ring at the outer disk edge with a shallow broad gap around a central bright region, similar to the one we obtained in Model~2, is observed in Elias~20, Sz~129, and FT~Tau.

\subsection{Properties of icy mantles}
\label{sec:properties}

The evolution of refractory grains and volatiles is closely linked. Icy mantles control the fragmentation velocity and thus the grain size and dynamics. On the other hand, grains of sufficiently large size (and Stokes number) drift radially and redistribute volatiles within the disk. An important parameter for the volatile transport is the ice-to-rock mass ratio in a grain. It can vary both in space and time, as well as depend on the grain size. In this subsection, we focus on water ice, specifically for the temperatures close to water freeze-out temperature ($\approx$150\,K). We also consider the abundances of other molecules in the icy mantles and the evolution of ice composition.

Ice deposition on a grown grain can occur either via direct freeze out or in the coagulation process. In the first case, the mass of icy mantle should be proportional to the grain surface area, so the ice-to-rock mass ratio is inversely proportional to the grain size. If ice is deposited to dust grains solely by the coagulation with other grains, the ice-to-rock ratio will be constant for any grain size and equal to the initial one. In real disk conditions, both processes are in action, so the self-consistent treatment of dust and volatile evolution is required to properly assess the ice-to-rock ratio. This is especially important near the condensation fronts, where bidirectional transport of volatiles and grains may significantly redistribute the ice between grains \citep{1988Icar...75..146S,2004ApJ...614..490C,2017A&A...608A..92D}.

\begin{figure}
\includegraphics[width=\columnwidth]{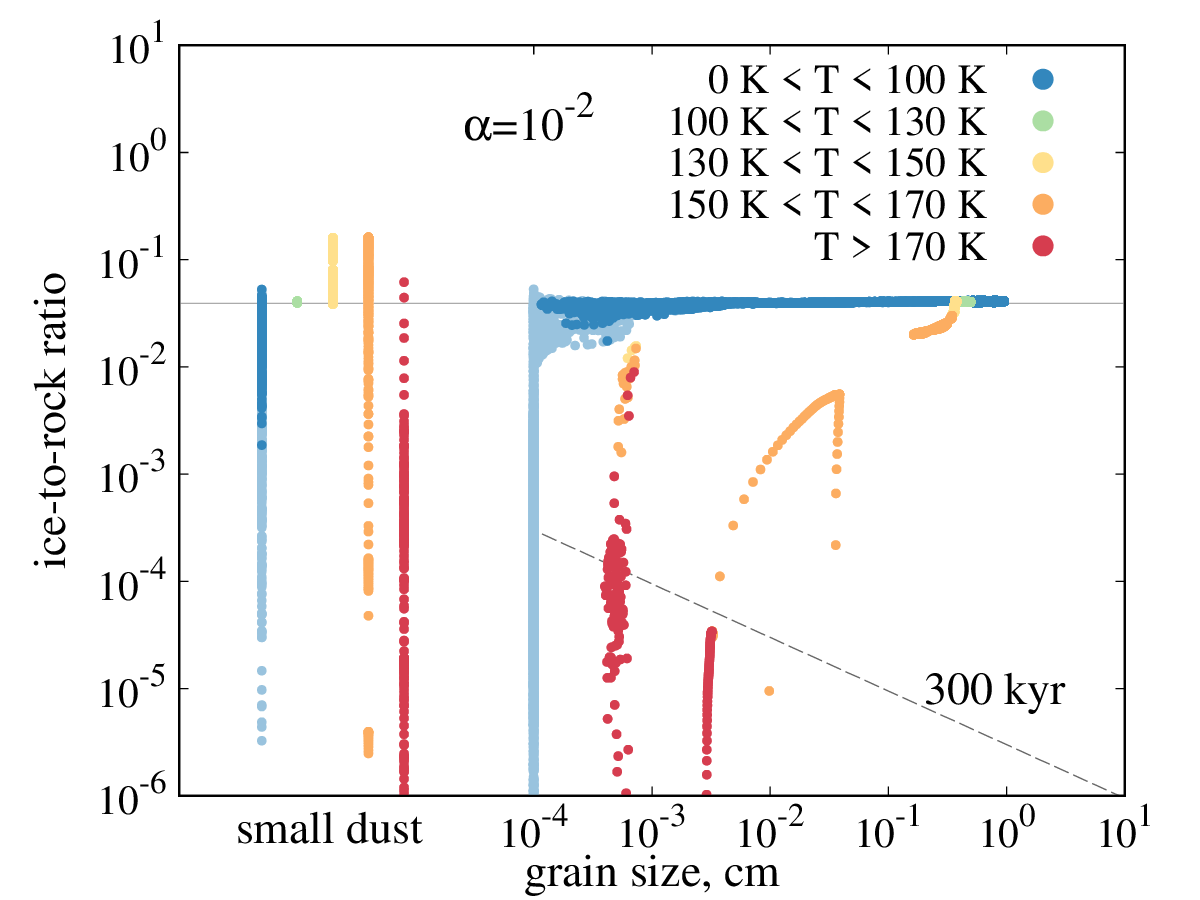}
\includegraphics[width=\columnwidth]{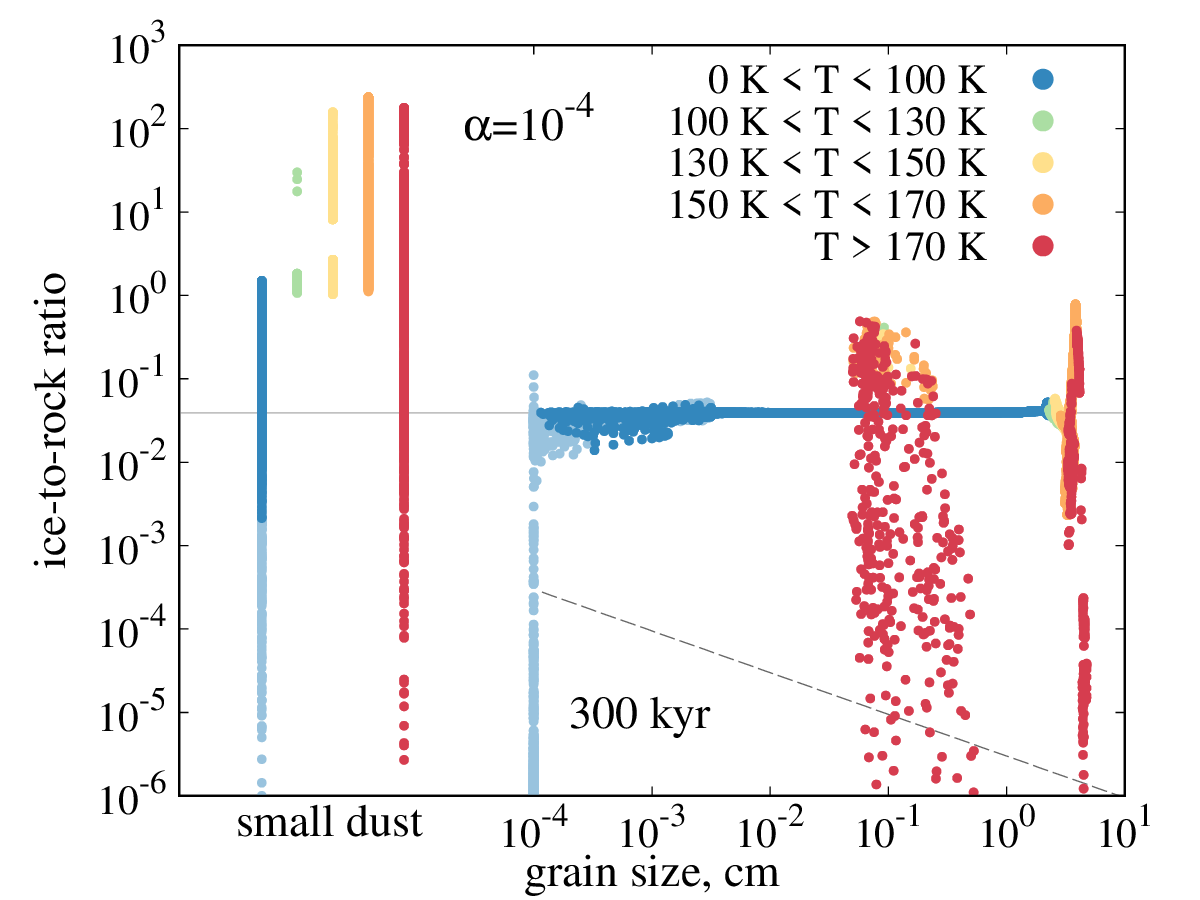}
 \caption{Water ice-to-rock mass ratio for small ($\Sigma_{\rm H_2O}^{\rm sm}$/$\Sigma_{\rm d,sm}$) and grown ($\Sigma_{\rm H_2O}^{\rm gr}$/$\Sigma_{\rm d,gr}$) dust. Models with $\alpha=10^{-2}$ (upper plot) and $\alpha=10^{-4}$ (lower plot) at 300\,kyr. The point color represents the local temperature, pale points are for corresponding dust surface density $<10^{-4}$\,g\,cm$^{-2}$. The initial water ice-to-rock ratio $3.9\times10^{-2}$ is marked with solid gray line. For grown dust, $a_{\rm max}$ is specified along horizontal axis. Small dust is shown on the left, grouped by temperature, the size of all small grains range between $5 \times 10^{-7}$ and $10^{-4}$\,cm. The dashed line shows the amount of ice corresponding to one monolayer mantles (see Eq.~\ref{eq:threshold}).}
  \label{fig:temps}
\end{figure}

Figure~\ref{fig:temps} shows the ice-to-rock mass ratio for water ice as a function of $a_{\max}$ for Model~1 (upper plot) and Model~2 (lower plot) at 300\,kyr.
The ratios for small ($\Sigma_{\rm H_2O}^{\rm sm}$/$\Sigma_{\rm d,sm}$)  and grown ($\Sigma_{\rm H_2O}^{\rm gr}$/$\Sigma_{\rm d,gr}$) dust are included.
Each dot on the plots represents some spatial location within the disk, while the dot color encodes the local disk temperature. The locations with low ice surface densities ($<10^{-4}$\,g\,cm$^{-2}$) are marked by pale dots. In both models there is a significant population of grown dust grains that are relatively cold (blue dots; $T<100$\,K) and have mantles with the initial ice-to-rock mass ratio $\Sigma_{\rm H_2O}^{\rm gr}/\Sigma_{\rm d,gr}=3.9 \times 10^{-2}$ (except for the regions where the dust content is very low, marked by pale blue dots). This indicates that the dominant process of ice deposition is coagulation rather than condensation on already grown grains. However, there is a prominent fraction of grains with ice-to-rock ratio significantly higher or lower than the initial one both for small and grown populations. Their spatial location corresponds to either the water condensation front (Model~1) or prominent rings in dust distribution (Model~2). The dashed line on both plots shows a threshold corresponding to one monolayer of ice. Grains below the dashed line are not entirely covered with ice and thus have a lower fragmentation barrier. 

In Model~1 with a higher $\alpha$-parameter, the grown grain size at the water snowline is governed by collisional fragmentation, so the change of fragmentation velocity due to the presence of ice is of relevance. The grains just outside the snowline are hot, relatively small and have very low water ice-to-rock ratio. The typical ice-to-rock ratio in Model~1 disk corresponds to the cold grains well beyond the snowline and is equal to the initial one. In Model~2 with a smaller $\alpha$-parameter, the dust size at the water snowline is limited by the radial drift rather than by fragmentation. The largest icy grains in the Model~2 disk are located just outside the snowline and thus are among the warmest ones. There is an additional population of hot mm-size grains (red dots at $\approx10^{-1}-10^{1}$\,cm) that belong to the highly dynamical region around the hot dust ring at 2\,au. Both populations may have ice-to-rock ratio up to an order of magnitude higher that the initial one. Note also that the typical temperatures of icy grains can exceed 170\,K: due to high density the characteristic freeze-out temperature in the corresponding disk regions is higher.

Small icy grains demonstrate even a wider range of ice-to-rock ratios. The excess of water in their mantles results from the recondensation of water that is brought from inside the snowline. In our modeling, we assume that the icy mantle does not affect the grain cross-section. This assumption holds in most of the disk, but appears to break down in the vicinity of snowlines where the ice-to-rock ratio $\gg1$ due to the accumulation of ices on small dust. While in Model~1 the water ice-to-rock ratio does not exceed 1, in Model~2 it reaches $\sim300$ for small grains. For an ice-to-rock ratio of $\sim1$, and the ice density 3 times lower than that of rock, the presence of a mantle would increase the grain size only by a factor of $\sqrt[3]{4}\approx1.6$, which is still of the order of unity. For all grown dust the relative amount of ice does not exceed 3, so the condition of mantles not contributing to dust size is not severely violated. For small grains with ice-to-rock ratios of 20 and 300, the increase in size would be $\approx4$ and $\approx10$, respectively. Such increase is harder to ignore. However, the mass fraction of small dust is notably lower than the one of grown dust, so most of collisions occur with grown grains rather that with small grains.

The presence of thick icy mantles could affect the dynamics and growth rate of dust, which should be further investigated. Figure~\ref{fig:temps} only presents ice-to-rock ratios for water, but other volatiles may have noticeable variations in the amount of ice, too. It is especially true for carbon dioxide, for which the effect of accumulation of ices on small dust is also very strong. The regions where different ices dominate over rock would generally not overlap with one another, as the effect works in a limited space near the snowline, so the thick mantles are nearly exclusively composed of one molecule.

Other studies also reported ices mostly residing on smaller grains around the snowline. \citet{2017A&A...600A.140S} found that outside the CO snowline ice mantles become thicker, and that small grains contain a larger fraction of ice (ice exceeding rock by 2--3 orders of magnitude) than grown grains (ice-to-rock ratio of 0.5--10 depending on the dust size). They also show that the accumulation region is wider for small grains. Similarly, \citet{2018ApJ...864...78K} found that when accumulating outside the snowline CO ice resides on small dust grains, rather than on grown pebbles.

\begin{figure}
\includegraphics[width=\columnwidth]{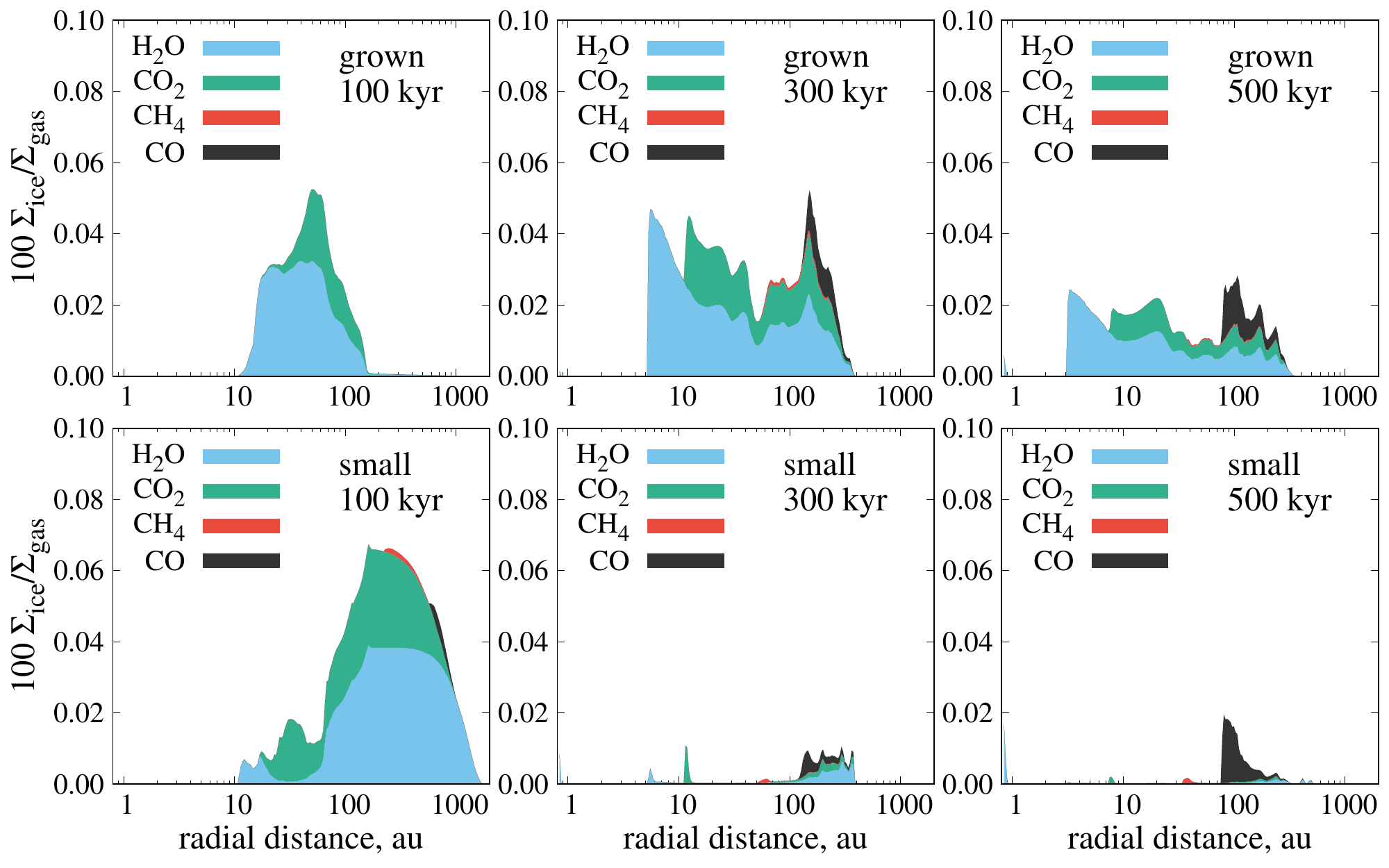}
\includegraphics[width=\columnwidth]{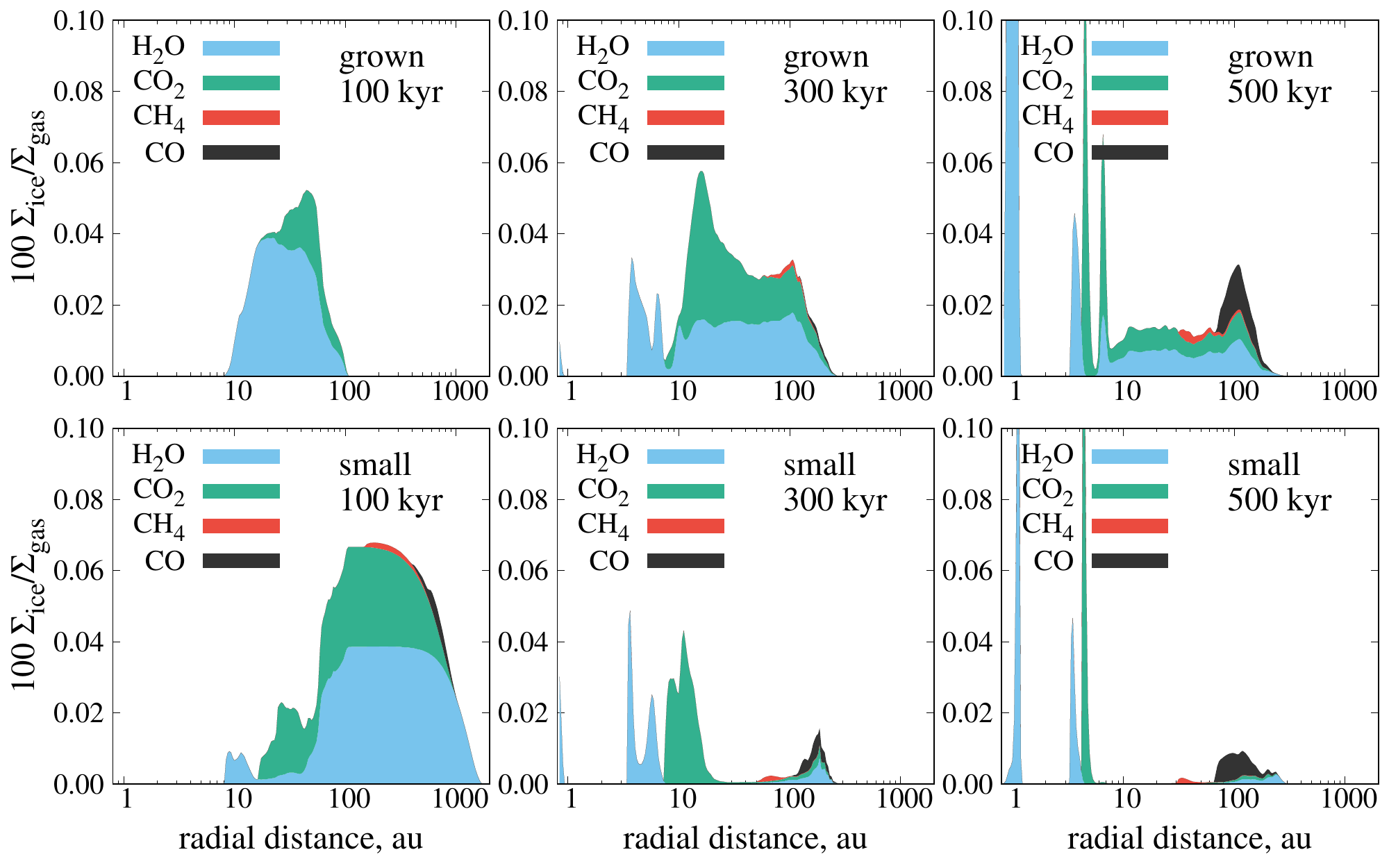}
 \caption{Cumulative radial distribution of ices relative to gas at different time moments. The upper six plots are for the $\alpha = 10^{-2}$ Model~1, the lower six plots are for the  $\alpha = 10^{-4}$ Model~2. At 500\,kyr in Model~2 water and CO$_2$ reach  peak values of $100 \times \Sigma_{\rm ice}/\Sigma_{\rm gas}\approx0.5$ (off the scale). }
  \label{fig:icebergs}
\end{figure}

The composition of icy mantles changes throughout the disk and evolves significantly compared to the initial one. The ratio of ice surface density to that of gas for different species is shown in Figure~\ref{fig:icebergs}. Despite the accumulation of ices on small dust, ice surface densities appear to be only a small fraction of matter in the disk, when normalized to gas surface density and azimuthally averaged. Typical ice-to-gas mass ratio is 0.02-0.08\,\%, only reaching $\approx1$\% in a region near inner rings (see the out-of-scale peaks in Model~2 at 500\,kyr in Figure~\ref{fig:icebergs}). The grains with thick mantles only present a very small sub-population of the dust, but a more sophisticated model accounting for the presence of these thick mantles is needed to understand their impact on dust evolution.

At the earlier stage (100\,kyr) ices on small and grown dust are notoriously separated in Figure~\ref{fig:icebergs}, ices on grown dust dominating inside $\approx$100\,au, and ices on small dust being abundant in the exterior disk parts. Initially deposited on surfaces of small grains, ices are transferred to grown dust mostly through coagulation, which is not efficient outside 100\,au at early evolutionary stages. With time, the disk spreads out and dust growth takes over in most of the ice-dominated space, so at later stages the ices predominantly reside on the grown grains. At 300~and 500\,kyr, only a fraction of ice remains on small dust in the regions beyond 100\,au. 

The total amount of ice in the disk decreases with time, as drifting grains bring the volatiles to the inner disk, where they turn to the gas and then accrete to the star. The disk global ice-to-gas mass ratio decreases by $\approx2$ times during the first 0.5\,Myr. At later times, most of the ice is on grown grains, as most of rock material has evolved into grown dust. On both small and grown dust, the amount of ice varies with the distance from the star.

The presence of (thermal) snowlines is reflected in the composition of ices: in the inner disk the mantles are made of water, then CO$_2$ is added, and CH$_4$ and CO in the outer regions. The ratios between ices evolve, too. If at early times they are close to the initial abundances from Table~\ref{tab:abundances}, later the fraction of CO$_2$ and CO grows and can exceed the one of water. For example, at 500\,kyr at 100--200\,au half of all the ice on grown dust is CO ice, and mantles on small grains consist almost entirely of this species. Carbon dioxide dominates in the composition of mantles between 10--30\,au in Model~2. The amount of CO ice increases with disk age with most of CO staying in the gas phase during first $\approx200-300$\,kyr of disk evolution, as also found in observations \citep{2020arXiv200808106V}.

There are studies considering the chemical evolution of ices in disks with dust dynamics and evolution. Detailed chemical modeling of quasi-stationary disks shows that at Myr timescales CO can be chemically depleted from both gas and ice phases by efficient reprocessing to CO$_2$, CH$_4$ and CH$_3$OH \citep{2018A&A...613A..14E,2018A&A...618A.182B} or under significant ionization by cosmic ray and UV radiation \citep{2018ApJ...856...85S,2019ApJ...877..131S}. \citet{2019MNRAS.487.3998B} show that in viscous disks ($\alpha>10^{-3}$) pebble drift dominates chemical depletion of CO, CH$_4$ and CO$_2$, although high cosmic ray ionization rate ($\zeta_{\rm CR}=10^{-17}$) is able to make chemical processing more efficient \citep[as also pointed out by][]{2018A&A...618A.182B}. Combination of chemical modeling with dust transport suggests CO depletion from the gas by two orders of magnitude \citep{2020ApJ...899..134K}. We follow the disk evolution only for several 100\,kyr, but it is possible that chemical reprocessing is already in action at these early stages. Adding chemical evolution of volatiles would be a natural development of the presented model.

\section{Conclusions}
\label{sec:conclusions}

We have incorporated the dynamics, adsorption, and desorption of four volatile molecules (H$_2$O, CO$_2$, CH$_4$, and CO) into the thin-disk hydrodynamic model FEOSAD of a protoplanetary disk with two evolving dust populations. We model the evolution of the gas-dust disk from its formation to an age of 500\,kyr, considering two values of the turbulent parameter $\alpha=10^{-2}$ and $10^{-4}$.

We demonstrate how the two-dimensional disk structure, dust growth, gas and dust dynamics, as well as time-dependent freeze-out and desorption can influence the distribution of volatiles in the gas phase and on the icy mantles of two considered dust populations (small dust with a sub-micron size and grown dust with a variable upper size). We also assess the impact of icy mantles on the dust evolution by means of a variable dust fragmentation velocity that depends on the presence or absence of icy mantles. Our main conclusions can be summarized as follows.
\begin{itemize}
    \item Each of the considered molecules can have multiple snowlines in the disk. In particular, the radial distribution of H$_2$O and CO$_2$ is characterized by several thermal desorption snowlines produced by radial variations in the disk density and temperature. In the envelope outside the disk ($\approx1000$\,au), the volatiles are  removed from the ice through photodesorption, resulting in the formation of a photodesorption snowline. However, this snowline can be less pronounced as  gas-phase volatiles in this region could be destroyed by photodissociation.
    
    \item As the disk evolves and cools down due to decreasing accretion and viscous heating, the thermal snowlines shift closer to the star. Between 100 and 500\,kyr, the distances from the star to the snowlines of H$_2$O, CO$_2$, and CH$_4$ decrease by a factor of 4--5, e.g., from 15 to 3\,au for water (model with $\alpha=10^{-2}$).
    The CO snowline appears at $\sim100$\,au after 200--300\,kyr of disk evolution. Before that CO resides almost entirely in the gas phase.
    
    \item Ices are delivered to grown dust mainly through coagulation with small icy grains. While in some disk regions small dust is sufficiently depleted through dust growth,
    the surface area of small grains still dominates the total surface area everywhere, so that freeze-out of volatiles from the gas phase occurs predominantly on small dust. This finding suggests that grown grains consist of sub-micron grains individually covered with agglutinate icy mantles, rather than a single rocky grain covered with a consolidated icy mantle.
    
    \item The change of fragmentation velocity  $v_{\rm frag}$ near the snowline notably affects dust size and surface density only in the $\alpha=10^{-2}$ model, in which the dust properties change sharply at the water snowline situated at 3--10\,au. In the inner disk region, where water ice is absent, the dust size is 10--100\,$\mu$m, which is $\sim$100 times lower than outside of the snowline. The fraction of small dust in this region is two orders of magnitude higher than in the rest of the disk, and the dust-to-gas ratio is elevated by a factor of several. In the $\alpha=10^{-4}$ model, the dust size is mostly determined by radial drift and is not affected by the change in $v_{\rm frag}$.
    
    \item Volatile species accumulate at the snowlines both in the ice and gas phases, as also shown in  \citet{1988Icar...75..146S,2003Icar..166..385C,2004ApJ...614..490C,2017A&A...608A..92D}. The amplitude of the effect and the width of the affected region varies for different snowlines and $\alpha$ values. This accumulation is caused by multiple mechanisms, including azimuthal variations in gas and dust radial velocities, and should be further investigated.
    
    \item The presence of non-axisymmetric spiral structures in the disk leads to a complex shape of the thermal desorption snowlines, especially for H$_2$O and CO$_2$, which cannot be described by a single radial position \citep[also previously noted by][]{2017MNRAS.472..189I}. The snowlines extend farther from the star due to the warming effect of spiral arms. The propagation of spiral density waves through the disk causes sublimation of ices followed by their recondensation on small dust, which creates thick mantles on small grains with masses comparable to or exceeding those of the rocky grains.
    
    \item The icy mantles on small grains become up to 300 times more massive than the rocky grains on which they reside in the vicinity of thermal snowlines of all volatiles and in the highly dynamic dust rings.  The contribution of ice into the dust mass and size, and the effect of icy mantles on the dynamical properties of dust need to be further investigated.
    
    \item The composition of icy mantles evolves significantly. The total amount of ices relative to gas decreases by a factor of 2 throughout the disk evolution. As water and carbon dioxide are efficiently transported into the inner disk, carbon monoxide and methane start to dominate in the icy mantles. At later evolutionary stages, the icy mantles on small grains are nearly entirely composed of single species in the vicinity of their thermal snowlines.
\end{itemize}

We are thankful to the anonymous referee for constructive comments that helped to improve the manuscript.
The research was carried out in the framework of the project ``Study of stars with exoplanets'' under a grant from the Government of the Russian Federation for scientific research conducted under the guidance of leading scientists (agreement N 075-15-2019-1875). E.I.V., V.A., A.S. and D.W. acknowledge the support of Ministry of Science and Higher Education of the Russian Federation under the grant 075-15-2020-780 (N13.1902.21.0039; Sections 2, 2.1, 2.3, 3.4 and 3.5.). The computational results presented have been achieved using the Vienna Scientific Cluster (VSC).

\vspace{5mm}
\software{FEOSAD \citep{2018A&A...614A..98V}}

\bibliography{thebibliography}{}
\bibliographystyle{aasjournal}

\appendix

\section{Analytic solution for the evolution of volatiles}
\label{sec:solutions}

The system of Equations~\ref{eq:sig1}--\ref{eq:sig3} has an analytic solution. If ${\Sigma_{\rm0}}_{s}^{\rm gas}$, ${\Sigma_{\rm0}}_{s}^{\rm sm}$ and ${\Sigma_{\rm0}}_{s}^{\rm gr}$ are the initial values of species surface densities, then after a time $\Delta t$ their values will be
\begin{equation}
\Sigma_{s}^{\rm gas}={\Sigma_{\rm0}}_{s}^{\rm gas} +\left(\frac{\eta_s}{\lambda_s}-{\Sigma_{\rm0}}_{s}^{\rm  gas}\right)\left(1-e^{-\lambda_s\Delta t}\right),
\label{eq:sig6}
\end{equation}
\begin{equation}
\Sigma_{s}^{\rm sm}={\Sigma_{\rm0}}_{s}^{\rm sm} + \frac{\lambda_s^{\rm sm}}{\lambda_s}\left({\Sigma_{\rm0}}_{s}^{\rm gas}-\frac{\eta_s}{\lambda_s}\right)\left(1 - e^{-\lambda_s\Delta t}\right),
\label{eq:sig7}
\end{equation}
\begin{equation}
\Sigma_{s}^{\rm gr}= {\Sigma_{\rm0}}_{s}^{\rm gr} + \frac{\lambda_s^{\rm gr}}{\lambda_s}\left({\Sigma_{\rm0}}_{s}^{\rm gas} -\frac{\eta_s}{\lambda_s}\right)\left(1 - e^{-\lambda_s\Delta t}\right).
\label{eq:sig8}
\end{equation}
Here, we use the fact that the ratio between both adsorption and desorption rates of two ice populations equals to the ratio of total surface areas of the corresponding dust populations $\eta_s^{\rm sm}/\eta_s^{\rm gr}=\lambda_s^{\rm sm}/\lambda_s^{\rm gr} = \sigma_{\rm tot}^{\rm sm}/\sigma_{\rm tot}^{\rm gr}$.

Asymptotic or equilibrium behavior of these solutions at infinite time is the following. From Equation~\ref{eq:sig6}, the equilibrium value of $\Sigma_{s}^{\rm gas}$ at $\Delta t \rightarrow \infty$ is $\eta_s/\lambda_s$. As it can be derived from Equations~\ref{eq:sig1}, \ref{eq:sig2}, and \ref{eq:sig3}, it is either $\eta_s/\lambda_s$, $\eta_s^{\rm sm}/\lambda_s^{\rm sm}$, or $\eta_s^{\rm gr}/\lambda_s^{\rm gr}$, which are all equal. Using Equations~\ref{eq:sig7} and \ref{eq:sig8} one can find the equilibrium solutions for ices at $\Delta t \rightarrow \infty$ being $\Sigma_{s}^{\rm gr}= {\Sigma_{\rm0}}_{s}^{\rm gr} + \lambda_s^{\rm gr}/\lambda_s\left({\Sigma_{\rm0}}_{s}^{\rm gas} -\eta_s/\lambda_s\right)$ and $\Sigma_{s}^{\rm sm}= {\Sigma_{\rm0}}_{s}^{\rm sm} + \lambda_s^{\rm sm}/\lambda_s\left({\Sigma_{\rm0}}_{s}^{\rm gas} -\eta_s/\lambda_s\right)$.

For certain initial conditions, which are present in protostellar disks, it is possible for equilibrium surface density of a species vapor $\eta_s/\lambda_s$ to be larger than the total amount of these species $\Sigma_{s}^{\rm total}$. This is a nonphysical effect of the selected approach of zeroth-order desorption, when desorption rate does not depend on the amount of ice present in the volume. The solution also allows for negative surface densities of ice.
To circumvent these inconsistencies, we modify the solution, so that there is no negative surface densities and gas surface density never exceeds total amount of the species $\Sigma_{s}^{\rm total}={\Sigma_{\rm0}}_{s}^{\rm gas}+{\Sigma_{\rm0}}_{s}^{\rm sm}+{\Sigma_{\rm0}}_{s}^{\rm gr}$. Once the amount of any ice goes below a very small positive value $\epsilon_{\rm tiny}=10^{-15}\Sigma_{s}^{\rm total}$, it stops there, and the other ice component is found using conservation of total species amount. If both ices of a species are below $\epsilon_{\rm tiny}$, then the species abundance in gas is set to $\Sigma_{s}^{\rm total}$, and ices are set to $\epsilon_{\rm tiny}$.

\begin{figure*}
\includegraphics[width=0.247\columnwidth]{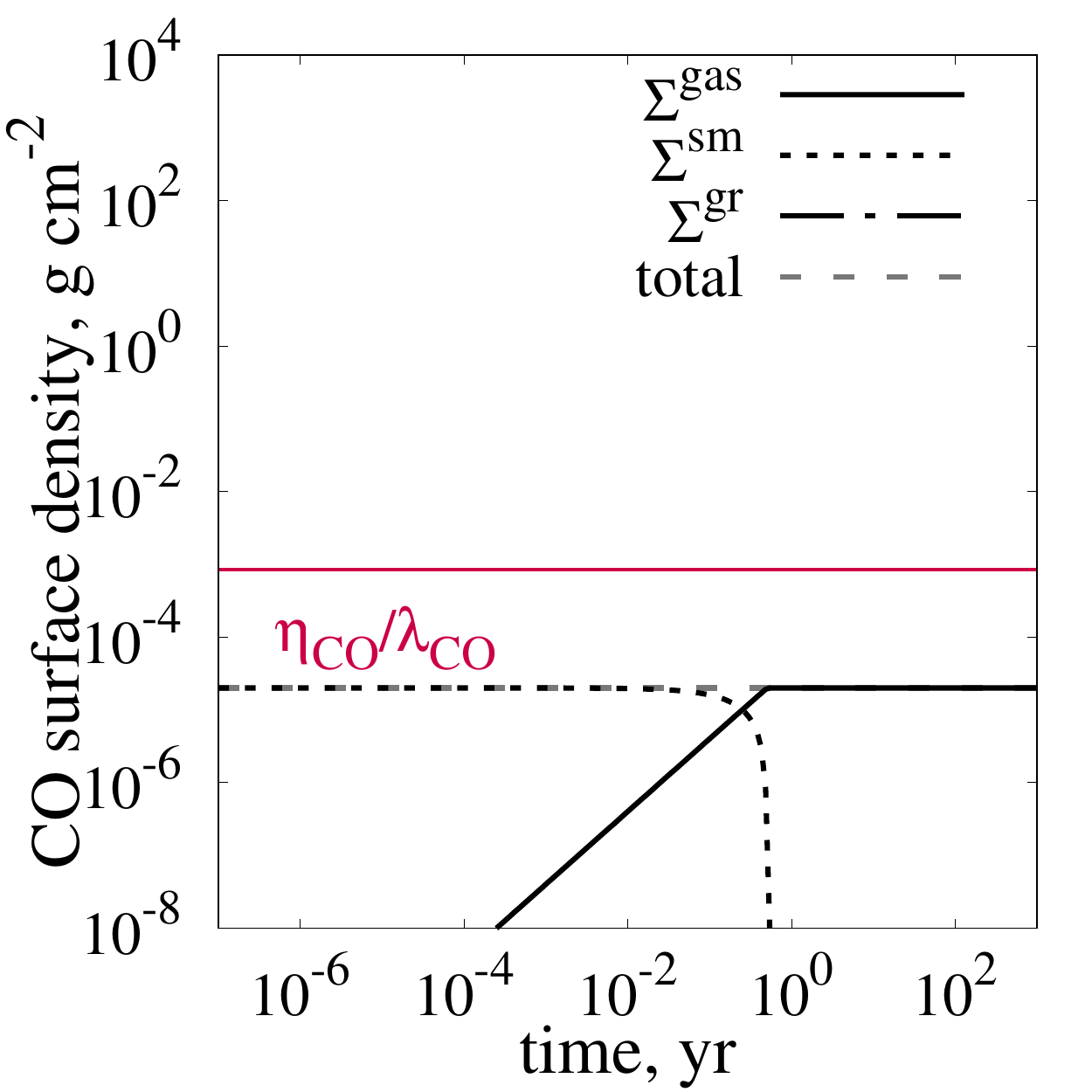} 
\includegraphics[width=0.247\columnwidth]{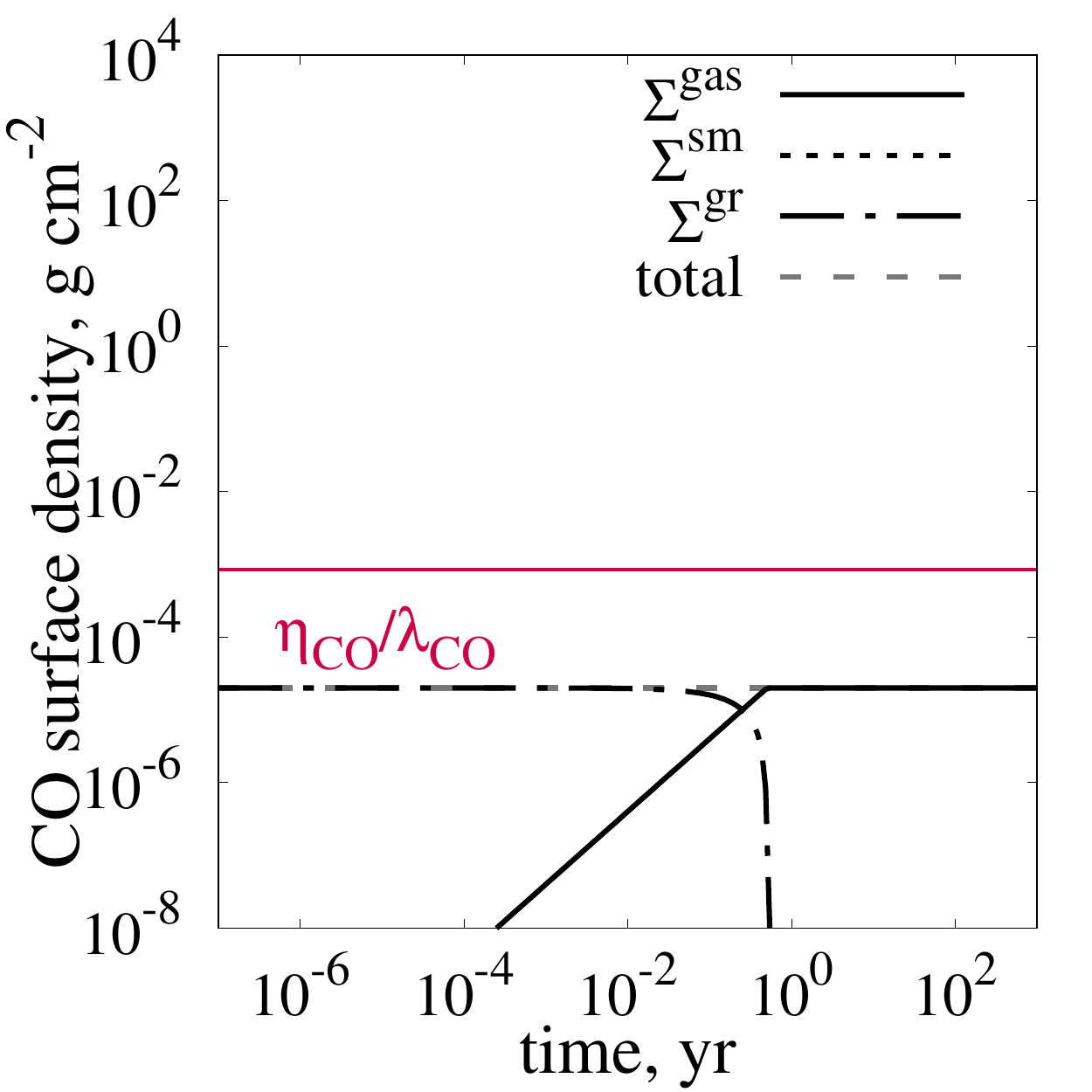} 
\includegraphics[width=0.247\columnwidth]{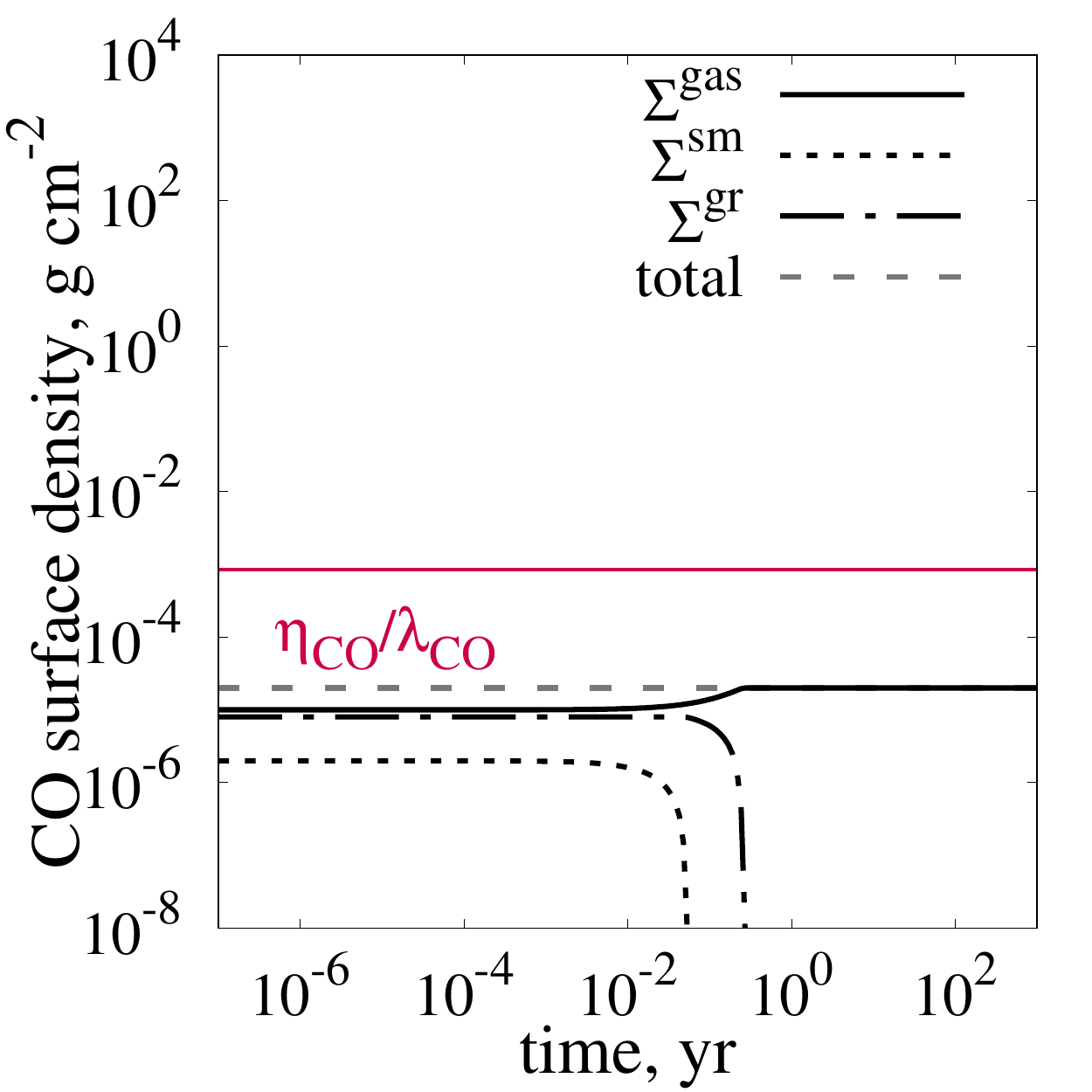} 
\includegraphics[width=0.247\columnwidth]{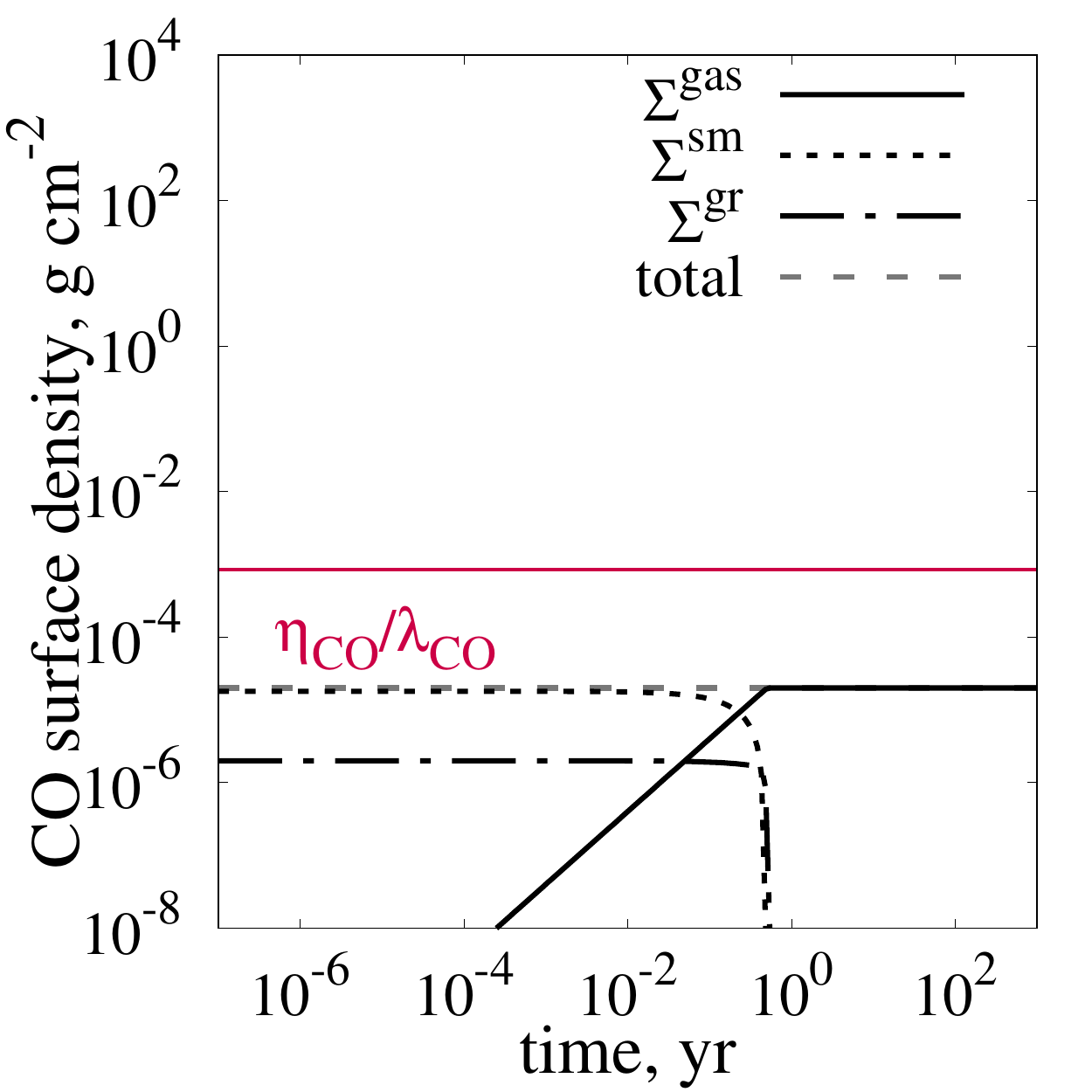}
\includegraphics[width=0.247\columnwidth]{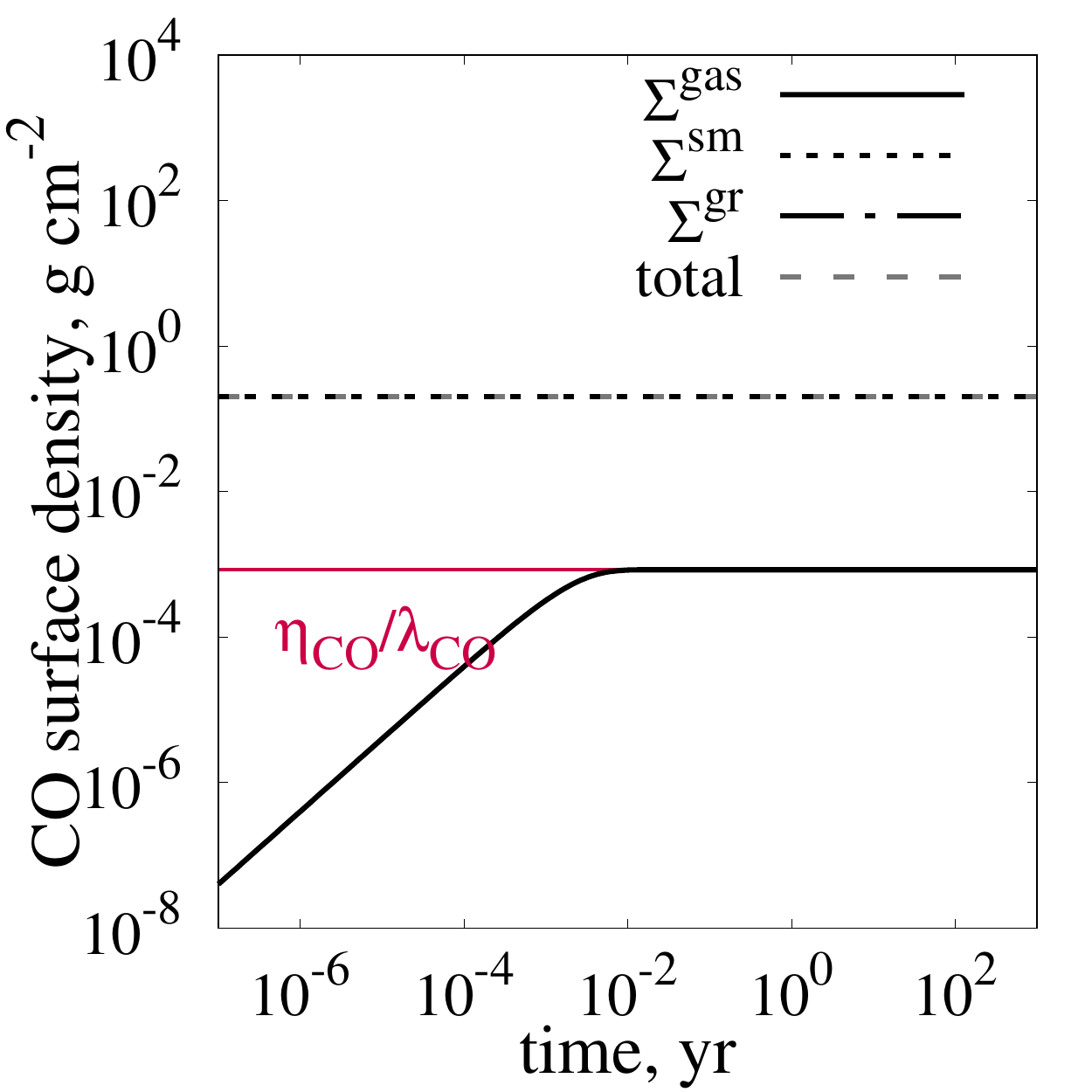} 
\includegraphics[width=0.247\columnwidth]{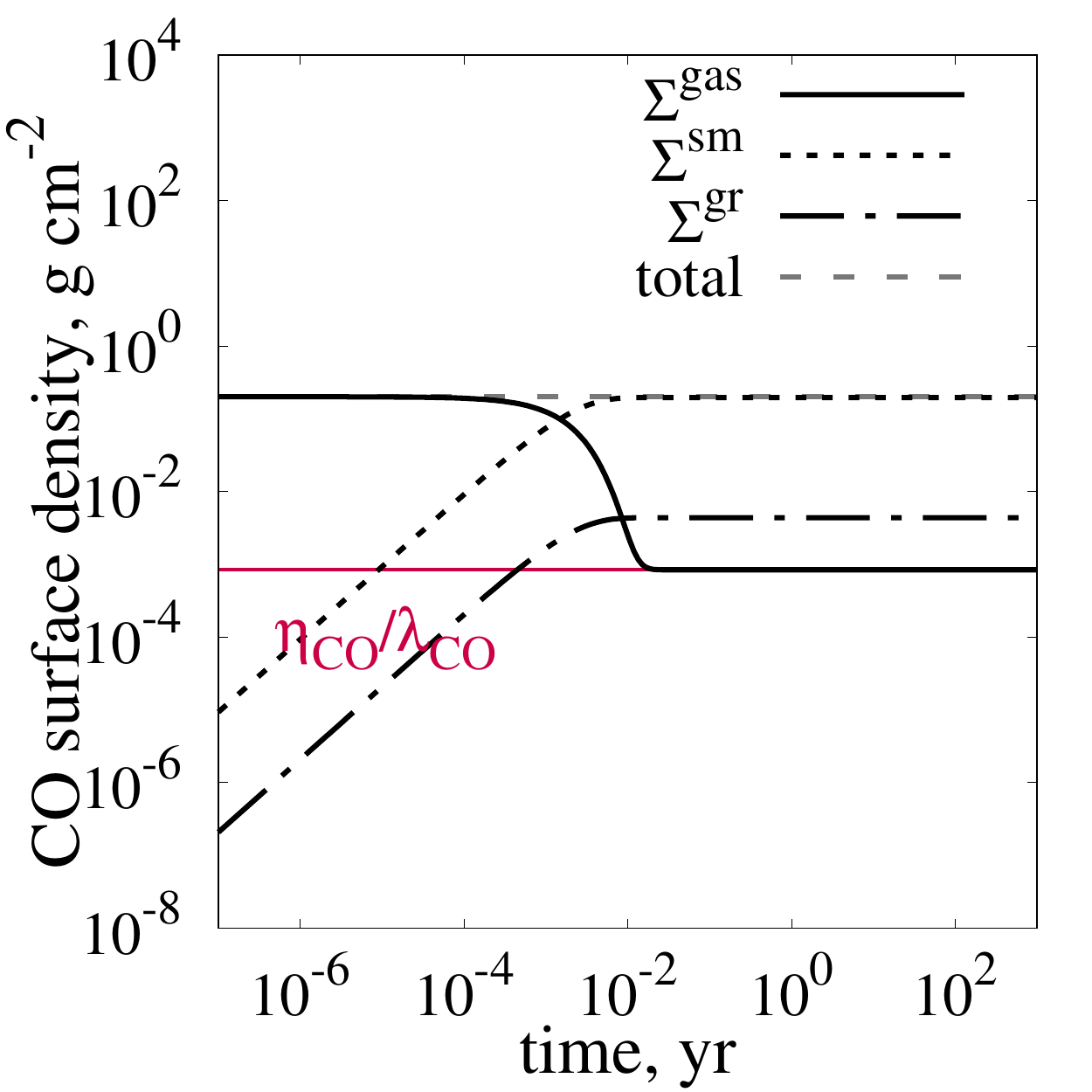} 
\includegraphics[width=0.247\columnwidth]{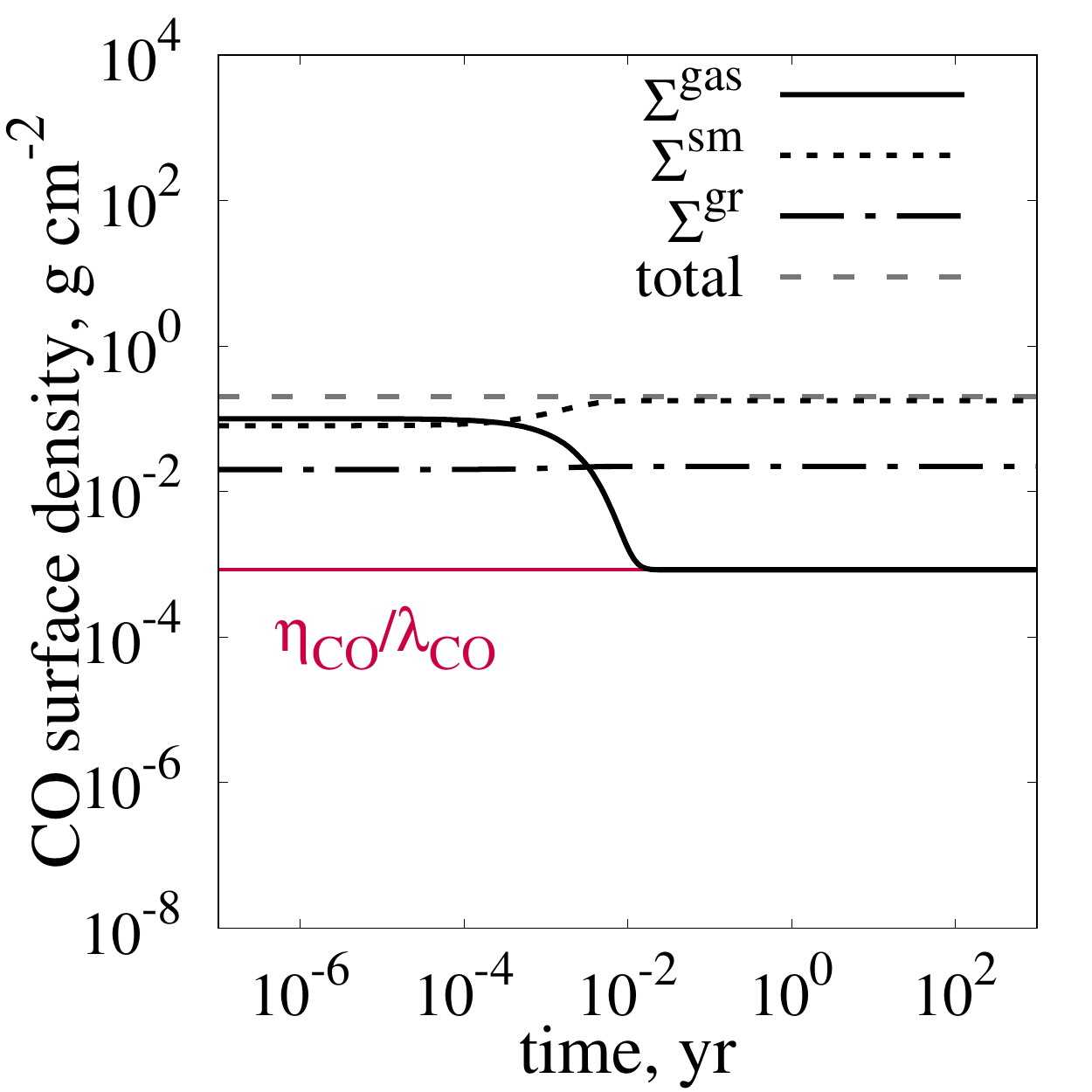} 
\includegraphics[width=0.247\columnwidth]{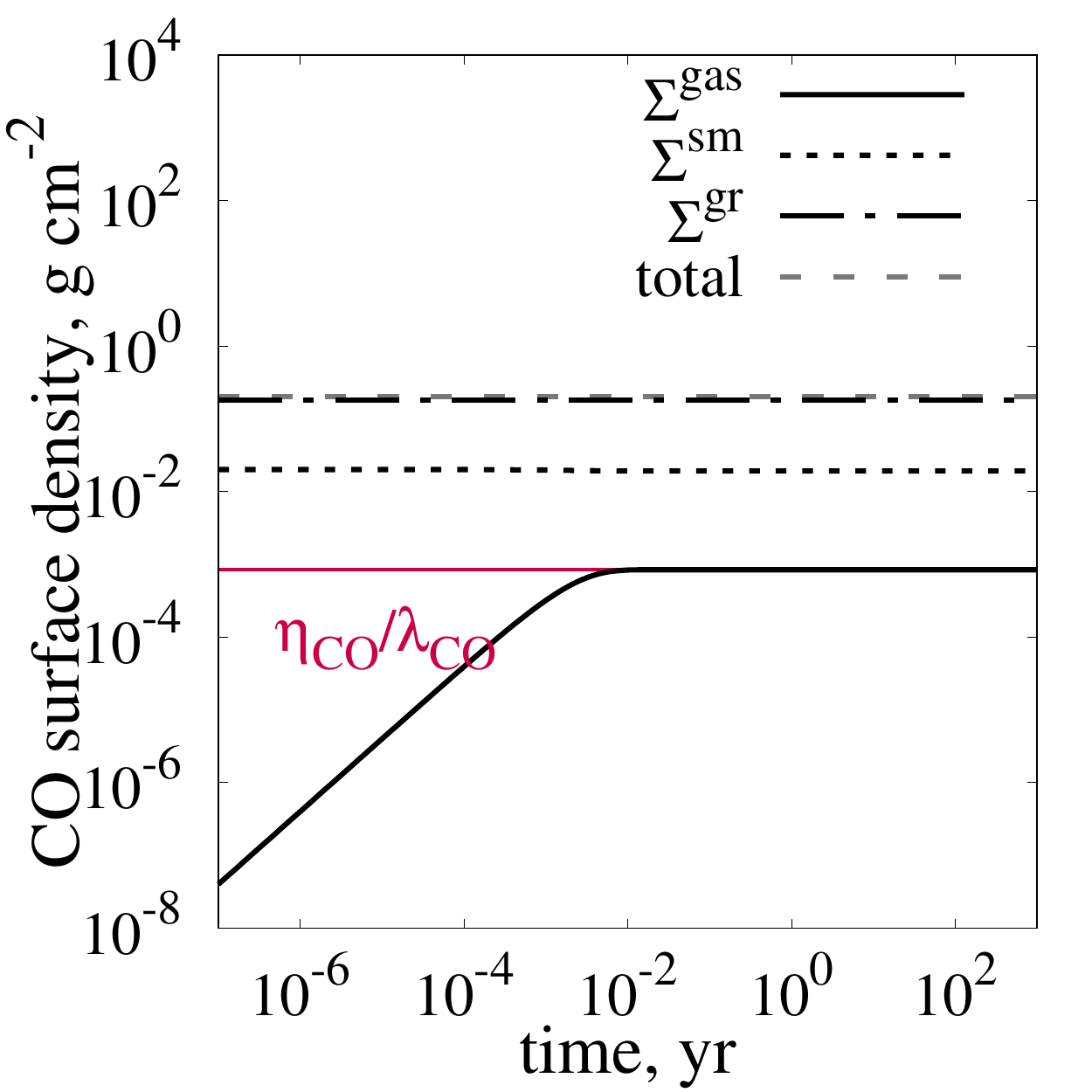} 
 \caption{Sample calculations for single-point model of adsorption/desorption of CO with different initial surface densities. The surface density of CO in the gas, on small dust grains, and on grown dust grains is shown in black lines, gray dashed line is the total surface density of CO in the three phases, thin red line shows the equilibrium gas-phase abundance $\eta_{\rm CO}/\lambda_{\rm CO}$.}
  \label{fig:tests}
\end{figure*}

Our approach is illustrated by the behavior of the solution for a simple single-point model. The evolution of the three surface density components is shown in Figure~\ref{fig:tests} for an example of CO molecule for different initial surface densities of the three phases. Only thermal desorption is included. In these test models, we assume the temperature of 21\,K, minimum small dust size $a_{\rm min} = 5\times10^{-7}$\,cm, maximum small dust size and minimum grown dust size $a_{*} = 10^{-4}$\,cm, as in the main calculation, and maximum grown dust size $a_{\rm max} = 10^{-3}$\,cm as an example of slightly evolved dust. Scale height is selected as $h=1$\,au. Surface densities of small and grown dust are equal and take values of $\Sigma_{\rm d,sm} = \Sigma_{\rm d,gr} = 10^{-4}$, $10^{-2}$, and 1\,g\,cm$^{-2}$, which are typical for outer, intermediate, and inner regions of protostellar disks. The desoprtion energy $E^{s}_{\rm des}$ and molecular mass are taken for CO (see Table~\ref{tab:abundances}).

We pick three main groups of models that are framed by the ratio between the total initial amount of CO and the equilibrium surface density of CO in the gas: $\Sigma^{\rm total}_{\rm CO} =\Sigma^{\rm gas}_{\rm CO} + \Sigma^{\rm sm}_{\rm CO} + \Sigma^{\rm gr}_{\rm CO} < \eta_{\rm CO}/\lambda_{\rm CO}$ and $> \eta_{\rm CO}/\lambda_{\rm CO}$, shown in the upper and lower row in Figure \ref{fig:tests}, correspondingly. These cases represent tenuous gas too hot for the presence of ices and dense gas tending to freeze-out, respectively. 
If freeze-out temperature is defined as one that corresponds to an equilibrium between gas and ice for given conditions plus equal abundances in the gas and in the ice, then we could say that in the first case, the assumed 21\,K temperature is above the freeze-out temperature and in the second case 21\,K is below freeze-out temperature.

For these three cases we calculate the evolution of abundances in the gas and the two ice phases. The upper row in Figure~\ref{fig:tests} shows the examples for undersaturated gas, $\Sigma^{\rm total}_{\rm CO} > \eta_{\rm CO}/\lambda_{\rm CO}$. Here ices sublimate and all the material transforms into the gas, going to the equilibrium solution, but never reaching it, as not enough material is available. The individual timescales of freeze-out for ices on small and grown dust are the same, as they have the same exponent in the Equations~\ref{eq:sig7} and~\ref{eq:sig8}. When both populations are present, the ice on small dust declines faster, because it is proportional to $\lambda_{\rm CO}^{\rm sm}$, which is larger, as small dust mostly has larger total surface area. For some conditions not uncommon in evolving protoplanetary disks, it can be vice versa, if small dust is extremely depleted and most of solid rocky material is transformed into  grown dust.

If the total amount of CO in the three phases exceeds $\eta_{\rm CO}/\lambda_{\rm CO}$, then various ratios between $\Sigma_{\rm CO}^{\rm sm}$ and $\Sigma_{\rm CO}^{\rm gr}$ are possible, depending on the initial conditions.

The timescales of freeze-out and sublimation are set by total adsorption rate $\lambda_{\rm CO}$. In the examples shown in Figure~\ref{fig:tests}, they vary from days to years, which is much faster than typical dynamical times in protostellar disks. However, in colder and more tenuous  medium, they can reach  hundreds of years, while in warm and dense regions they are as short as minutes. Some examples for water, as well as the first implications of the presented model are shown in \citet{2019INASR...4...40M}.

\begin{figure*}
\includegraphics[width=0.247\columnwidth]{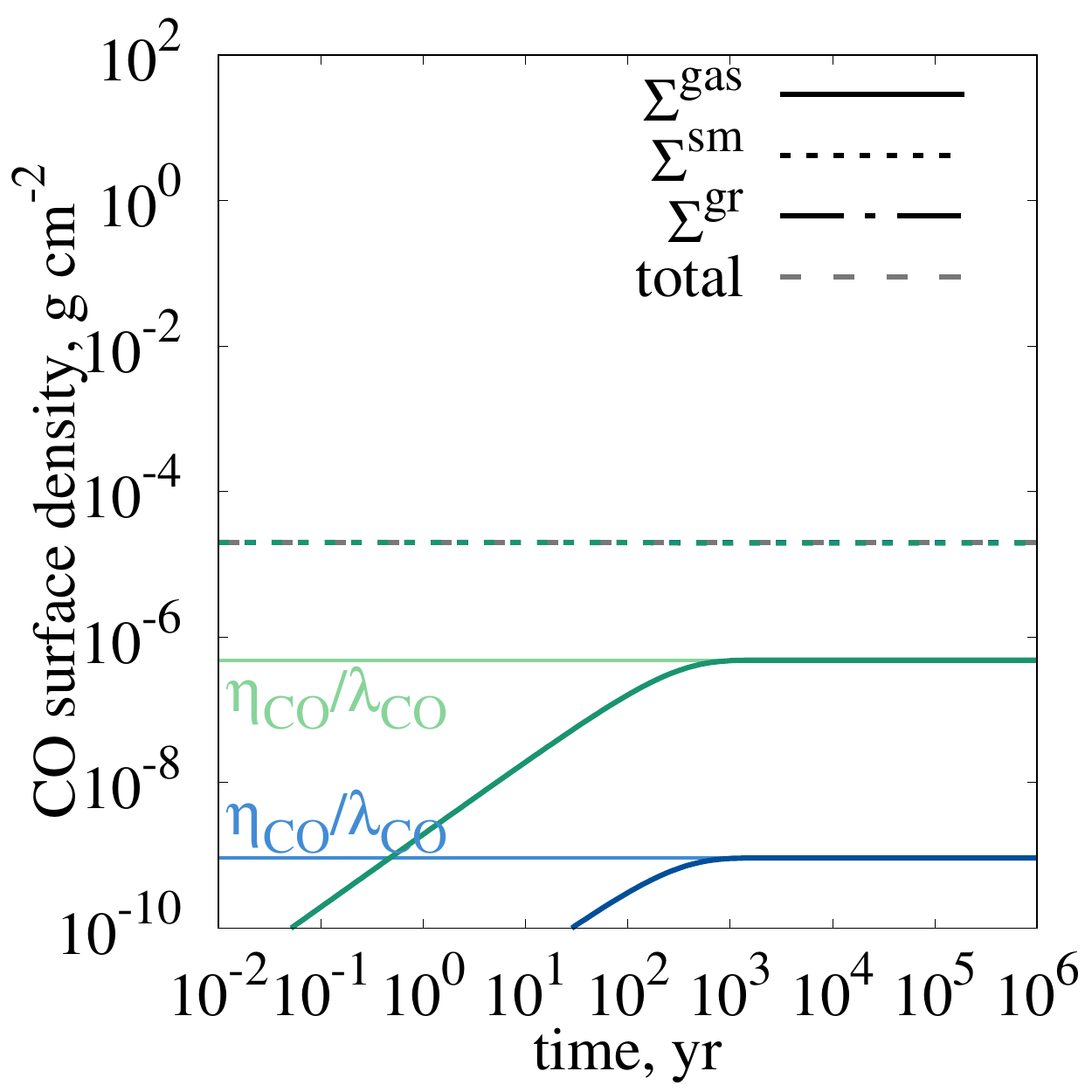} 
\includegraphics[width=0.247\columnwidth]{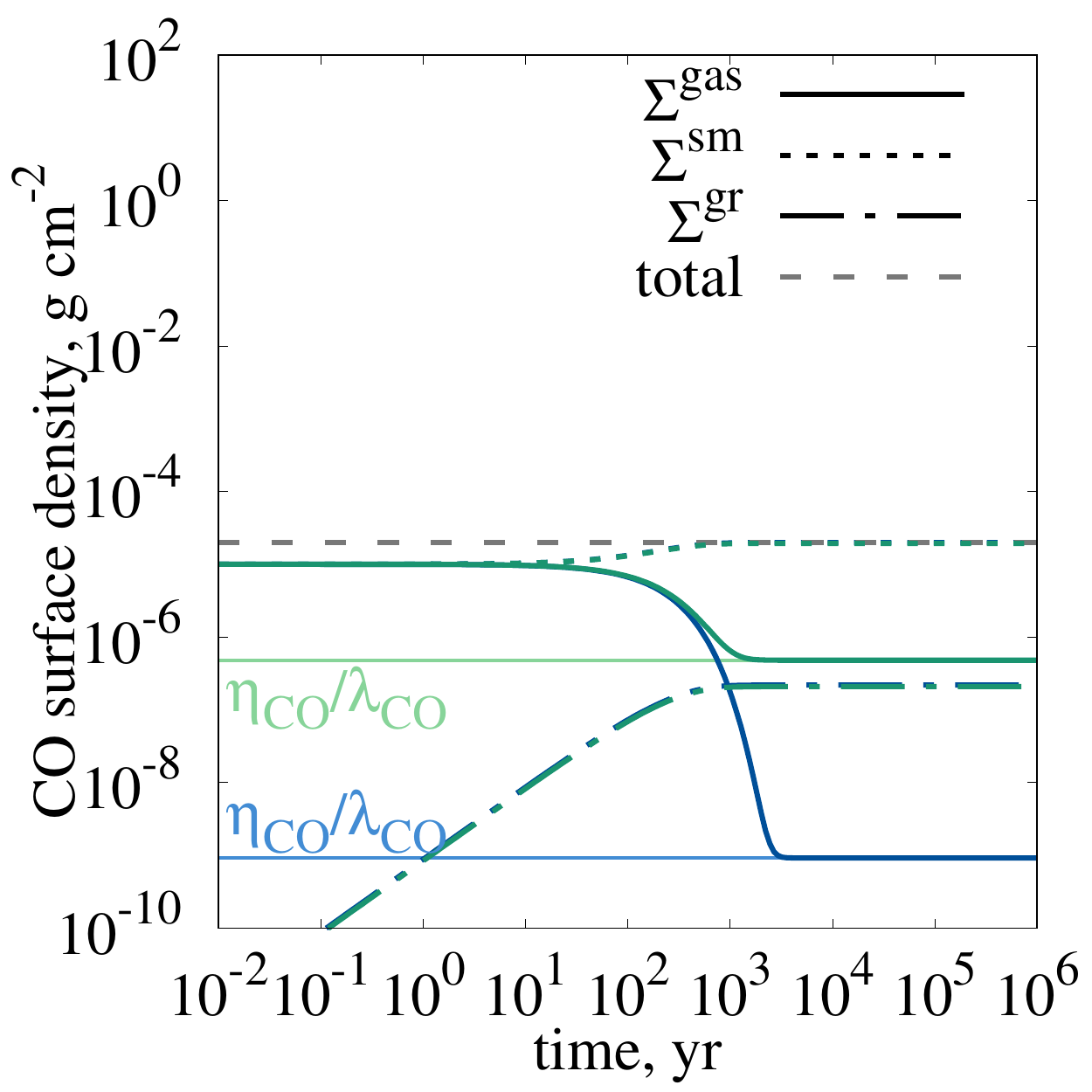} 
\includegraphics[width=0.247\columnwidth]{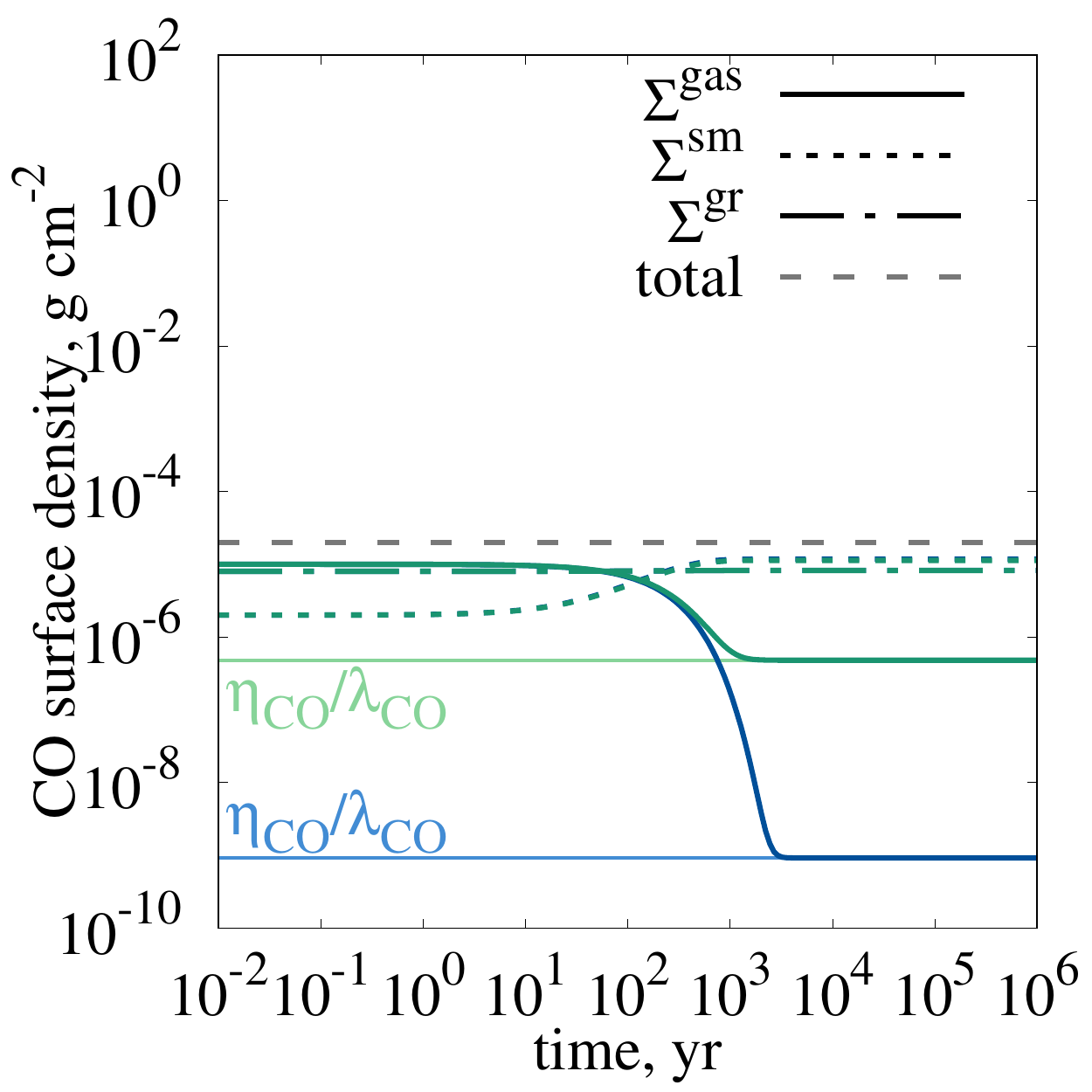} 
\includegraphics[width=0.247\columnwidth]{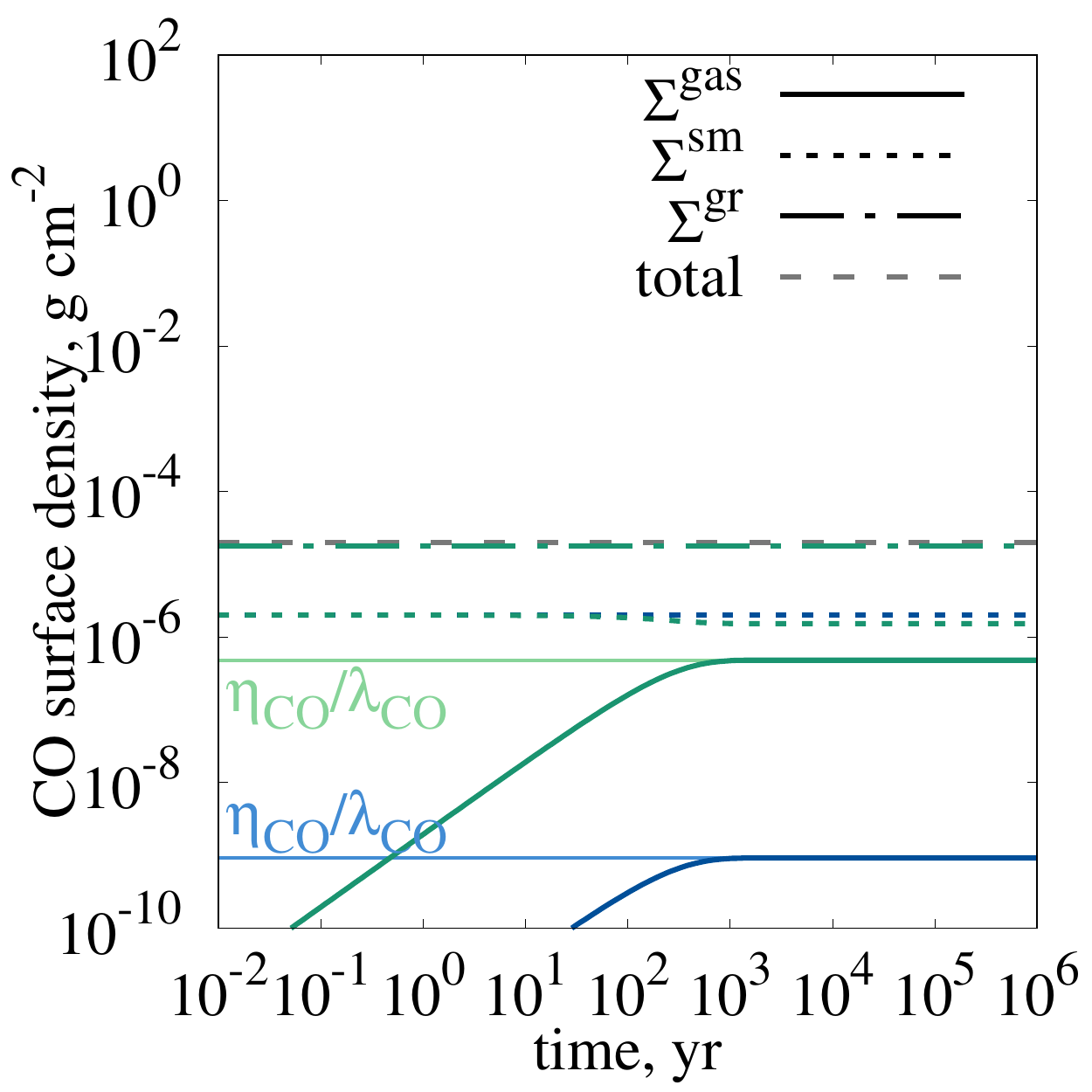}
 \caption{Sample calculations for models with (green) and without (blue) photodesorption. The scale height in the models is 10\,au, temperature is 15\,K, surface densities of small and grown dust are equal to $10^{-4}$\,g\,cm$^{-2}$.}
  \label{fig:tests_ph}
\end{figure*}

Figure~\ref{fig:tests_ph} shows test solutions for CO in single-point model with and without photodesorption. Different initial conditions are presented. Photodesorption is only important in the regions with low temperature and low dust surface densities. It is responsible for depletion of ices in tenuous outer regions of disk envelope.

\section{Azimuthal variations in gas and dust radial velocities}
\label{sec:radvels}

\begin{figure}
\centering
\includegraphics[width=0.5\columnwidth]{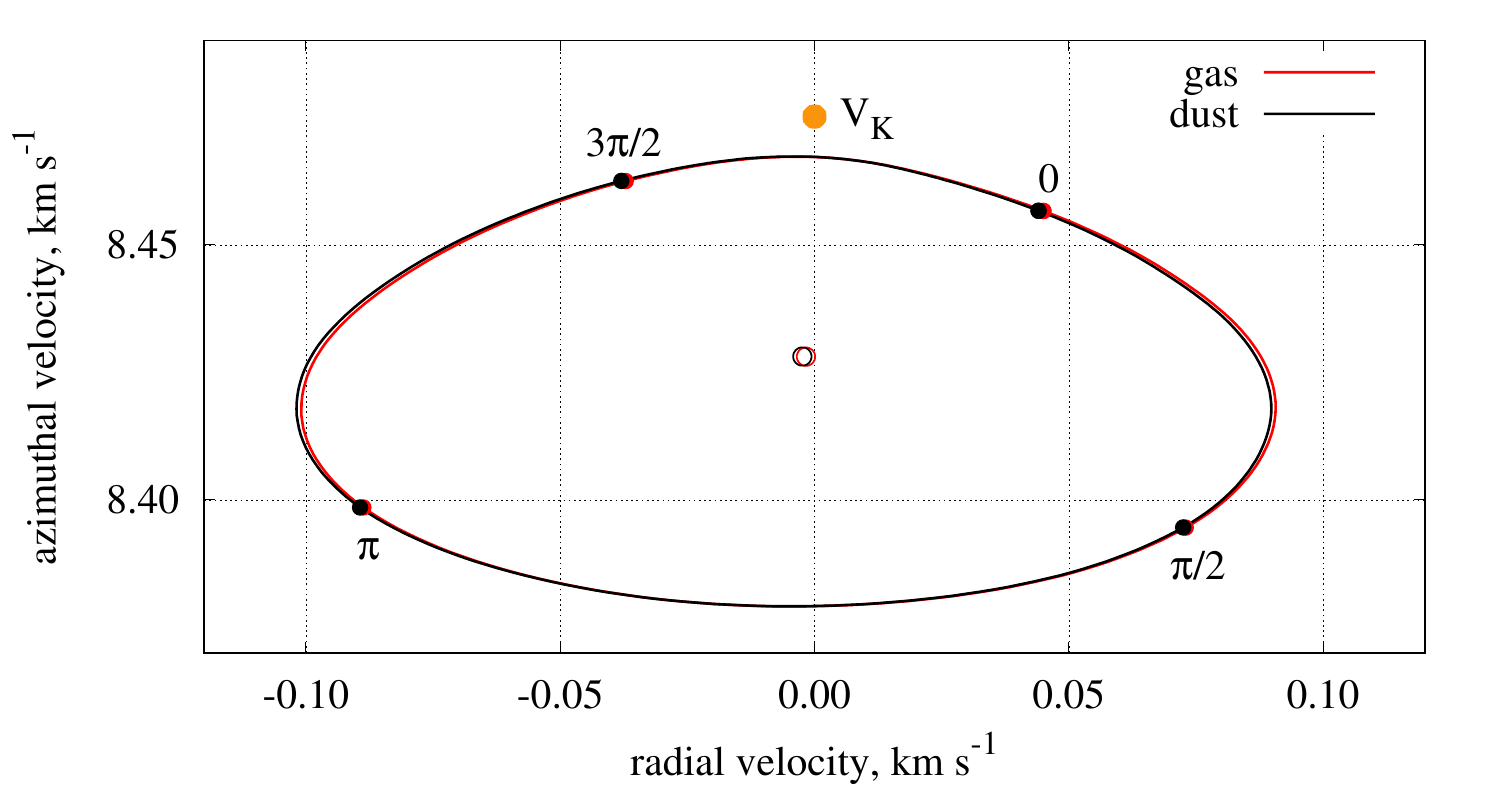}
 \caption{Radial and azimuthal components of the gas and dust velocities for different azimuthal angles $\phi$ at $r=5.3$\,au and  at 300\,kyr in Model~1 ($\alpha=10^{-2}$). Filled circles at the corresponding curves mark several azimuthal angles counted counterclockwise from the positive direction of the $x$-axis. Red and black open circles show the corresponding azimuthally averaged velocities. The orange circle marks Keplerian velocity at this distance taking all mass inside 5.3\,au into account, including the star, the sink cell and the disk.}
  \label{fig:vel}
\end{figure}

One of the features of our 2D model is the presence of azimuthal variations in the gas and dust radial velocities at a fixed radial distance. These variations are important for the dynamics of dust, gas, and volatiles, and act as effective diffusion allowing volatiles to move across their snowlines.

Figure~\ref{fig:vel} shows the gas and dust velocities in  Model~1 at $t=300$\,kyr along the circumference of a fixed radius ($r=5.3$\,au). This radial distance corresponds to the position just inside the thermal water snowline, where $v_{\rm frag}$ changes sharply between the values corresponding to bare and icy grains. The azimuthally averaged radial velocity is negative for both gas and dust (meaning the motion toward the star), but for almost half of the presented azimuthal points the radial velocity is positive. The difference between gas and dust velocities is small because grown dust is fragmented efficiently at this distance, and drifts very slowly relative to gas.

The negative values of the azimuthally averaged velocities seen in Figure~\ref{fig:vel} are characteristic of most of the disk, as both gas and dust accrete to the star. The amplitude of the azimuthal variations is more than an order of magnitude higher than the azimuthally averaged values. We also show the local Keplerian velocity in Figure~\ref{fig:vel} to demonstrate that the associated variations in azimuthal velocities are comparable with the deviation from the Keplerian rotation. Such velocity variations appear at all radial distances and time instances, with a varying amplitude and pattern. The amplitude of the variations generally decreases with the distance and with time.
\end{document}